\documentclass{emulateapj}
\newcommand\Mj{\mbox{$M_{\rm Jup}$}}

\newcommand\Msun{\mbox{$M_\sun$}}

\newcommand\Lsun{\mbox{$L_\sun$}}
\newcommand\masyr{{\rm mas~yr\mbox{$^{-1}$}}}
\newcommand\maspix{mas~pix\mbox{$^{-1}$}}

\newcommand\Teff{\mbox{$T_{\rm eff}$}}

\begin{document}

\shortauthors{Metchev \& Hillenbrand}
\shorttitle{Adaptive Optics Survey of Young Solar Analogs}

%\title{The Frequency of Wide Sub-Stellar Companions to Young Solar Analogs}
%\title{Sub-Stellar and Stellar Companions to Young Solar-Mass Stars: 
\title{The Palomar/Keck Adaptive Optics Survey of Young Solar Analogs:
Evidence for a Universal Companion Mass Function}

\author{Stanimir A.\ Metchev\footnote{Author's current address is: Department of Physics \& Astronomy, State University of New York, Stony Brook, New York 11794--3800; metchev@astro.sunysb.edu}}
\affil{Department of Physics and Astronomy, 430 Portola Plaza, University of California, Los 
Angeles, California 90095--1547}
\email{metchev@astro.ucla.edu}
\and
\author{Lynne A.\ Hillenbrand}
\affil{Department of Physics, Mathematics \& Astronomy, MC 105--24, California Institute of Technology, Pasadena, California 91125}

\begin{abstract}
We present results from an adaptive optics survey for substellar and stellar companions to Sun-like stars.  The survey targeted 266 F5--K5 stars in the 3~Myr to 3~Gyr age range with distances of 10--190~pc.  Results from the survey include the discovery of two brown dwarf companions (HD~49197B and HD~203030B), 24
 new stellar binaries, and a triple system.  We infer that the frequency of 0.012--0.072~$\Msun$ brown dwarfs in 28--1590~AU orbits around young solar analogs is $3.2^{+3.1}_{-2.7}$\% (2$\sigma$ limits).  The result demonstrates that the deficiency of substellar companions at wide orbital separations from Sun-like stars is less pronounced than in the radial velocity ``brown dwarf desert.'' 
We infer that the mass distribution of companions in 28--1590~AU orbits around solar-mass stars follows a continuous $dN/dM_2 \propto M_2^{-0.4}$ 
relation over the 0.01--1.0~\Msun\ secondary mass range.  While this functional form is similar to the that for $<$0.1~\Msun\ isolated objects, over the entire 0.01--1.0~\Msun\ range the mass functions of companions and of isolated objects differ significantly.  Based on this conclusion and on similar results from other direct imaging and radial velocity companion surveys in the literature, we argue that the companion mass function follows the same universal form over the entire range between 0--1590~AU in orbital semi-major axis and $\approx$0.01--20~\Msun\ in companion mass.  In this context, the relative dearth of substellar versus stellar secondaries at {\it all} orbital separations arises naturally from the inferred form of the companion mass function.
\end{abstract}

\keywords{stars: binaries: visual---stars: imaging---stars: low-mass, brown dwarfs---stars: mass function}

\section{INTRODUCTION}

The properties of brown dwarf companions to stars are important for understanding the substellar companion mass function (CMF), the formation of brown dwarfs, and the formation and evolution of low-mass ratio binary systems.  Widely-separated brown dwarf companions, in particular, are an important benchmark for studying the properties of substellar objects because of their accessibility to direct spectroscopic characterization and their relative ease of age-dating---from assumed co-evality with their host stars.    

However, brown dwarf companions have been an elusive target for direct imaging.
The main challenge has been the need to attain sufficient imaging contrast to detect secondaries that are $>$10$^3$ fainter than their host stars at angular separations spanning solar system-like scales ($<$40~AU~$=0\farcs4$ at 100~pc).  

The problem is alleviated at young ages when brown dwarfs are brighter.  In addition, nearby stars offer an additional advantage because the relevant angular scales are correspondingly wider and more accessible to direct imaging.  Young nearby stars are thus the preferred targets for substellar companion searches through direct imaging.

Nevertheless, early surveys for substellar companions, performed with seeing-limited or first-generation high-contrast imaging technology \citep{oppenheimer_etal01, hinz_etal02, mccarthy_zuckerman04} had very low detection rates, $\lesssim0.5$\%.  This low brown dwarf companion detection rate was similar to that inferred from precision radial velocity surveys \citep[$<0.5\%$ over 0--3~AU;][]{marcy_butler00}, and prompted \citet{mccarthy_zuckerman04} to conclude that the so-called ``brown dwarf desert'' extends far beyond the orbital separations probed by radial velocity surveys, out to at least $\approx$1200~AU.

Over the past few years, advances in adaptive optics (AO) technology and high-contrast imaging methods have improved the chances for the direct imaging of substellar secondaries.  Modern AO systems, with $>$200 corrective elements spread across the beam of a 5--10~m telescope, are able to deliver high order rectification ($<$250~nm r.m.s.\ residual error) of wavefronts perturbed by Earth's turbulent atmosphere
at up to kHz rates.  In addition, our empirical appreciation of the local young stellar population has improved over the past decade, as demonstrated by the recent discoveries of a large number of young ($<500$~Myr) stellar associations within 200~pc from the Sun \citep[e.g.,][and references therein]{kastner_etal97, mamajek_etal99, zuckerman_webb00, zuckerman_etal01a, montes_etal01a, zuckerman_song04}.  These have allowed us to select more suitable targets for direct imaging searches for substellar companions.  

Several recent direct imaging surveys of nearby young stellar associations conducted on high-order AO-equipped telescopes (\citealt{neuhauser_guenther04}, 25 A--M stars; \citealt{chauvin_etal05a, chauvin_etal05b}, 50 A--M stars) or with the {\it Hubble Space Telescope} ({\it HST}; \citealt{lowrance_etal05}, 45 A--M stars) have enjoyed higher detection rates (2--4~\%) than the first generation of surveys.  In addition, at very wide ($>$1000~AU) separations, where the detection of brown dwarf companions to solar-neighborhood stars is not hindered by contrast, \citet{gizis_etal01} have found that the frequency of substellar companions to F--K dwarfs is fully consistent with that of stellar companions to G dwarfs \citep{duquennoy_mayor91}.  Thus, while the radial velocity ``brown dwarf desert'' remains nearly void within 3~AU even after the discovery of numerous extra-solar planets over the past decade, brown dwarf secondaries at $>$100--1000~AU separations seem to not be as rare.  

The precise frequency of substellar companions in direct imaging surveys remains controversial. Several highly sensitive surveys performed with the {\it HST} (\citealt{schroeder_etal00}, 23 A--M stars; \citealt{brandner_etal00}, 28 G--M stars; \citealt{luhman_etal05e}, 150 B--M stars) and with high-order AO (\citealt{masciadri_etal05}, 28 G--M stars; \citealt{biller_etal07}, 54 A--M stars) have reported null detections of substellar companions.  Given the low (few per cent) detection rate of substellar companions in direct imaging surveys, it is now clear that, with $<50$ targets per sample, some of these surveys were too small to expect to detect even a single brown dwarf companion.  However, the non-detection of substellar secondaries in two largest surveys \citep{luhman_etal05e, biller_etal07} is potentially significant.

Given current understanding of the importance of stellar mass for (1) stellar multiplicity rates \citep[see review in][]{sterzik_durisen04} and (2) binary mass ratio distributions \citep[see review in][]{burgasser_etal07}, it is imperative that any study of the substellar companion frequency is considered in the context of the mass distribution of primary stars in the sample.  Indeed, a large survey sample comprising primaries with identical masses is ideal.

The problem of the brown dwarf companion frequency is perhaps most comprehensively dealt with in the context of solar mass primaries.  For these 
a uniquely large body of stellar and substellar multiplicity data exist on all orbital scales.  On one hand, the exhaustive
spectroscopic and imaging study of G dwarf multiples by \citet{duquennoy_mayor91} provides an important anchor to the properties of 0.1--1.0~\Msun\ stellar companions to Sun-like stars.  On the other hand, the results from more than a decade of precision radial velocity surveys for planets around G and K stars allow a comparison with the planetary-mass end of the substellar companion mass range.  

A large uniform sample of young Sun-like stars has been compiled by the Formation and Evolution of Planetary Systems (FEPS) {\it Spitzer} Legacy team.  The purpose of the FEPS Legacy campaign with {\sl Spitzer} was to study circumstellar disk evolution in the mid-IR.  However, the sample is also well-suited for a high-contrast imaging survey for substellar companions.  Seventy percent of the FEPS stars are younger than $\sim$500~Myr, and all are within 200~pc.

As an auxiliary component to the FEPS program, we imaged most %(95\%) %(228 stars) 
of the northern FEPS sample with the high-order AO systems on the Palomar 5~m and the Keck 10~m telescopes.  We further expanded our AO survey by observing several dozen additional nearby and mostly young solar analogs.  Preliminary results from the project were published in \citet{metchev_hillenbrand04} and in \citet{metchev_hillenbrand06}, including the discoveries of two brown dwarf companions: HD~49197B and HD~203030B.  The survey has now been completed, and no further brown dwarf companions have been found.  The results were analyzed in \citet{metchev06}.  Here we present the AO survey in its entirety and focus on the statistical interpretation of the data.  

The paper is organized as follows.  A full description of the survey sample is given in \S~\ref{sec_target_sample}.  The Palomar and Keck AO observing campaigns and the data reduction and calibration techniques are described in \S~\ref{sec_obsredcal}.  The candidate companion detection approach and the survey detection limits are discussed in \S~\ref{sec_detlim}.  The various methods used for bona fide companion confirmation are presented in \S~\ref{sec_candidates}.  Section~\ref{sec_results} summarizes the results from our survey, including all of the newly-discovered and confirmed substellar and stellar secondaries.   Section~\ref{sec_incompl_bias} contains a brief discussion of the various sources of incompleteness and a full discussion of the biases in the survey.  (A full-fledged incompleteness analysis is presented in the Appendix.) 
%(\S~\ref{app}).  
In \S~\ref{sec_anal} we estimate the frequency of wide substellar companions to young solar analogs, and present evidence for trends in the companion mass and companion frequency with semi-major axis and primary mass.  In \S~\ref{sec_discussion} we consider the results of the current investigation in the broader context of stellar multiplicity, and suggest the existence of a universal CMF.  Section~\ref{sec_conclusion} summarizes the findings from our study.

\section{TARGET SAMPLE}
\label{sec_target_sample}

The main criteria used for selecting stars for the survey were Sun-like mass, youth, proximity, and visibility from the Northern hemisphere.  In this Section we describe how they were applied to generate our AO survey sample.

\subsection{Selection} 
\label{sec_selcrit}

The selection of the AO survey sample was largely based on the target selection criteria employed in the construction of the FEPS program sample \citep{meyer_etal06}.  The FEPS selection criteria will not be reproduced here.  The final FEPS target list comprises 328 F5--K5 stars within 200~pc distributed uniformly in logarithmic age intervals between 3~Myr and 3~Gyr.  Approximately a third of
these are members of open clusters and stellar associations, and the remainder are field stars.  We observed 228 of the 240 FEPS stars north of $\delta=-30\degr$ with AO at Palomar or Keck.  

A further 38 solar analogs were added to the AO survey toward the end of the first epoch of observations to bolster the sample size, mirroring FEPS target-selection policy.  The additional stars were selected from three sources: (1) the broader compilation of FEPS candidate targets, including stars that had been eliminated from the final FEPS sample based on infrared background or age redundancy considerations; (2) the compilations of nearby young stars by \citet{montes_etal01a} and \citet{wichmann_etal03}; and (3) our own Palomar echelle survey of nearby stars \citep{white_etal07}.
The final set of 266 targets in our AO sample has spectral type and age distributions similar to those of the FEPS sample. 

\subsubsection{Deep and Shallow Samples}

To optimize sensitivity to substellar companions, we chose to observe a portion of the youngest and nearest AO sample stars with deep coronagraphic exposures.  We applied the following additional guidelines to select stars for the deep coronagraphic sub-sample:
\begin{enumerate}
\item stellar activity and lithium levels indicating ages less than 500~Myr;
\item no $\Delta K_S < 4$ objects between 0$\farcs$8 and 13$\farcs$0, as determined from real-time flux ratio measurements during survey observations; \label{crit_dK}
\item nearby stars were given priority over more distant stars;
\item isolated stars, not belonging to one of the young open clusters or stellar assocciations, were
given priority for deep observations. 
\end{enumerate}
The first criterion was motivated by the fact that substellar
companions should be intrinsically brightest at the youngest ages.
The second constraint was aimed at avoiding the loss
of sensitivity to faint objects over a large portion of the detector field of view (FOV) because of the presence of a bright neighboring star.\footnote{Following more accurate post-reduction photometry, a 3$\farcs$1 companion to one of the stars in our deep sample, HD~31950, was found to be only $\Delta K_S=3.70$~mag fainter (Table~\ref{tab_deep_companions}).  Although this companion violates criterion \ref{crit_dK}, we have chosen to keep HD~31950 as a member of the deep sample.}
Binaries with separations $\leq0.8\arcsec$ had both their components
sufficiently well-covered by the 1$\arcsec$ coronagraphs in the Palomar and Keck AO cameras that they were allowed in the deep sample.  The motivation for the third constraint was to
optimize sensitivity to substellar companions at the smallest
physical separations.  The last criterion was applied to
avoid duplication with previous sensitive high-angular resolution 
studies of open clusters: \citet[][the Pleiades, AO]{bouvier_etal97}, \citet[Upper Scorpius;][speckle]{kohler_etal00}, and \citet[][$\alpha$~Persei, speckle]{patience_etal02}.  

Based on the additional criteria outlined above, 84 of the 228 stars selected from the final FEPS sample and 16 of the 38 additional targets were included in our deep sample.  The deep sub-sample thus consists of 100 F5--K5 stars with ages less than 500~Myr.  

All remaining stars were observed primarily in short sequences of non-coronagraphic
images to establish stellar multiplicity.  These will be referred to as
the ``shallow'' sample.  The shallow sample includes 11 stars older than 500~Myr that were also observed with long coronagraphic exposures: 2 Hyades
($\sim$600~Myr) members and 9 other stars whose subsequent age-dating 
showed that they
were older than originally estimated.  Although these 11 stars were observed coronagraphically, for the purpose of limiting our deep sample only to the observations with the highest sensitivity to substellar mass, they are not considered as part of the deep sample.

The deep and shallow sample stars and their characteristics are listed in Tables~\ref{tab_deep_sample} and \ref{tab_shallow_sample}.  Median age, distance, and spectral-type statistics for the deep,
shallow, and complete (deep+shallow) samples are given in Table~\ref{tab_sample_stats}.  

\subsection{Stellar Properties} 
\label{sec_stellar_props}

\begin{figure}
%\plottwo{teff_hist.eps}{mass_hist_bincor.eps}
\plottwo{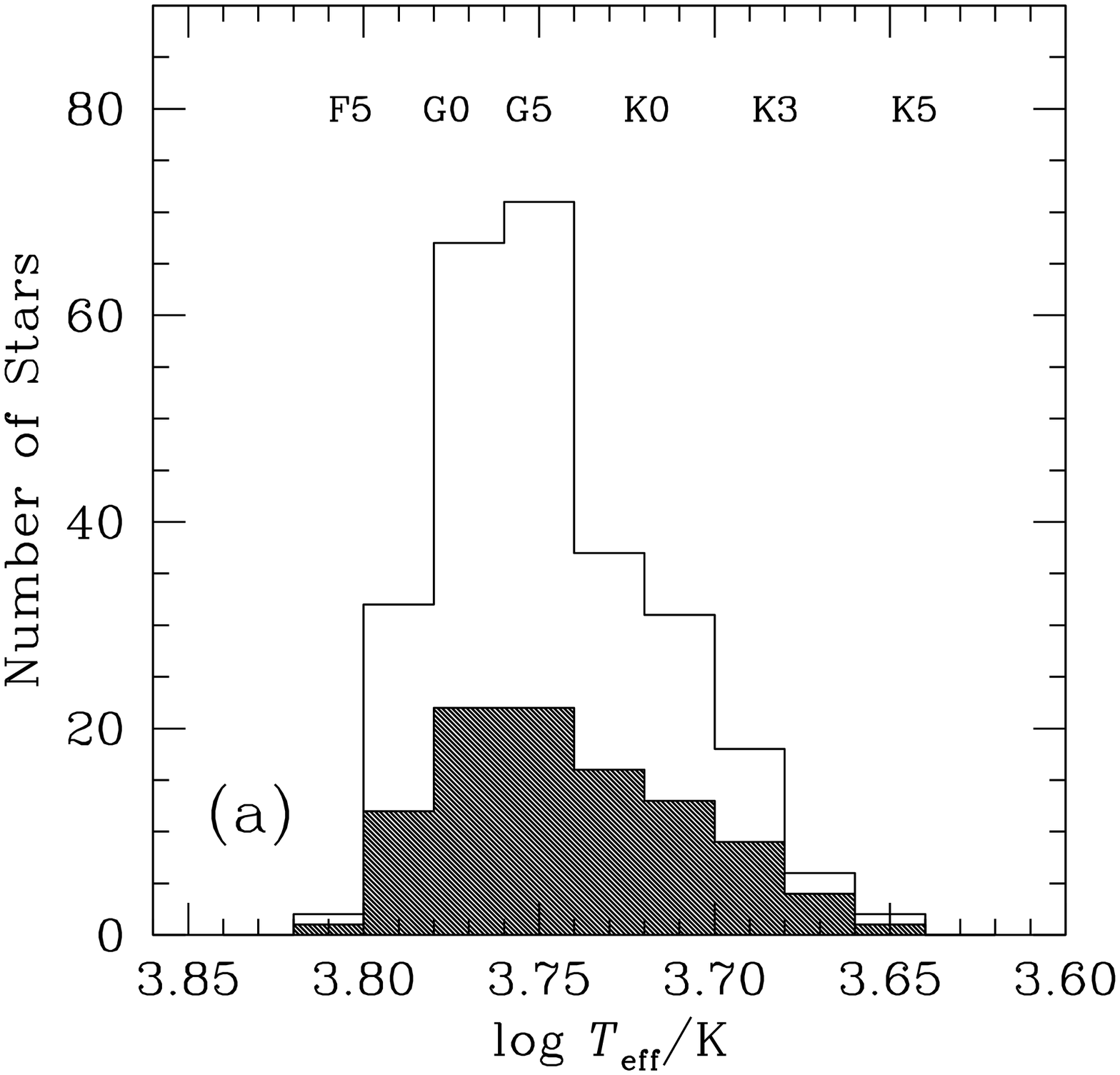}{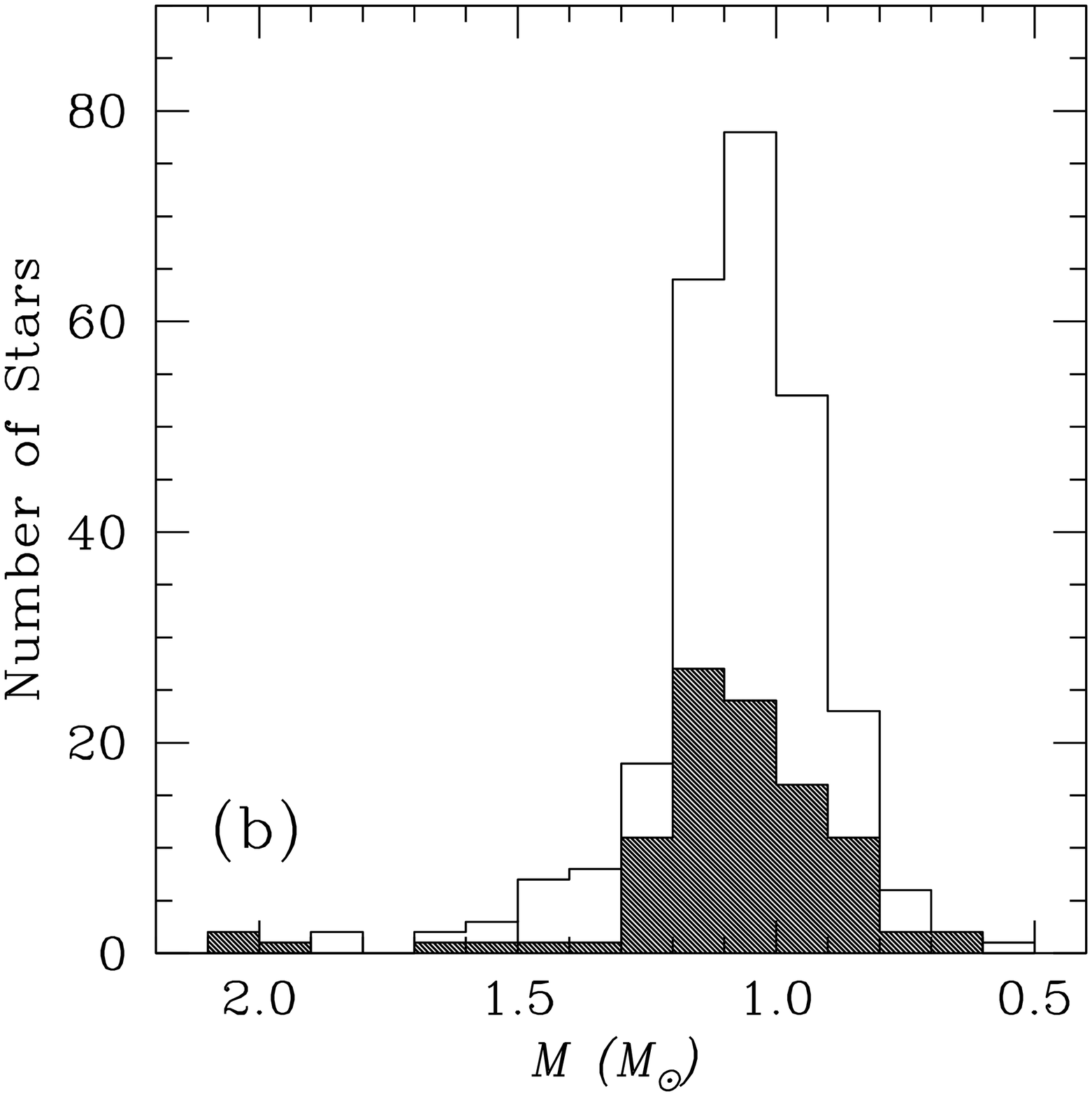}
\figcaption{Distribution of the sample stars as a function of
effective temperature (a) and mass (b).  The non-shaded histograms refer to
the entire sample of 266 stars, whereas the shaded histograms refer to the
deep and young sub-sample of 100 stars.  All stars fall in the F5--K5 range
of spectral types and the majority are between 0.7~\Msun\ and 1.3~\Msun.  
\label{fig_sptype_mass_hist}}
\end{figure}

Our sample stars are near-solar (G2~V) analogs, ranging in spectral type between F5 and K5 (6300~K~$> T_{\rm eff} > 4400$~K) and, depending on stellar age,
between IV and V in luminosity class ($3.4 < \log g \leq 4.5$ in cgs units).  The corresponding mass range,
based on dynamical mass estimates in binary systems and on stellar
thermodynamic models \citep{baraffe_etal98, dantona_mazzitelli94} is approximately 0.7--1.3~$M_\sun$, following the design of the FEPS sample \citep{meyer_etal06}.  For greater detail in the assignment of spectral types, effective temperatures, and surface gravities to FEPS sample stars we refer the reader to \citet{carpenter_etal08}.  Histograms of the distribution of stellar effective temperatures and masses of all stars in our AO survey sample are shown in Figure~\ref{fig_sptype_mass_hist}.  

\begin{figure}
%\plotone{age_hist.eps}
\plotone{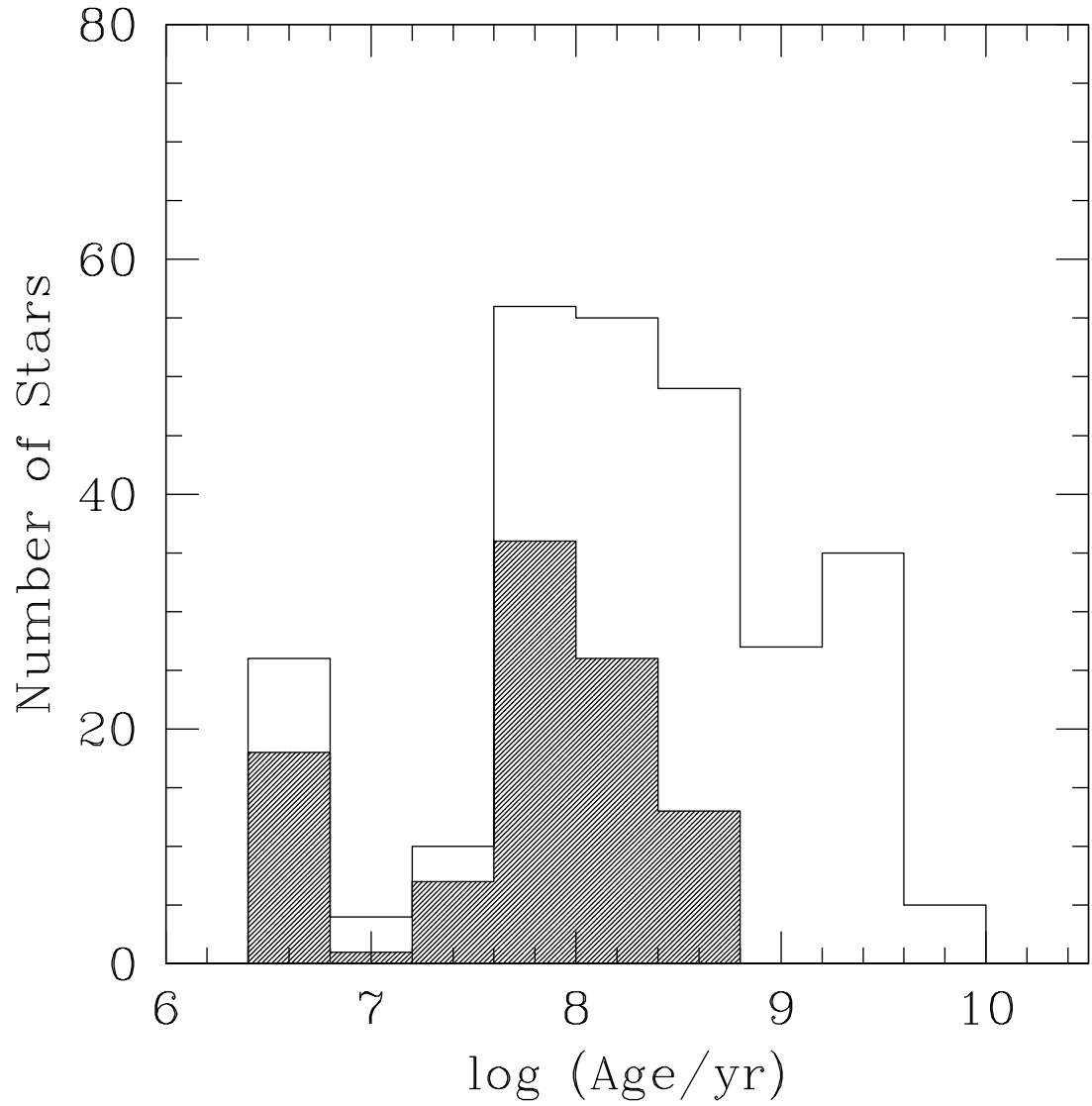}
\figcaption{Age distributions of the complete survey sample (non-shaded
histogram) and of the deep sub-sample (shaded histogram).
\label{fig_age_hist}}
\end{figure}

Seventy-nine of our sample stars are members of known young stellar associations: Upper Scorpius, $\alpha$~Persei, the Pleiades, and the Hyades.  For these we have adopted ensemble ages from the literature: 
%$5\pm1$
5~Myr for the Upper Scorpius OB association \citep{preibisch_etal02}, 
%$80\pm10$
80~Myr for $\alpha$~Persei \citep{ventura_etal98},
%$120\pm10$
120~Myr for the Pleiades \citep{ventura_etal98}, and
%$600\pm50$
600~Myr for the Hyades \citep{perryman_etal98}.  Sample stars that %overlap with the FEPS program and 
do not belong to any known associations were age-dated following one of two approaches: (1) based on the strength of the chromospheric \ion{Ca}{2} H and K (3968\AA\ and 3933\AA) line emission for $>30$~Myr-old stars, and using the recent activity-age relation of \citet{mamajek_hillenbrand08}, or (2) isochrone fitting for $<30$~Myr-old pre-main sequence stars using the tracks of \citet{baraffe_etal98}.  Where data from high-dispersion optical spectra were previously available \citep{strassmeier_etal00, white_etal07}, these were also reviewed for the strength of the lithium 6708~\AA\ absorption line to put additional constraints on the ages.  
All chromospheric and isochronal ages were assumed uncertain to within a factor of two.  
%A more detailed discussion of the age-dating of the FEPS sample is presented in \citet{hillenbrand_etal08}.  
Ages for a remaining set of 20 stars not present in the extended FEPS sample were taken from the literature \citep{barradoynavascues_etal97, gutierrez_etal99, montes_etal01a, wichmann_etal03, nordstrom_etal04}.
Histograms of the age distribution of the complete survey sample and of
the deep sub-sample are shown in Figure~\ref{fig_age_hist}.

Distances to 166 sample stars with individual {\it Hipparcos} parallaxes were taken from the
{\it Hipparcos} catalog \citep{perryman_etal97}.  For a further 55 known members of 
young open clusters and OB associations, we adopted the corresponding mean cluster distance,
calculated from a combination of trigonometric, orbital, secular, and cluster
parallaxes in the literature, as inferred from {\it Hipparcos} and Tycho-2 \citep{hog_etal00} 
astrometry, long-baseline interferometry, or high-resolution spectroscopy.
The adopted distances for open cluster members were:
$133\pm6$~pc for stars in the Pleiades (a weighted mean of the distances to
seven members presented in \citealp{pan_etal04}, 
\citealp{munari_etal04}, \citealp{zwahlen_etal04}, 
\citealp{southworth_etal05}, and assuming $\sim1\degr$ cluster angular extent
from \citealp{adams_etal01}), and 
$190\pm11$~pc for stars in $\alpha$~Persei \citep[][assuming 1$\degr$ cluster
radius]{vanleeuwen99}.  For stars belonging to the Upper Scorpius association  
we adopted $145\pm40$~pc \citep{dezeeuw_etal99, mamajek_etal02}.  All of
these distances agree with estimates from main-sequence fitting for the
corresponding clusters.  For 18 
%was 37 
more stars we adopted secular parallaxes from
\citet{mamajek_etal02} and \citet{mamajek04, mamajek07}.
Finally, for 27
%was six
remaining $>$30~Myr-old stars we obtained approximate distances
%by comparing their $B-V$ colors, chromospheric ages, and absolute magnitudes to the predictions of the stellar evolution models of \citet{baraffe_etal98}.  Since we did not use isochronal fitting to infer the ages of these stars, we used the evolutionary tracks to infer their distances instead.  
based on a combination of moving group association, secular parallax, and spectroscopic parallax, with care to avoid redundancy in distance and age derivation.
The errors on the distances in these cases were conservatively assumed to be 25\%--50\%.  More refined distance and age estimates for these stars will be included in a future publication from the FEPS program.

Accurate proper motions for the sample stars are essential in identifying bona fide companions through multi-epoch astrometry.  Proper motions for the 166 stars with individual {\it Hipparcos} parallaxes were
taken from the {\it Hipparcos} database.  For the remaining 100 stars proper
motions were adopted from The Second U.S.\ Naval
Observatory CCD Astrograph Catalog \citep[UCAC2;][]{zacharias_etal04} and
from the Tycho--2 Catalog \citep{hog_etal00}.  The three catalogs provided
similar astrometric accuracy ($\pm1.0$~\masyr) for the sample stars, although
the UCAC2 and Tycho--2 catalogs went deeper.  

\begin{figure}
%\plottwo{dist_hist.eps}{pm_hist.eps}
\plottwo{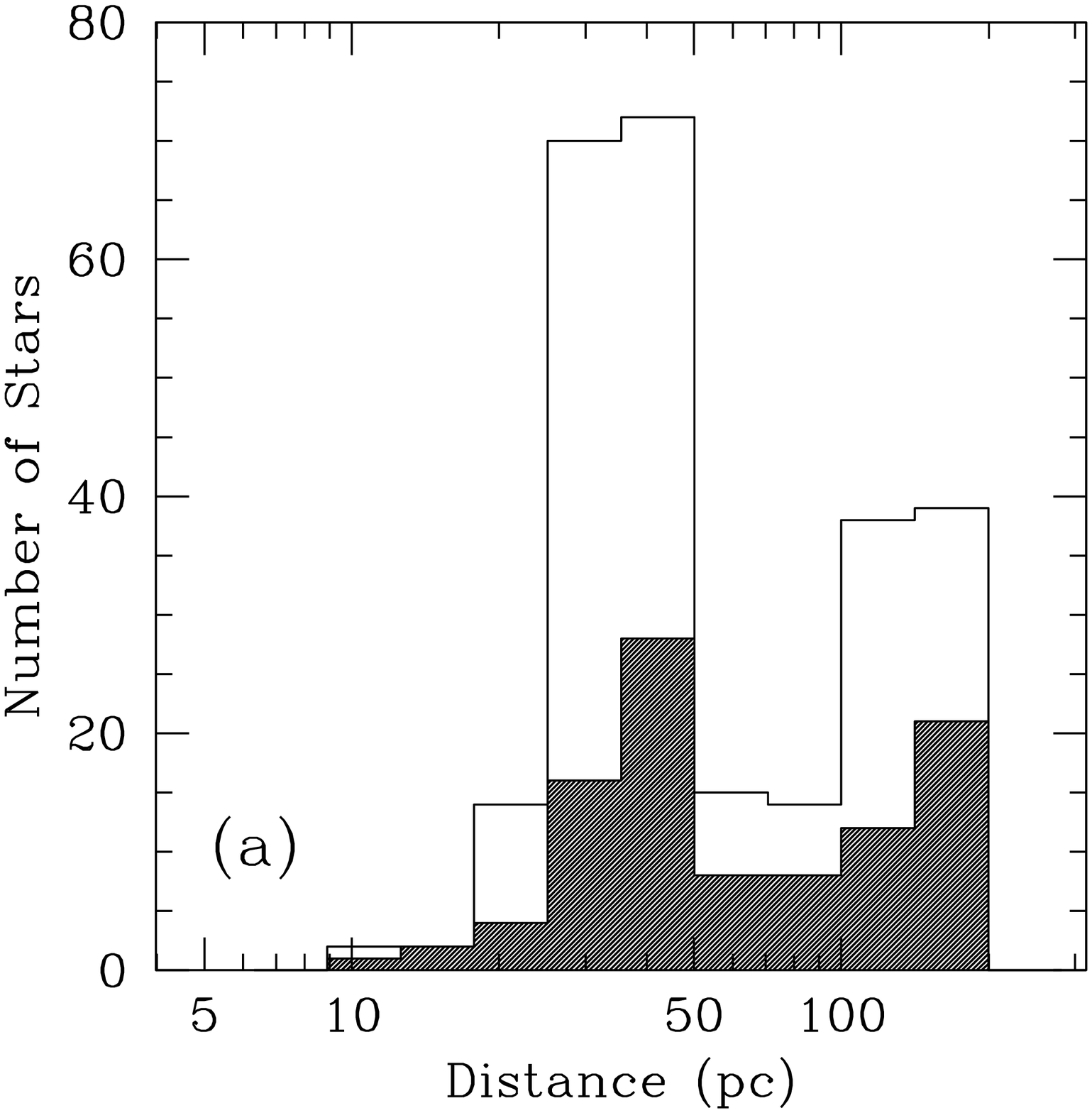}{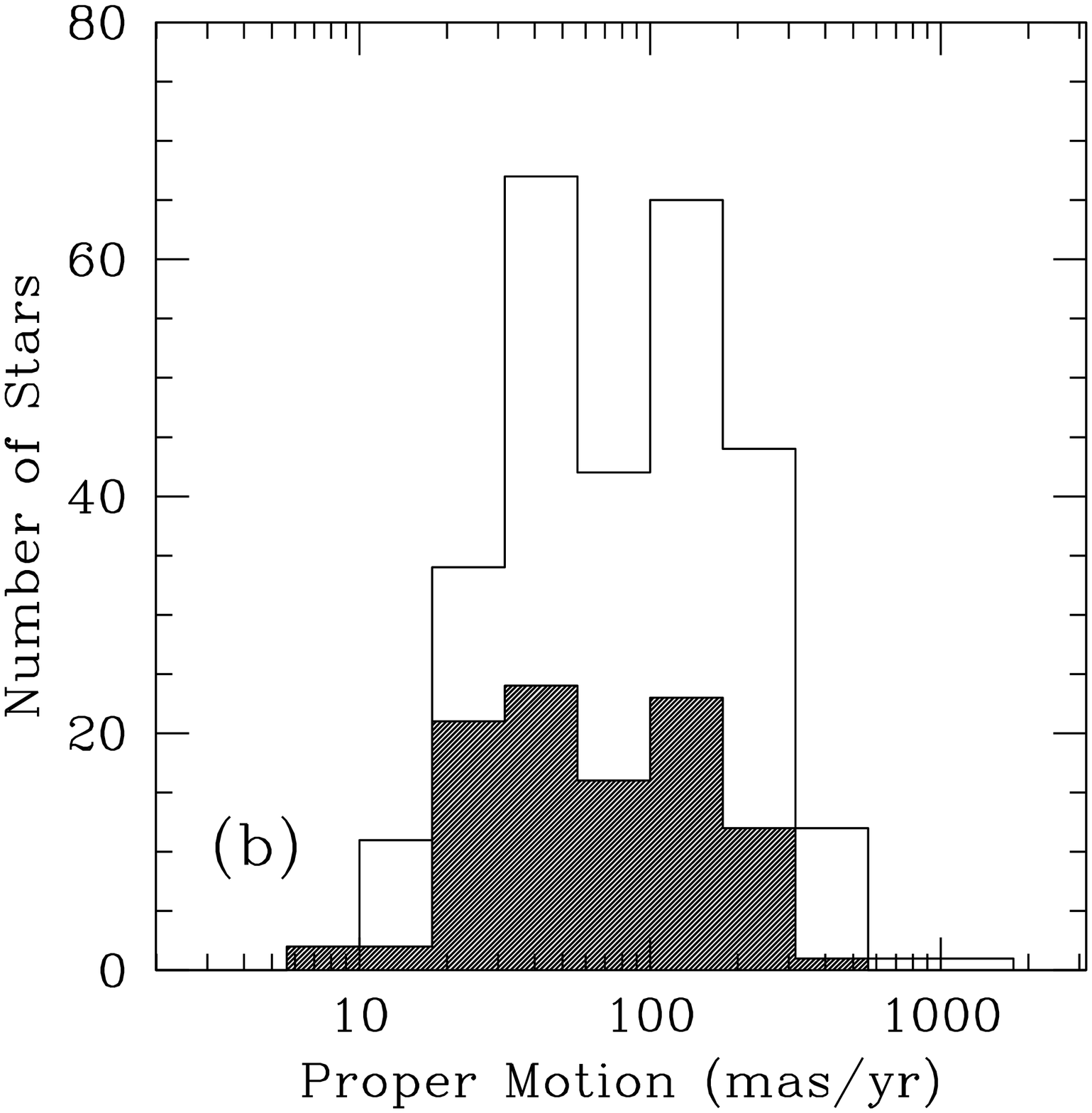}
\figcaption{Heliocentric distance (a) and proper motion (b) distributions of surveyed
stars in the complete sample (non-shaded histograms) and in the deep
sub-sample (shaded histograms).
\label{fig_dist_pm_hist}}
\end{figure}

Figure~\ref{fig_dist_pm_hist} presents histograms
of the heliocentric distances (panel a) and total proper motions  ($\sqrt{(\mu_\alpha \cos \delta)^2 + \mu_\delta^2}$; panel b) of the stars in the complete sample and in the deep  
sub-sample.  The bi-modal distribution of the distances is a combined effect of
the large heliocentric distances (130--190~pc) of the youngest (3--120~Myr)
stars in the sample, and of the preference given to closer systems at older
ages.

\section{OBSERVATIONS, DATA REDUCTION, AND CALIBRATION}
\label{sec_obsredcal}

\subsection{Observing Strategy} 
\label{sec_observations}

A complete description of the observing strategy of our AO survey was given in \citet{metchev_hillenbrand04} and in \citet{metchev06}.  Here we briefly review the approach and summarize the survey observations.

The full sample of 266 stars was observed in the near-IR with AO at the Palomar and Keck~II telescopes on 47 clear nights over the course of 3 years: between 2002 January 31 and 2005 January 24.  Additional astrometric follow-up was obtained during 2006 and 2007 in a few individual cases.  

We opted to conduct the entire survey in the $K_S$ band to take advantage of the much better AO performance at $>$2\micron.  Although cool T-type brown dwarfs \citep[$\Teff\lesssim1400$~K;][]{golimowski_etal04, vrba_etal04} are faintest at $K$ band in the near-IR, warmer (potentially younger) L-type brown dwarfs are brightest at $K$.  Thus, given superior imaging contrast and the relative youth of our deep sample, the 2$\micron$ region was seen as the best choice for optimizing sensitivity to close-in young substellar companions.

The majority of the science targets were observed first at Palomar.  Only seven of the targets (all belonging to the deep sample) were observed initially and only at Keck. The Palomar campaign was conducted with the PALAO system
\citep{troy_etal00} and the PHARO near-IR camera
\citep{hayward_etal01} in its 25~\maspix\ mode, providing a
$25.6\arcsec\times25.6\arcsec$ FOV.  At Keck, we used the
facility AO system \citep{wizinowich_etal00} on Keck~II and the NIRC2
near-IR camera %(Matthews et al., in preparation) 
in its 40~\maspix\
mode, offering an FOV of $40.6\arcsec \times 40.6\arcsec$.  To improve overall
sensitivity and contrast the 100 targets in the deep sample were observed coronagraphically with the opaque 0$\farcs$97-diameter occulting spot in PHARO and the partially transmissive 1$\farcs$0-diameter occulting spot in NIRC2.  All of the sample stars were sufficiently bright to allow use of the AO systems in NGS mode, i.e., to have the wavefront
sensing performed on the primaries themselves.

\subsubsection{First-Epoch Observations at Palomar} 
\label{sec_1stepoch}

We spent 24~min of net exposure time per target during first epoch at Palomar, attaining an imaging depth of $K_S\approx19.7$~mag on average for stars in the deep sample.  The 24~min of exposure were divided in 4 sets of 6~min taken at different orientations of the telescope Cassegrain ring (CR) rotator.  The 6~min of net exposure at each CR rotator angle consisted of two sets of three one-minute on-target exposures, with three one-minute sky exposures in between.  The purpose of the CR rotation approach was to improve the quality of point-spread function (PSF) subtractions for data taken with an equatorial-mount telescope (Palomar), in a manner similar to that attained with angular differential imaging (ADI) on altitude-azimuth-mounted telescopes \citep{marois_etal06}.  Stacking images taken at different CR angles also averages out detector and sky noise, much like mosaicking dithered images. Unfortunately, in addition to being less efficient, the CR rotation approach was later found to also produce notable smearing of the PSF in the co-added de-rotated images at $\gtrsim5\arcsec$ from the star, leading to degradation both in imaging depth and in astrometric precision \citep{metchev06}.  We have since demonstrated that judicious matching of nearby science targets to use as PSFs for one another---a suitable approach for surveys of target-rich young stellar associations---enhances the contrast attainable with PALAO by 0.5--1.0~mag over 
%$0\farcs5$--3$\arcsec$ from bright stars 
the one reported here 
%(see \S~\ref{sec_byeye}) 
without incurring the overhead of CR rotations \citep{tanner_etal07}.

We used two different undersized Lyot stops to block the secondary obscuration and the secondary mirror support structure at Palomar: the ``medium'' and the
``big'' cross, obscuring 40\% and 76\% of the total telescope aperture,
respectively \citep{hayward_etal01}.  The use of an appropriately sized
Lyot stop was expected to noticeably improve the
dynamic range achievable in high-order AO coronagraphy by suppressing
light diffracted by the edge of the coronagraph
\citep{sivaramakrishnan_etal01}.  Early experiments with the PALAO/PHARO
system by \citet{oppenheimer_etal00} had suggested that the big cross
provided the best contrast in single
exposures of up to several seconds, outperforming the medium and ``standard''
(no undersizing) Lyot masks by up to 0.5~mag between 
0.5$\arcsec$--2.0$\arcsec$
from bright stars.  However, our experience from observing each star in
multiple longer exposures was that the less oversized Lyot stops
allowed better real-time monitoring of the star-coronagraph alignment
and more accurate post-processing image registration and astrometry.  With
the medium and the standard Lyot stops the position of the star behind
the coronagraph could be monitored by the location of a Poisson-like
spot within the dark area of the coronagraph: %(Fig.~\ref{fig_corona}b): 
the result of constructive interference 
of high spatial frequency light pushed by the coronagraph to the
periphery of the Lyot plane.  The big Lyot stop
likely shutters incoming starlight too aggressively to allow the formation of a
sufficiently bright Poisson spot. %(Fig.~\ref{fig_corona}a).  
Because image registration of multiple exposures
was crucial for obtaining greater overall exposure 
depth, we stopped using the big cross after March 2002.  Given the adopted technique of rotating the Cassegrain ring to four
mutually orthogonal orientations during the imaging of each star, the 
medium cross provided the best 
compromise between registration ability for the final images and
consistency with which it would obscure the telescope secondary mirror support at each
CR orientation.  At the end of our survey, only seven of the 100 stars in the deep sample had their deepest observations obtained with the big Lyot cross setup.  

In addition to the long coronagraphic $K_S$-band exposures, we also observed each deep sample target in short (1.4--10~s) unocculted exposures.  These were taken to check for stellar multiplicity within the 0$\farcs$5 coronagraph radius and to allow relative photometric calibration of the deep occulted exposures.  The short exposures were obtained at $J$, $H$, and $K_S$ bands, using a 1\% neutral density (ND) filter to prevent saturation whenever necessary.  For these observations we used a five-point dither pattern as is standard for infrared imaging.

The 166 targets in the shallow sample were observed only in short dithered $JHK_S$ exposures at Palomar to check for stellar multiplicity.  The imaging depth of the shallow survey varied greatly from star to star, depending on whether the ND filter was used or not, and was generally in the $12 < K_S < 17$~mag range.

\subsubsection{Follow-up Observations} 
\label{sec_followup}

After an examination of the initial Palomar images, target stars which contained other objects in the same image---candidate companions---were followed up with
additional imaging at later epochs to test for common proper motion between the candidate companions and the host stars.
%Confirmation was obtained by applying the common proper motion test, i.e., by requiring that the primary and the companion share the same apparent motion with respect to field stars between imaging epochs. 
Upon establishing common proper motion,  candidate companions were observed spectroscopically to confirm their physical association with the primary.  
%Any proper motion companions that displayed a spectral type consistent with equidistance with their respective primary were assumed to be physical companions.

The imaging and spectroscopic follow-up was performed at both Palomar and Keck.  
Imaging at Keck was done in $6\times1$~min coronagraphic integrations per target, with an additional $3\times1$~min spent on sky.  We used the ``inscribed circle'' NIRC2 pupil mask (90.7\% throughput) to occult the telescope mirror outer edge.  (None of the available NIRC2 pupil masks occult the Keck segment edges and the secondary support structure.) 
In most cases the 6~min-long exposures at Keck were $\approx$0.5~mag deeper ($K_S\approx20.2$~mag) than the 24~min Palomar exposures, and occasionally revealed new candidate companions.  Nearly half (48/100) of our deep sample stars were observed at Keck in addition to at Palomar, including the seven targets observed only at Keck.  Because of the marginal difference in the depths of the Keck and Palomar components of the deep survey, and for the sake of preserving the integrity of our well-defined 100-star deep sample, we analyze the Palomar and Keck AO campaigns together as a single survey.

We obtained near-IR spectra of several bona fide and candidate companions for the purposes of further confirmation of their physical association and characterization of their photospheres.  The spectroscopic observations and data reduction were described in \citet{metchev_hillenbrand04, metchev_hillenbrand06}.  Here we present spectroscopy of only one additional companion candidate, to ScoPMS~214.  The observations and data reduction for that are briefly described in \S~\ref{sec_scopms214b_spt}.

\subsection{Imaging Data Reduction} 
\label{sec_datareduction}

The imaging data reduction procedure for the survey, including flat-fielding, sky-subtraction, bad-pixel correction, image registration, and image stacking was detailed in \citet{metchev_hillenbrand04}.  We have since augmented the procedure to include a correction for the non-linear flux response of the PHARO and NIRC2 detectors.  Near-infrared detectors often have non-linear response even at small flux levels, that is important to take into account when seeking accurate photometry.  We measured the non-linearity of the PHARO and NIRC2 detectors from series of variable-length exposures of the brightly illuminated telescope dome interiors, interspersed with multiple dark frames to mitigate charge persistence effects.
The response of the PHARO camera, which employs an HgCdTe detector that does not support multiple non-destructive reads, was found to be $>$1\% non-linear beyond 10,000 counts/pix and $>$5\% non-linear beyond 45,000 counts/pix.  The InSb detector on NIRC2, which supports non-destrictive read-outs, was found to be $>$1\% non-linear beyond 3000 counts/pix/read and $>$5\% non-linear beyond 7000 counts/pix/read.
We created custom {\sc IDL} routines\footnote{The PHARO and NIRC2 detector linearization routines are available at \url{http://www.astro.caltech.edu/palomar/200inch/palao/Pharo/pharo.html}} to linearize the PHARO and NIRC2 flux response.  The linearization was applied to all images before any other data reduction steps.  %Following linearization, the individual science exposures were sky-subtracted, flat-fielded, bad pixel-corrected, and then de-rotated (if taken with CR rotations at Palomar), aligned, and median-combined to obtain a deep exposure of each science target.  

To enhance our ability to detect faint candidate companions in the deep coronagraphic exposures we attempted various methods of PSF removal, including: (1) subtracting a median-combined PSF of the star formed from the individual
images taken at all four CR angles at Palomar, (2) subtraction of a 180$\degr$-rotated version of the image centered on the star from itself, (3) high-pass filtering by subtracting a Gaussian-smoothed (Gaussian FWHM~= 1--3$\times$PSF FWHM) version of the image from itself, and (4) simple subtraction
of an azimuthally medianed radial profile.  We found that (1--3)
gave comparable results, while (4) did not perform as well as the rest
because of the four-cornered or six-spoked symmetry of the PALAO or Keck AO PSFs.  Even though (3) is arguably the most widely used method for PSF subtraction when separate PSF
observations are not available and when the observations were not taken using ADI, we found that because of the central
symmetry of the brightest AO speckles \citep{boccaletti_etal02,
bloemhof03} method (2) worked almost as well.  Method (2) also did not
alter the photon statistics of the PSF-subtracted image in the spatially
correlated manner incurred by Gaussian smoothing.  Therefore, for PSF subtraction we relied on
method (2) the most.  

\subsection{Precision Astrometry} 
\label{sec_astrometry}

Multi-epoch astrometry is essential for demonstrating physical association of bound pairs.  This is the principal method employed for candidate companion confirmation here.  Below we discuss the calibration steps that we undertook to ensure self-consistent astrometric measurements throughout our campaign.

We calibrated our astrometry by obtaining repeated measurements of the positions of well-known visual binaries at each observing epoch.  We selected binary stars with well-known ephemeris from the Sixth Orbit Catalog \citep{hartkopf_etal01, hartkopf_mason03}, combining binaries with grade~1 (accurately determined, short-period) and grade~4 (less accurately known, longer-period) orbital solutions, as recommended for astrometric calibration by \citet{hartkopf_mason03}.  Despite the lower quality of the orbital solutions for the grade 4 binaries, their periods are generally much longer, so that their motions are predicted with sufficient accuracy for many years into the
future.  The selected calibration binaries and their orbital parameters are given in Table~\ref{tab_calbin}.

The above astrometric calibration was adequate for detecting astrometric signals $\delta\rho/\rho\gtrsim1\%$ with PHARO.  Such accuracy allowed the confirmation of the first brown dwarf companion in our survey, HD~49197B \citep{metchev_hillenbrand04}.  However, that initial calibration assumed that the pixel scale and field orientation over the entire PHARO
detector were well-determined from measurements taken near the center of
the array, ignoring possible image distortion in the focal plane.  In reality, the PHARO beam is known to be distorted \citep{hayward_etal01}.
%, by up to 0.4\% at f/29 (25~\maspix) and 0.8\% at f/18 (40~\maspix). 
Accurate characterization of this distortion was necessary before considering the
results from our complete survey, which focused on stars with small proper motions (10--100~\masyr; Fig.~\ref{fig_dist_pm_hist}b) and detected candidate companions over the
entire $25.6\arcsec\times25.6\arcsec$ PHARO FOV.  

We arrived at an improved astrometric calibration of the PHARO 25~mas~pix$^{-1}$ camera in \citet[][\S4]{metchev06}, where we determined the full extent of the focal plane distortion over the entire array and solved for its dependence on telescope hour angle, declination, and orientation of the CR rotator.  For that calibration we used a custom-made astrometric mask with pinholes distributed on a rectangular grid that we inserted in the telescope beam path at the Cassegrain focus.  From exposures taken with the mask in place we measured the variations in the spacing among the pinhole images with changes in the instrument gravity vector.  We found that the PHARO pixel scale varied by up to $\delta\rho/\rho =0.9\%$ from the center to the corner of the array in the 25~mas~pix$^{-1}$ camera.  After fitting two-dimensional polynomials to the distortion, we calibrated the variation to within 0.15\% over the entire chip.  
The polynomial fits to the focal plane distortion on the PHARO 25~mas~pix$^{-1}$ camera and its dependence on telescope pointing are given by Equations 4.1--4.4 and 4.7--4.11, and Tables~4.4--4.5 in \citet{metchev06}.  An {\sc IDL} program that corrects for the distortion at an arbitrary coordinate on the PHARO 25~mas~pix$^{-1}$ camera is available at the PHARO instrument web page \url{http://www.astro.caltech.edu/palomar/200inch/palao/Pharo/pharo.html}.\footnote{The PALAO/PHARO astrometric calibration was performed in March 2005.  The optics on the PALAO system have since been realigned to accommodate recent and future science instrument upgrades. The astrometric calibration presented here is not applicable to PALAO data taken since 2007.}

A similar astrometric calibration has
already been performed for all three NIRC2 cameras during the pre-ship
testing of the instrument \citep{thompson_etal01}.  Because NIRC2 sits on the Keck~II Nasmyth platform and thus has a constant gravity vector, the distortion of the camera pixel scales does not change with telescope pointing.  We implemented the existing astrometric calibration of the NIRC2 cameras in the analysis of our Keck AO imaging data.\footnote{A more precise astrometric calibration of the NIRC2 cameras has since been obtained by \citet{cameron_etal08}.}

\subsection{Photometry} 
\label{sec_photometry}

We used 1--2$\times$PSF FWHM-diameter apertures for object photometry, with the smaller apertures used on fainter sources for higher signal-to-noise measurements.  The diffraction-limited FWHM of the $K_S$-band PSF of PALAO was consistently $\approx$$0\farcs1$ while for the Keck AO system it was $\approx$$0\farcs05$.  The local background was measured around each object in an annulus with a wide enough inner radius so that the halo of the point source did not affect the background measurement.  The inner radius was as small as 1.5$\times$PSF FWHM for faint sources embedded in the halos of bright stars, or as large as 25--30$\times$PSF FWHM for the target primaries.  The variations in the sizes of the apertures and of the background annuli resulted in photometric uncertainties on the order of 0.10--0.30~mag.  Uncertainties of $\geq$0.5~mag were found in a few isolated cases involving very faint point sources and/or point sources near the edges of the FOV, where the PSF was noticeably distorted by anisoplanatism and circular apertures did not produce accurate photometry.

PSF-fitting, rather than aperture photometry was used to measure the fluxes of closely-separated point sources.  The photometric uncertainties in such cases were generally $\leq0.20$~mag.

For absolute calibration we relied on the 2MASS fluxes of the primaries.  Photometric measurements were always obtained relative to the fluxes of the target primaries, as measured from the unocculted, short exposures, often taken with the PHARO ND 1\% filter in place.  We calibrated the near-IR extinction of the ND filter from photometric measurements of three program stars on images taken with and without the filter in place.  Images with Keck/NIRC2 were obtained only in
coronagraphic mode, using predominantly the 1$\arcsec$-diameter
spot, although the 2$\arcsec$-diameter spot was used during 16--18 May
2003.  Unlike the PHARO coronagraphic spots,
the NIRC2 spots are transmissive,
offering the possibility to obtain relative photometry with respect to the primary.  A measurement of the throughput of the 2$\arcsec$ spot was given in
\citet{metchev_hillenbrand04}.  Subsequent observations showed that
such measurements were dependent on the quality of the AO correction,
possibly because of the amount of additive background caused by light from the stellar halo diffracted within the area of the coronagraph.  Thus, approximate relative photometry with the NIRC2 coronagraph
is likely feasible only with good AO correction (usually at $H$ or
$K$ bands), when the amount of scattered (``spill-over'') light within the area of the coronagraph is minimized.
Table~\ref{tab_extinctions} lists the measured near-IR extinction in magnitudes
for the PHARO ND~1\% filter and for the 1$\arcsec$ and 2$\arcsec$ NIRC2 coronagraphs.  The large apparent
difference in the $J$-band transmissivity of the two NIRC2 coronagraphic spots
is a probable effect of spill-over (more significant for the smaller spot), aggravated by poorer AO performance at $J$.

\section{Object Detection and Detection Limits} 
\label{sec_detlim}

\subsection{Object Detection} 
\label{sec_detection}

Object detection is a straightforward matter to automate in point-source-rich images where the PSF is radially symmetric, approximately constant in time, and
has a well-characterized dependence on image location.  Unfortunately, none of these qualifications describe the sparsely populated high-contrast images in our deep survey, in which the main (and frequently only) point source is occulted by the coronagraph.  In addition, automated source finding in AO images of bright stars is hindered by large numbers of speckles.  Speckles are individual images of the star that form from uncorrected and/or induced (by the telescope optics) aberrations in the wavefront, and appear indistinguishable
from point sources to automated detection routines.  
%As AO-corrected wavefront changes on time scales of seconds or less, speckles have variable intensities and locations between successive images.  
As a result, even though certain source detection algorithms
have been developed \citep[{\sc StarFinder};][]{diolaiti_etal00}, or adapted 
\citep[{\sc DAOPHOT II, IDAC};][]{stetson92, jeffries_christou93}, for diffraction-limited image restoration, they did not produce
satisfactory results on our images.  Our experiments with
{\sc DAOPHOT}, {\sc WAVDETECT} \citep{freeman_etal02}, and {\sc StarFinder} produced large numbers of spurious detections, the vast majority of which
could be identified with speckles around the coronagraph.  If the
signal-to-noise threshold in the source-finding algorithms was adjusted
to a correspondingly higher level, the algorithms would miss bona fide point sources far from the central star.  The performance of the automated algorithms
did not change whether we used various methods of PSF
subtraction (\S~\ref{sec_datareduction}) or not.  Similar experiences and conclusions were drawn independently by \citet{carson_etal05}, who also used the PALAO/PHARO system for their substellar companion search.  Therefore, after some experimentation, and despite an understanding that automated source
detection has the potential to offer greater repeatability and conceptual
clarity, we abandoned the approach.

Instead, we opted for visual point source identification, which, barring subjective factors, produces
superior results compared to automated detection.  We carefully inspected all of the final coronagraphic images for candidate companions.
%and measured their positions and fluxes with the {\sc PHOT} task in {\sc IRAF}.  
The visual inspection was repeated multiple times during the steps of image reduction, photometry, and astrometry to reduce the effect of subjective factors to a minimum.  

The high-contrast imaging literature abounds with examples where the authors have resorted to by-eye identification of candidate companions \citep[e.g.,][]{tokovinin_etal99, brandner_etal00, schroeder_etal00, luhman_jayawardhana02, mccarthy_zuckerman04, masciadri_etal05, luhman_etal05e, lafreniere_etal07}.  A notable exception is the study of \citet{lowrance_etal05}, who apply a
rigorous custom-made automated  detection
scheme to their {\it HST}/NICMOS data. However, the \citeauthor{lowrance_etal05} survey benefits
from the well-behaved PSF of space-borne {\it HST} imaging.  In a
separate instance, \citet{shatsky_tokovinin02} use {\sc daophot ii}
for their non-coronagraphic AO data.  Still, they do not discuss an application of the approach to their set of coronagraphic data, which are likely to be speckle-dominated.

\subsection{Determination of Detection Limits} 
\label{sec_byeye}

We quantified our ability to visually detect faint objects by introducing artificial point sources in the Palomar and Keck $K_S$ images of one of our targets, HD~172649,  for which data were taken under good observing conditions with Strehl ratios of $\approx$50\%.   The method was first described in \citet{metchev_etal03} and developed more fully in \citet{metchev06}.  We summarize it here briefly.

We introduced 1000--5000 artificial point sources of constant brightness at random locations over the entire 25$\farcs6\times25\farcs6$ area of the image and counted the fraction of them that were retrievable by eye in 0.25$\arcsec$--1.0$\arcsec$-wide concentric annuli centered on the star.  We recorded both the minimum point source $K_S$ magnitude, at which 100\% of the artificial point sources were visible at the given angular separation, and the maximum $K_S$ magnitude, at which only a few artificial point sources were visible.  We took the mean of the $K_S$ magnitude range as the representative limiting magnitude at the given separation.  We repeated the experiment for a range of artificial star magnitudes, at steps of 0.5~mag, on both the coronagraphic and the non-coronagraphic images of HD~172649.  The PSF for artificial stars in the coronagraphic image was obtained from a fit to the brightest field object ($\Delta K_S = 6.4$~mag), whereas in the unocculted image the PSF was obtained from a fit to HD~172649 itself.

\begin{figure}
%\plotone{detlimits_eye_4sigma.eps}
\vspace*{-14mm}
\plotone{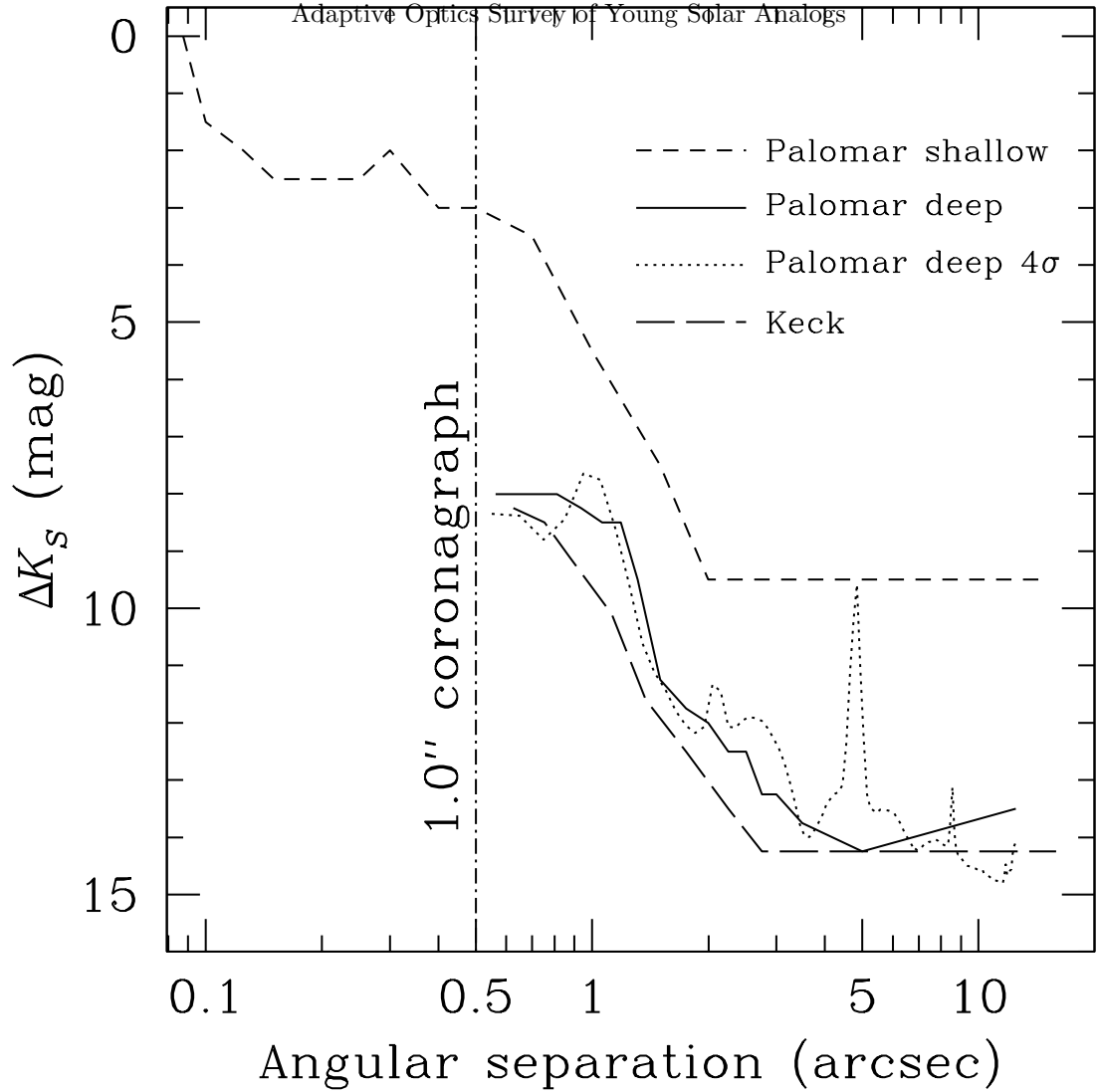}
\figcaption{Empirical $K_S$-band contrast limits as determined from artificial star experiments in images of the program star HD~172649 ($V=7.5$~mag), taken under good AO performance ($\approx$50\% Strehl ratio).  The solid and long-dashed curves delineate coronagraphic observations
at Palomar (24~min) and Keck (6~min), respectively.  The short-dashed line shows the non-coronagraphic component of the Palomar survey.
The dotted line represents the 4$\sigma$ r.m.s.\ deviation of counts in
the PSF halo as a function of separation, normalized to an aperture with
radius 0$\farcs$1: equal to the FWHM of the $K_S$-band PALAO PSF.  The vertical dash-dotted line shows the edge of the occulting spot at Palomar and Keck.  The slight decrease in contrast in the
Palomar coronagraphic  limits at $>$5$\arcsec$ separations is due
to an additive parameter used to model the decreasing
exposure depth toward the edge of the PHARO field, because of image
mis-registration among the different CR angles (\S~\ref{sec_1stepoch}).  The contrast degradation
is set to vary %linearly 
between 0~mag and 0.75~mag in the 4$\farcs$0--12$\farcs$5 separation range.
The bumps and spikes in the r.m.s.\ limits correspond to bright features in the image of HD~172649, such as the corners
of the waffle pattern at 1.0$\arcsec$ and projected companions to
the star at 2.1$\arcsec$, 4.8$\arcsec$, and 8.6$\arcsec$.
\label{fig_detlimits}}
\end{figure}

The inferred detection limits based on the artificial point source experiments are shown in Figure~\ref{fig_detlimits}.  We see that the 6~min long Keck AO coronagraphic images offered 0.5--1.5~mag higher contrast and up to 0.5~mag greater depth than the 24~min
PALAO images.  The greatest difference in contrast is in the
1$\farcs$0--1$\farcs$5 angular separation range, where the presence of waffle-mode
distortion in the PALAO PSF limits the attainable contrast.

For the purpose of estimating the completeness of our survey, it was important to determine sensitivity limits on a per-star basis.  However, the above approach was too tedious to apply to all observations.  Instead, we employed a simpler strategy based on the r.m.s.\ scatter of the pixel counts in the radial profile of each sample star.
To match the approximate spatial correlation scale in the PALAO and Keck AO images, we normalized the r.m.s.\ scatter to an aperture with radius equal to the 0$\farcs$10 FWHM of the PALAO PSF.  That is, we multiplied the r.m.s.\ profile by the square root of the number $N$ of pixels in the photometry aperture; $N=50.3$~pix for PALAO/PHARO with the 25~\maspix camera and $N=20.0$~pix for Keck AO/NIRC2 with the wide (40~\maspix) camera.  This procedure imposed a more stringent requirement on the significance of the detection of a candidate point source by raising the effective multiple of the pixel-to-pixel r.m.s.\ scatter used as a threshold by an additional factor of 4.5--7.1.  We show the thus-obtained 4$\sigma$ aperture-normalized r.m.s.\ noise profile of the halo for our PALAO coronagraphic image of HD~172649 in Figure~\ref{fig_detlimits}.   We found that the 4$\sigma$ line best approximated the visually determined PALAO detection limits.  The strongest systematic deviation of the 4$\sigma$ r.m.s.\ profile from the visually-determined contrast limits is at angular separations $>$7$\arcsec$.  This is to be expected, to some extent, because in this region we have adjusted the visual detection limits to account for CR angle image mis-registration (\S~\ref{sec_1stepoch}).

The agreement between the detection limits from visual inspection
and from r.m.s.\ statistics is dependent on a number of factors, such
as the radius of the normalization aperture, the PSF pixel sampling, the treatment of point source photon statistics (ignored in our r.m.s.\ analysis), and the appropriate functional treatment of non-Gaussian sources of error (speckles, shape of the PSF core and halo; also ignored here).  As a check on whether the adopted r.m.s.\ detection limit approach was a valid approximation across the range of PSF and image characteristics encountered in our survey, we repeated the artificial point source experiment on six additional images of targets observed both at Palomar and at Keck.  These images were taken under a range of seeing conditions, resulting in PSF Strehl ratios between 10\% and 50\%.  We found that, on average, the by-eye detection limits varied between three and five times the level of the aperture-normalized r.m.s.\ noise profile over the entire range (0$\farcs$5--$12\farcs5$) of probed angular separations.  Thus, the additional experiments confirmed that our choice of the 4$\sigma$ level was an adequate detection threshold.  

In closing, we note that because our image noise statistics in the contrast-limited regime are not Gaussian, the adoption of a 4$\sigma$ threshold does not carry the statistical significance of a confidence level at which 99.997\% of random fluctuations are rejected.  
Only recently have \citet{marois_etal08} shown that quasi-Gaussian behavior of the PSF subtraction residuals can be obtained with the ADI technique, allowing such formal estimates on the detection limits.  Because our data were not taken in ADI mode, the same formalism can not be applied here.

\subsection{Illustrative Detection Limits for the Deep Sample} 
\label{sec_survey_detlims}

%The empirical detection limits in Figure~\ref{fig_detlimits} were obtained for observations that benefited from good AO correction (Strehl ratio of $\approx$50\%) on a bright star ($V=7.5$~mag) observed under median atmospheric conditions for the survey (1$\farcs$2 seeing, scattered cirrus).   While high-quality AO performance on bright stars at Palomar and Keck is now consistently achieved under good to average atmospheric conditions, fainter stars and worse weather incur a toll on the quality of the AO correction, resulting in poorer contrast and shallower imaging depth.  Therefore, for the purpose of estimating the completeness of our survey, it is important to determine the attained contrast on a per-star basis.

Table~\ref{tab_deep_obs} lists the attained point-source magnitude sensitivity for each star in the deep sample at angular separations of 1$\arcsec$, 2$\arcsec$, and 5$\arcsec$.  Beyond 5$\arcsec$ the detection limits are constant to within 0.5~mag.  In the cases where multiple
images of the same star were taken at different epochs, we have listed the sensitivity only for the epoch with the deepest image.  We thus formed a set of 58 Palomar and 42 Keck images that represented the deepest observations
of the 100 stars in the deep sample.

\begin{figure}
%\plottwo{ensemble_limits_dKs_fixed.eps}{ensemble_limits_Ks_fixed.eps}
\plottwo{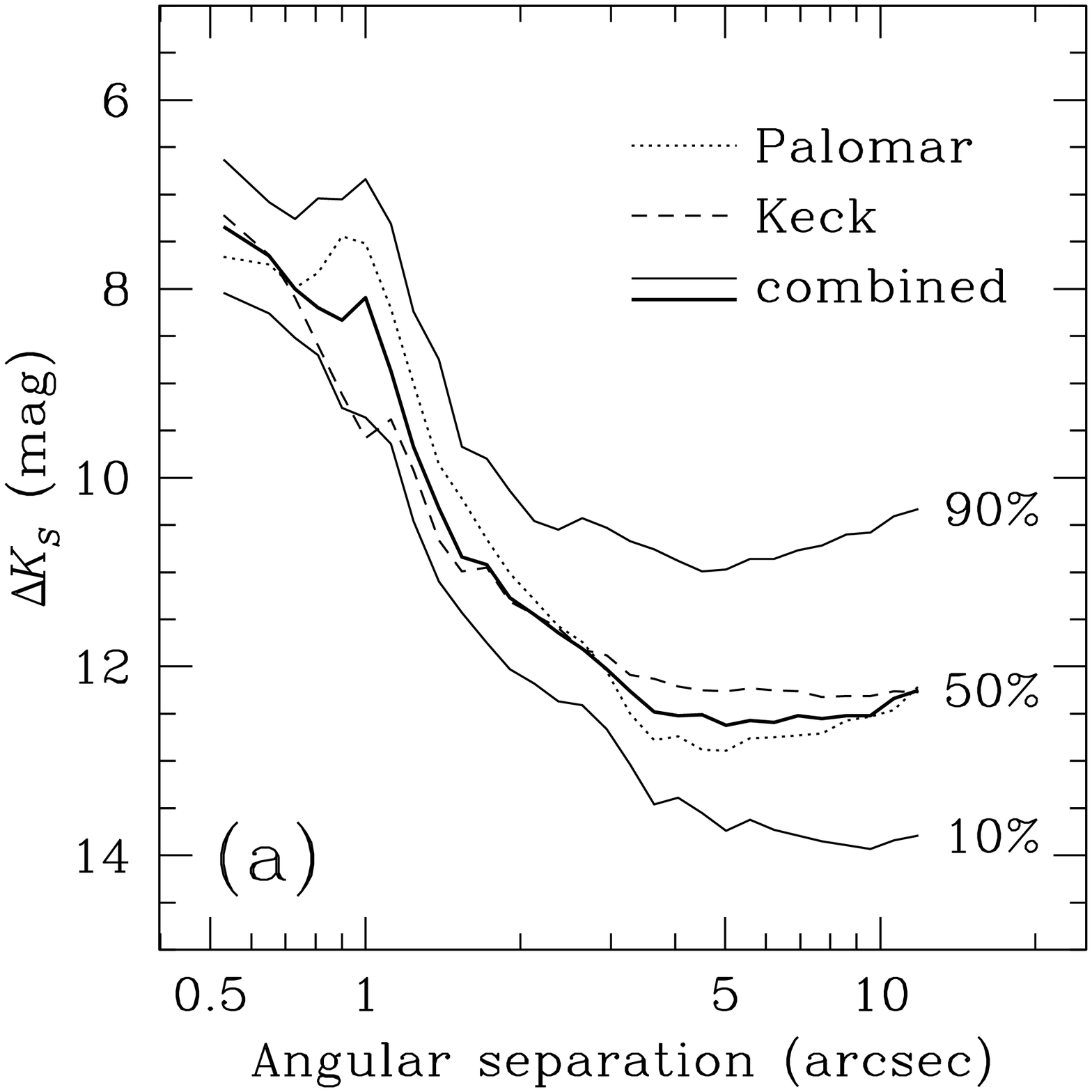}{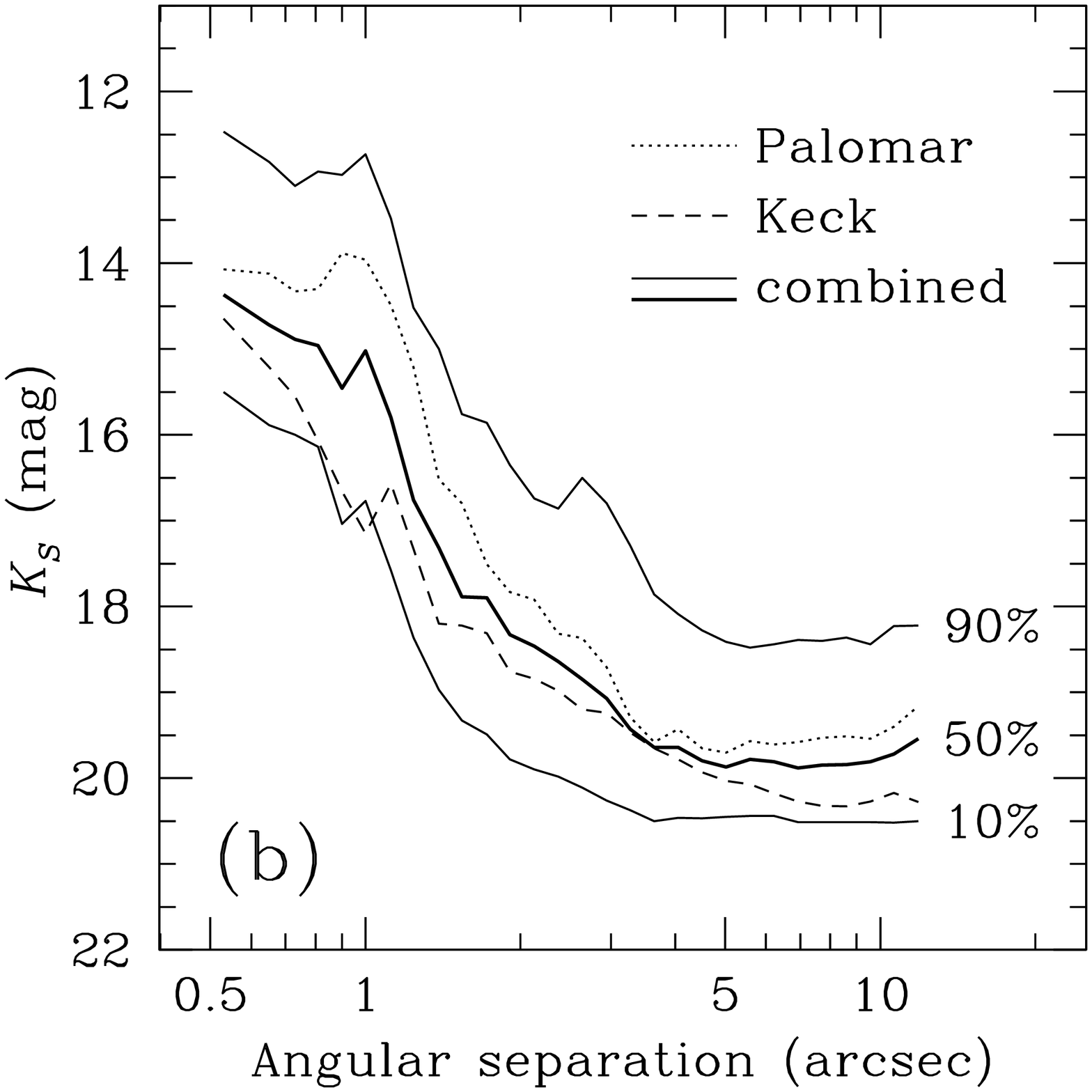}
\figcaption{Contrast (a) and depth (b) of the deep survey at $K_S$.  The solid lines 
represent the 10\%, 50\% (thick), and 90\% completeness of the combined Palomar + Keck AO survey.  The median (50\%) sensitivities of the Palomar (dotted line) and Keck (dashed line) surveys are also shown.  The gradual decrease in imaging contrast and depth at Palomar between 4$\arcsec$--12$\farcs$5 is partially due to mis-registration of images taken at different CR angles (\S~\ref{sec_datareduction}), and partially to the sometimes smaller depth of observations at 11$\arcsec$--12$\farcs$5 separations because of a 0$\farcs$5--1$\farcs$5 offset of the coronagraphic spot from the center of the PHARO array.
\label{fig_ensemble_contrast}}
\end{figure}

Figure~\ref{fig_ensemble_contrast}a depicts the range of
attained $K_S$-band contrast for the coronagraphic observations in the
entire survey (thick solid line), and from the Palomar
(dotted line) and Keck (dashed line) portions of it.  Figure~\ref{fig_ensemble_contrast}b uses the same notation to depict
the imaging depth of the survey in
terms of apparent $K_S$ magnitude (i.e., with the magnitude of the
primary added in each case).   
The median sensitivities of the combined survey range from $\Delta
K_S=8.4$~mag at 1$\arcsec$ to $\Delta K_S\approx12.5$~mag over
4$\arcsec$--12$\farcs$5
in contrast and from $K_S=15.4$~mag at 1$\arcsec$ to $K_S\approx19.7$~mag
in depth.  These detection limits will be used in the Appendix
to estimate the completeness of the deep survey
to substellar and stellar companions.
%(\S~\ref{app}).

We obtained the detection limits for the shallow sample in a manner similar to that used for the deep sample: from the 4$\sigma$ dispersion of the radial profile of each star, normalized to an aperture with radius equivalent to the FWHM of the PALAO PSF.  The shallow sample detection limits are given in Table~\ref{tab_shallow_obs}, where we have in addition listed the sensitivity at 0$\farcs$5.  

In some cases close binary companions elevate the dispersion in the radial profile of the primary, resulting in unusually low sensitivities at certain angular separations: e.g., for HD~172649 at 5$\arcsec$ in Table~\ref{tab_deep_obs} and for HD~224873 at 2$\arcsec$ in Table~\ref{tab_shallow_obs}.  We have retained these lower sensitivities in Tables~\ref{tab_deep_obs} and \ref{tab_shallow_obs} as an indication that part of the images around the sample stars in question were compromised by a nearby bright companion.

\section{CONFIRMATION OF CANDIDATE COMPANIONS} 
\label{sec_candidates}

\subsection{Detected Candidate Companions} 
\label{sec_detected_comps}

In the course of the three year survey we
discovered 287 candidate companions brighter than $K_S=20.6$~mag within
12.5$\arcsec$ of 130 from the 266 sample stars.  Of these candidate
companions 196 were around 61 of the 100 stars in the deep sample.
The remaining 91 were in the vicinity of 70 of the 166 shallow-sample
targets.  All candidate companions around stars in the
deep and shallow samples are listed in Tables~\ref{tab_deep_companions}
and \ref{tab_shallow_companions}, respectively.  Figure~\ref{fig_dmag_sep} shows all detected candidate companions as a
function of magnitude difference $\Delta K_S$ and angular separation $\rho$.  Thirty-nine stars in the deep sample and 96 in the shallow sample showed no projected companions within 12.5$\arcsec$.  

\subsection{Deciding Physical Association} 
\label{sec_phys_assoc}

The physical association of each candidate companion was decided based
on one of the following criteria: (1) common proper motion with the candidate
primary, (2) a combination of the position on a $J-K_S$ versus $M_{K_S}$ color---absolute magnitude diagram (CAMD; assuming equi-distance with the primary)
and background star density arguments, 
or (3) extent
of the radial profile of the candidate companion beyond that of a
point-source PSF (which suggests an extragalactic object).  Candidate substellar companions that satisfied the common proper motion test were also observed spectroscopically to confirm that their spectral types were in agreement with their projected substellar masses.

\subsubsection{Proper Motion} 
\label{sec_assoc_pm}

Proper motion 
%(\S~\ref{sec_assoc_pm}) 
is usually the criterion of choice in companion studies, as it provides nearly unambiguous evidence of association between two objects: whether as components of a gravitationally bound system or as members of a multi-star moving group sharing a common origin.  We used the common proper motion criterion through the combined application of two requirements:
(i) that the change in the position of the candidate companions relative
to the primaries was within 3$\sigma$ of zero in all of right ascension ($\alpha$), declination ($\delta$), angular separation ($\rho$), and position angle ($\theta$) , 
%XX Or chi-square?  but that significantly decreases the power of proper motion discrimination!
and (ii) that the expected change in
relative positions of the candidate companions, had they been stationary
background objects, was more than 3$\sigma$ discrepant in either $\alpha$, $\delta$, $\rho$, or $\theta$ from the observed
change.  Often in cases of candidate close ($\lesssim$20~AU) binaries,
criterion (i) was not satisfied because of appreciable orbital motion.  In such situations we instead made
sure that (iii) the observed change in relative position was much smaller
(and less significant) than the expected change if the components of the
candidate binary were not gravitationally bound.  A detailed example 
of the implementation of the above astrometric criteria is worked out in \citet[][\S5.4.1]{metchev06}.

When a relatively bright field star (4~mag $<\Delta K_S\lesssim8$~mag) was present
in the deep coronagraphic exposures at Palomar, its position in the shallow non-coronagraphic images was used as an additional
astrometric reference.  In cases where the subsequent astrometric
measurements with respect to the primary and to fainter field objects
showed such bright field stars to be approximately stationary, they could be
used to bootstrap the association of other candidate companions with the
primary, and thus circumventing the somewhat higher positional uncertainty associated with
locating the primary behind the opaque PHARO coronagraph.  This technique was particularly important in determining the association of systems in the distant
Upper Scorpius (145~pc) and $\alpha$~Persei (190~pc) regions, where the primaries
have small proper motions ($\lesssim40$~\masyr) and the images contain multiple background
stars because of the low galactic latitude ($5\degr<\vert b \vert<25\degr$).

\subsubsection{Near-IR CAMD and Background Object Density} 
\label{sec_assoc_cmd}

Systems with bright ($\Delta K_S < 5$~mag) close-in candidate secondaries often lacked
dual-epoch astrometry in our survey.  Such systems were given lower priority
in follow-up observations because the companions were considered to be stellar and almost certainly bound.  Multi-epoch astrometric analysis was inapplicable in these cases.  However,
the candidate stellar secondaries in these systems were bright
enough to be seen in the shallow non-coronagraphic $JHK_S$ exposures of
our targets.  Hence, for the majority of
the candidate stellar systems lacking astrometric confirmation, physical
association could be estimated based on the near-IR colors and expected
absolute magnitudes of the components.  
%This approach can also be applied to fainter candidate companions, for some of which single-epoch $J$-band coronagraphic data were also taken.

In evaluating the association of a candidate companion based on its
near-IR photometry, we placed it on a $J-K_S$ versus $M_{K_S}$ CAMD, and checked whether it laid on the same isochrone as the primary.  
In the substellar regime, especially near the L/T transition ($12<M_K<14$) where the isochrones are not well-constrained, we relied on the
empirical main sequence as traced by nearby M--T dwarfs \citep{leggett_etal02, reid_etal04} with known parallaxes \citep[from][]{dahn_etal02, vrba_etal04}. 
All candidate companions with available $J$-band photometry for which
the astrometry was inconclusive had their physical associations with
their candidate primaries evaluated in this manner
(Fig.~\ref{fig_candidates_mkjk}).  
To limit the probability of misclassifying field stars as bona fide
companions, positive associations were adopted only for candidate
companions within a 5$\arcsec$-radius field of interest from the primary.

This approach was successful mostly for stellar-mass companions bluer
than $J-K_S=0.8$~mag, i.e., earlier than spectral type M0.  The main sequence for redder M0--M6 dwarfs is nearly degenerate in $J-K_S$ over nearly 4~mag in $M_{K_S}$ (see Fig.~\ref{fig_candidates_mkjk}) and does not allow reliable association estimates from the $J-K_S$ color.  At even later spectral types, potentially representative of young brown dwarf companions, the
higher photometric uncertainties and the larger
empirical color scatter at substellar masses prevented the
conclusive determination of physical association in all but a handful
of faint projected companions.  $H$-band photometry, where available,
did not improve the analysis because of the smaller wavelength range sampled by the $H-K_S$ vs.\ $J-K_S$ colors.  Thus, no candidate substellar companions were confirmed through near-IR photometry.  However, a few could be rejected.

In addition to using near-IR colors, it was also possible to obtain
a probabilistic
estimate of the physical association for a candidate companion to its
corresponding primary by
comparing the number of detected objects within the 12$\farcs$5
survey radius to the
surface density of stars at the relevant galactic coordinates down to
the limiting magnitude of the survey.
Because of the lack of large-area deep ($K_S\lesssim20$~mag) near-IR
survey data, we limited this type of analysis only to candidate
companions in the shallow survey.  Although
the depth of the shallow survey varied depending on the use of the ND
filter at Palomar, it was roughly comparable to the 99\% completeness
limit of the 2MASS catalog: $K_S<14.3$~mag in unconfused regions of the
sky.  Therefore, for all candidate companions brighter than $K_S=14.3$~mag,
an empirical estimate of the association probability was possible based on 2MASS.
Given that the faintest primaries in the sample have $K_S$ magnitudes of
9.6, such a probabilistic analysis could be performed on all
candidate companions with $\Delta K_S\leq4.7$~mag.

To estimate the contamination from $K_S\leq14.3$~mag field stars,
we counted the number of 2MASS objects within a 5$\arcmin$-radius circular
area offset by 12$\arcmin$
from each sample star (to avoid bright artifacts), and from that obtained the expected number of
background objects in the 5$\arcsec$-radius field of interest.  We used
this as an estimate of the purely geometrical chance alignment
probability (CAP):
\begin{equation}
\mbox{CAP } = \mbox{ (number of 2MASS sources within } 5\arcmin) \times
\frac{\pi (5\arcsec)^2}{\pi (5\arcmin)^2}.
\label{eq_cap}
\end{equation}
Table~\ref{tab_color_comp} lists the separations, $\Delta K_S$ and $K_S$
magnitudes,
and the CAPs for all sample stars
with color companions (i.e., the ones with ``yes(c)'' entries in
Tables~\ref{tab_deep_companions} and \ref{tab_shallow_companions}).
Most color companions have chance alignment
probabilities $\lesssim$2\%, with the exception of HD~155902B and
HE~935B.  However, both of these are very close ($<0\farcs1$) to their candidate primaries, and are thus almost
certainly physical companions.  (These two systems are in fact below the
resolution limits of the 95~mas PALAO $K_S$-band PSF.  Their
binarity was only appreciated from PALAO $J$-band images, where the PSF is
50~mas wide.)
The ensemble probability of at least one of the 17 color companions being
a false positive is 33\%, or 16\% if HD~155902 and HE~935 are excluded.

\subsubsection{Source Extent} 
\label{sec_assoc_ext}

Any bona fide companions to our sample stars were expected to be point sources.  Apparent source extension could in principle be used to exclude background galaxies seen in projection.  However, the determination of source extent
is not a trivial task when the quality of the AO correction and hence, the size and shape of the PSF change throughout the course of a single night depending on guide star
brightness and on atmospheric stability.  In addition, anisoplanatism may radially elongate point sources PSFs far away from the central AO guide star.  Therefore, departures from the nominal, diffraction-limited PSF size and from a centrally symmetric PSF shape were regarded with caution.
These were used to classify an object as an extended source only when
they were in disagreement with the size and radial behavior of the profiles
of other point sources in the same image, if such were present.

\subsubsection{Physical Association Summary} 
\label{sec_assoc_summary}

Using the above criteria, we were able to determine the physical association for 198 of the 287 companion candidates.  The proper motion criterion (\S~\ref{sec_assoc_pm}) was used to establish the majority of associations or non-associations: 166 out of 197 (84.3\%).  These included 55 bona fide common proper motion companions and 111 non-common proper motion background objects seen in projection. 

The CAMD and chance alignment criteria (\S~\ref{sec_assoc_cmd}) are not conclusive in proving physical association.  They were invoked
only when astrometric follow-up was not obtained or the proper
motion data were ambiguous, and when additional $J$-band
images were taken (\S~\ref{sec_assoc_cmd}).
These criteria were used in tandem to establish with high fidelity the physical association of candidate
{\it stellar} companions in the 18 cases listed in Table~\ref{tab_color_comp}.  The CMD criterion alone was used in seven cases to exclude background
interlopers.  

The source extent criterion (\S~\ref{sec_assoc_ext}) was used to weed out faint galaxies, which may
otherwise have red near-IR colors, partially due to line-of-sight extinction, and may thus pose as candidate substellar objects for the
preceding criterion.  This criterion was applied in four cases. 

None of the above criteria were applicable to
89 candidate companions (31.0\% of the total) that remained ``undecided.''
The vast majority of these were faint objects in the
fields of distant ($>100$~pc) stars with small apparent proper motions
($<50$~\masyr), often
at low galactic latitudes ($b<15\degr$).  These were often discovered only
in follow-up deeper imaging with Keck and thus lack the full
time-span of astrometric observations.  Judging by the large number of such
candidate companions per star, and based on expectations of the 
background star
contamination rate at low galactic latitudes, probably none of these 
candidate companions are associated.   Throughout the rest of the analysis, we shall assume that all 89 of the undecided candidates are unassociated field stars.

\section{SURVEY RESULTS} 
\label{sec_results}

Preliminary results from the survey were already published in \citet{metchev_hillenbrand04, metchev_hillenbrand06}, including the discovery of two brown dwarf companions,
HD~49197B and HD~203030B.  In this paper we report the results from the full survey.  We found no more substellar companions in our sample.  We summarize the findings on the two previously discovered substellar companions in \S~\ref{sec_results_bds}.  We also report 21 new stellar companions, in addition to three (HD~129333B, HE~373B, and RX~J0329.1+0118B) already announced in \citet{metchev_hillenbrand04}.  Four of the newly-discovered stellar companions, HD~9472B, HE~373B, HD~31950B, PZ99~J161329.3--231106B, in addition to RX~J0329.1+0118B announced previously, have masses of only $\approx$0.1~\Msun, and reside in very low mass ratio $q=M_2/M_1\approx0.1$ systems.  The results on the stellar companions are detailed in \S~\ref{sec_binaries}.

A proper motion companion to the star ScoPMS~214, considered to be a brown dwarf based on its apparent $K_S$ magnitude in \citet{metchev06}, was found to most probably be an unasociated foreground M star after spectroscopic follow-up.  This companion, ScoPMS~214``B'', is discussed in \S~\ref{sec_scopms214} as an example of a pathological case where spectroscopic analysis argues against the physical association in an apparent common proper motion system.

Independently of the unbiased survey for substellar companions, we also observed and established the physical association of a previously known \citep{bouvier_etal97} candidate
companion to HII~1348.  The estimated mass of HII~1348B is near the limit for sustained hydrogen burning.   Because of our deliberate inclusion of HII~1348 in our observing program based on known binarity, it is excluded from the present analysis.  HII~1348B will be the subject of a future publication. 

\subsection{Brown Dwarf Companions} 
\label{sec_results_bds}

Both brown dwarf companions, HD~49197B and HD~203030B, were found in the 100-star deep survey.   The observed photometric and astrometric properties of the two and their inferred masses are listed in Table~\ref{tab_stellar_comps} alongside those of the stellar secondaries observed in our survey.  The spectral types of HD~49197B and HD~203030B are L$4\pm1$ and L$7.5\pm0.5$, respectively, and their masses are estimated at $0.060^{+0.012}_{-0.020}$~\Msun\ \citep{metchev_hillenbrand04} and $0.023^{+0.008}_{-0.011}$~\Msun\ \citep{metchev_hillenbrand06}.

Because of their association with main sequence stars, the ages of HD~49197B and HD~203030B are known with relative certainty.  Their moderate youth (250--500~Myr) makes them valuable as benchmarks for substellar properties at $\log g \approx 5$ surface gravities, $\sim$0.5~dex lower than the gravities expected of $\sim$3--5~Gyr-old brown dwarfs in the field.  

At the time of its discovery, HD~49197B was only the fifth known L dwarf younger than 1~Gyr.  At a projected separation of only 43~AU from its host star, HD~49197B was also one of the closest-in resolved substellar companions, second only to HR~7672B \citep[14~AU;][]{liu_etal02}.  Both HR~7672B and HD~49197B provided early indication that the brown dwarf desert may not extend much outside of 3~AU \citep{metchev_hillenbrand04}.

HD~203030B was the first young brown dwarf with a spectral type unambiguously as late as the L/T transition.  Its surprising underluminosity, by $\approx$0.5~dex compared to theoretical predictions for $\sim$1400~K brown dwarfs at its age, indicated that its effective temperature was $\approx$200~K cooler (i.e., $\approx$1200~K) than expected at the L/T transition.  That is, the spectrophotometric properties of HD~203030B indicated that either the effective temperature at the L/T transition had a heretofore unappreciated dependence on surface gravity, or that the entire population of field substellar objects had had their effective temperatures and ages significantly overestimated.  

In fact, underluminosity and $<$1400~K expected effective temperatures are observed in all known substellar companions near the L/T-transition \citep{metchev_hillenbrand06}, including both the recently discovered T2.5 dwarf HN~PegB \citep{luhman_etal07} and the presumed planetary mass 2MASS~J1207334--393254B \citep[L5--L9;][]{chauvin_etal05a, mohanty_etal07}.  With the mean ages of their respective primaries ranging between 8~Myr and 2~Gyr, all six known L/T-transition companions (GJ~584C, \citealt{kirkpatrick_etal01}; GJ~337CD, \citealt{wilson_etal01, burgasser_etal05b}; 2MASS~J1207334--393254B, \citealt{chauvin_etal05a}; HD~203030B, \citealt{metchev_hillenbrand06}; HN~PegB, \citealt{luhman_etal07}) are likely younger than the 2.9~Gyr model-dependent mean age of L/T-transition dwarfs in the solar neighborhood \citep{allen_etal05}.  Therefore, the theory may indeed be overestimating the ages of field brown dwarfs, by a factor of at least 1.5.  This hypothesis has now been independently reinforced by the first measurement of the dynamical mass of a binary field T dwarf. \citet{liu_etal08} find that the components of the T5.0+T5.5 dwarf binary 2MASS~J15344984--2952274AB are about 100~K cooler than derived for similar field objects: a fact that they interpret as evidence for a factor of $\approx6\pm3$ overestimate in the adopted ages of field brown dwarfs.  Future high-contrast imaging and astrometric observations and discoveries of benchmark brown dwarfs with known ages and dynamical masses will shed important light on these surprising results.

\subsection{Stellar Secondaries} 
\label{sec_binaries}

The entire survey produced 24 
% was 24 pre-ScoPMS 214B.
new stellar companions, including the three already announced in \citet[][HD~129333B, HE~373B, and RX~J0329.1+0118B]{metchev_hillenbrand04}.  HD~129333 had previously been identified as a probable long-period single-lined spectroscopic binary by \citet{duquennoy_mayor91}, and was independently resolved by \citet{koenig_etal05}.  
%Two of the newly resolved stellar binaries, HD~9472A/B and HE~373A/B, have mass ratios $q=M_2/M_1\leq0.1$, with the masses of the secondaries in both cases estimated at 0.11~\Msun.  
Four other new binaries have since been independently confirmed in analyses of {\it Hipparcos} ``problem'' stars by \citet[][HD~26990 and HD~135363]{makarov_kaplan05}, \citet[][HD~152555]{goldin_makarov06}, and \citet[][HD~155902]{goldin_makarov07}.  In addition, PZ99~J161329.3--231106, resolved by us, has since been suggested as a possible spectroscopic binary by \citet{guenther_etal07}.  

In addition to the 24
% was 24
 new systems, the physical association of 51 known binary stars was confirmed astrometrically.  The star HD~91962, a previously known binary \citep[][and references therein]{mason_etal01b}, was resolved into a triple system.  No higher-order
multiples were resolved.  A higher fraction of multiple systems might have been expected, especially given the high rate of occurrence (34--96\%) of visual companions to close (spectroscopic) binary systems \citep{tokovinin_etal06}.  The visual companions in such multiple systems must have either fallen outside of our 12$\farcs$5 survey radius, or been removed from our AO sample by the design of the FEPS target list, which discriminates against visual companions (see \S~\ref{sec_biases}).

%However, some of the newly-resolved binaries are themselves members of composite higher-order systems with previously known wider ($>$12$\farcs$5) visual and/or closer-in spectroscopic companions.  For example, recent work on high order multiples by \citet{tokovinin_etal06} indicates that $\approx$63\% of spectroscopic binary systems with 1--30~day orbits also have wider visual companions.  Based on the known frequency and orbital period distribution of solar-mass binaries \citep{duquennoy_mayor91}, we would expect $\sim$9 spectroscopic binaries with 1--30~day orbital periods in our sample, of which $\sim$6 should have wide tertiary companions according to the \citet{tokovinin_etal06} results.  In all likelihood, systems with such visual companions have either been excluded from our survey by virtue of the design of our sample (\S~\ref{sec_selcrit}), or span projected separations that are too wide to fit within our 25$\farcs6\times25\farcs6$ FOV.

The majority (57 out of 74) 
%was 74)
of the binaries plus the triple system are members of
the shallow survey, as a result of the requirement that no $\Delta K_S < 4.0$~mag candidate companions were present at $>0\farcs8$ from deep sample stars (\S~\ref{sec_selcrit}).  Hence, the
binaries found in the deep survey either have very low mass ratios, such that
the secondary is $>$4~mag fainter than the primary at $K_S$, or
have high mass ratios but $<$0$\farcs$8 angular separations, so that both
components were fit under the 1$\arcsec$ coronagraph.

%$K_S$-band magnitudes and $J-K_S$ colors of the stellar companions were already included in Tables~\ref{tab_deep_companions} and \ref{tab_shallow_companions}.  From these 
We derive $K_S$-band absolute magnitudes $M_{K_S}$ for the companions using the known distances to the primaries (Tables~\ref{tab_deep_sample} and \ref{tab_shallow_sample}).  We estimate the stellar companion masses directly from $M_{K_S}$ and from the primary star age using stellar evolutionary models from \citet{baraffe_etal98}.  The mass ratios of the resolved stellar binaries ranged between 0.1 and 1.0.  Including the two substellar companions, the mass ratios covered the full 0.02--1.0 range.
Table~\ref{tab_stellar_comps} lists $M_{K_S}$ and the mass for each bona fide companion, along with projected separations and system mass ratios.  

\subsection{The Apparent Proper Motion Companion to ScoPMS 214} 
\label{sec_scopms214}

%In \citet{metchev06} we claimed the detection of three proper motion substellar companions in our survey sample: HD~49197B, HD~203030B, and ScoPMS~214B.  Near-IR spectra of the first two companions confirmed their cool temperatures and substellar masses \citep{metchev_hillenbrand04, metchev_hillenbrand06}.  However, our recent spectroscopy of ScoPMS~214``B'' indicates that its mass is probably above the hydrogen-burning limit, and it is most likely not a companion to ScoPMS~214 at all.  We discuss this pathological case in the following.  , as it provides a vivid example of the importance of spectroscopic confirmation of candidate proper motion companions.

%\subsubsection{A Bona-fide Proper Motion Companion to ScoPMS~214 or Not? 
%\label{sec_scopms214b_pm}}

We detected seven projected companions within 12$\farcs$5 of ScoPMS~214 (Fig.~\ref{fig_scopms214_image}; Table~\ref{tab_deep_companions}).  
%The astrometric measurements for all candidate companions to ScoPMS~214 are shown in Table~\ref{tab_scopms214_ccs}.
Among these, candidate companion 1 (CC1) is brightest and closest to the star, and shares the proper motion of ScoPMS~214 to within 3$\sigma$ limits over the course of 4.8 years (specifically, $\Delta\alpha / \sigma(\Delta\alpha)=0.7, \Delta\delta / \sigma(\Delta\delta)=2.4, \Delta\rho / \sigma(\Delta\rho) = 0.7, \Delta\theta / \sigma(\Delta\theta) = 2.3$).  The apparent proper motion of CC1 is significantly different from the remaining three candidate companions (2, 3, and 4) to ScoPMS~214 for which we have sufficiently precise astrometric solutions.  The proper motion of ScoPMS~214 is predominantly to the south ($\mu_\alpha \cos{\delta}=-5.6$~\masyr, $\mu_\delta=-22.1$~\masyr), and candidate companions 2--4 systematically lag behind in their declination motion $\Delta\delta$, at a level of 4.5--5.3$\sigma(\Delta\delta)$ over 4.8 years.  These three candidates, along with candidate companion 6, for which the astrometry is insufficiently precise to decide its proper motion association status, are consistent with being stationary distant objects seen in projection (Fig.~\ref{fig_scopms214_pm}).  Astrometric data for candidate companions 5 and 7 does not exist over the entire 4.8-year period, and hence they are not plotted on the proper motion diagram in Figure~\ref{fig_scopms214_pm}.  However,  they are also consistent with being background stars.

In summary, CC1 satisfies all of the proper motion association criteria established in \S~\ref{sec_assoc_pm}, whereas none of the other candidate companions to ScoPMS~214 do.  Therefore, CC1 has a high likelihood of being a bound companion to ScoPMS~214, although it could also be an unrelated member of Upper Scorpius---the parent association of ScoPMS~214---seen in projection.  

It is in principle possible to distinguish between the above two possibilities in a probabilistic manner, by following a two-point correlation function analysis, as done for Upper Scorpius by \citet{kraus_hillenbrand07}.  We find that the probability to find at least one chance alignment among our 23 deep sample targets that belong to young stellar associations with an unrelated $\geq$M4 dwarf ($\lesssim$0.1~\Msun) within the same association is $\sim$2.5\%.  That is, if CC1 is a member of Upper Scoprius, then there is a 97.5\% probability that it is physically bound to ScoPMS~214.  Similar reasoning lead us to conclude in \citet{metchev06} that CC1 was a bona-fide companion to ScoPMS~214, which we named  ScoPMS~214B.

However, as we shall see in \S~\ref{sec_scopms214_hr}, the spectroscopic evidence argues against membership of CC1 (ScoPMS~214``B'') in Upper Scorpius.

\subsubsection{Spectral Type and Effective Temperature of ScoPMS 214``B''}
\label{sec_scopms214b_spt}

We obtained a $R \approx 1200$ $K$-band spectrum of ScoPMS~214``B'' (CC1) with Keck AO/NIRC2 on 14 July 2005.  We used the 80~mas-wide slit and the medium (20~\maspix) NIRC2 camera.  We integrated for a total of 7.5~min on the companion, following an ABC pointing sequence with 2.5~min integrations per pointing.  We observed a nearby A0 star for telluric correction.  Exposures of Ne and Ar lamps were obtained for wavelength calibration.  The individual 2.5~min exposures were pair-wise subtracted and the spectrum of ScoPMS~214``B'' was traced and extracted from each exposure in a 280~mas-wide ($\approx5.6$ PSF FWHM) aperture.  The three individual spectra were median-combined and smoothed to the resolution set by the instrument configuration using a Savitsky-Golay smoothing algorithm.

The resultant $K$-band spectrum of ScoPMS~214``B'' is shown in Figure~\ref{fig_scopms214b_spec}, where it is compared to IRTF/SpeX $K$-band spectra of field M dwarf and M giant standards from the IRTF Spectral Library\footnote{\url http://irtfweb.ifa.hawaii.edu/\textasciitilde spex/spexlibrary/IRTFlibrary.html.} \citep{cushing_etal05, rayner_etal08}, smoothed to the same resolution.  The dominant atomic and molecular absorption features due to \ion{Na}{1}, \ion{Ca}{1}, and CO are identified.  
%We note that the $K$-band continuum of CC1 appears slightly redder than those of the comparison M dwarfs.  

The overall $K$-band continuum slope of ScoPMS~214``B'' is much closer to the continuum slopes of the M dwarfs than to those of the M giants, although ScoPMS~214``B'' is redder than both sets of standards.  With an extinction of $A_V\sim2$~mag towards Upper Scorpius the expected reddening of ScoPMS~214``B'' at $K$ band is negligible.  Instead, the discrepancy between the continuum slopes of ScoPMS~214``B'' and the M standards may be due to instrumental systematics between the Keck AO/NIRC2 and IRTF/SpeX spectra.  In particular, accurate continuum slopes are difficult to extract from classical AO spectroscopy \citep{goto_etal03, mcelwain_etal07} because of the chromatic behavior of the AO PSF and because of the narrow slits (here 80~mas) used to match the width of the AO PSF.  %Hence, continuum slopes from AO spectra needs to be considered with care.  
Nevertheless, an independent indication that ScoPMS~214``B'' has a dwarf-like surface gravity ($\log g \sim 5$) comes from the relatively shallow depth of the CO bandheads in the spectrum of ScoPMS~214``B'': comparable in strength to the CO bandheads of the M dwarfs and weaker than the CO bandheads of the M giants.  This is not unusual despite the possibility that ScoPMS~214``B'' may still be contracting toward the main sequence.  Even at 5~Myr ages M stars are expected to have surface gravities that are much more similar to those of dwarfs than to those of giants ($\log g\sim 1$).

From a visual examination of the spectra in Figure~\ref{fig_scopms214b_spec} we estimate that ScoPMS~214``B'' has an M3--M5 spectral type, based on the relative strengths of the \ion{Na}{1} and \ion{Ca}{1} absorption features compared to the other M dwarfs.  Unfortunately, a more accurate classification based on the $K$-band spectrum alone is not possible.  On one hand, the \ion{Na}{1} 2.21~$\micron$ doublet is known to be sensitive to both effective temperature and surface gravity \citep{gorlova_etal03}.  On the other hand, while the \ion{Ca}{1} 2.26~$\micron$ triplet is considered to be a good temperature indicator for G and K stars \citep{ali_etal95}, the scatter at early- to mid-M spectral types is significant \citep{gorlova_etal03}.  This is evidenced by the non-monotonic change in the depth of the \ion{Ca}{1} triplet in the M3--M6 spectral type sequence in Figure~\ref{fig_scopms214b_spec}.  Therefore, we adopt M3--M5 as our final estimate of the spectral type of ScoPMS~214``B.''

The effective temperature corresponding to the M3--M5 range is 3250--2800~K (within errors of $\pm$100~K), according to the field M dwarf temperature scale of \citet[][see their Table~4.1]{reid_hawley05}.  More recent work on M dwarf effective temperatures, supported by highly accurate photometric and interferometric measurements \citep{casagrande_etal08}, finds that the \citeauthor{reid_hawley05} scale systematically overestimates the temperatures of $<$3000~K field M dwarfs by about 100~K.  However, \citet{luhman99} finds that young M dwarfs specifically are significantly {\it hotter} than their older field counterparts.  \citeauthor{luhman99}'s conclusion is based on the requirement that all components of the GG~Tau quadruple system lie on the same 1~Myr theoretical isochrone from the NextGen models of \citet{baraffe_etal98}, and is supported by population age analyses in other young associations, such as IC~348 \citep{luhman99} and the Orion Nebular Cluster \citep{slesnick_etal04}.  Although \citeauthor{luhman99}'s conclusion relies on theoretical isochrones from \citet{baraffe_etal98}, the models in question have been shown to most successfully and, on average, fairly accurately predict the fundamental properties of pre-main-sequence stars \citep{hillenbrand_white04}.  The effective temperature range of 1~Myr M3--M5 dwarfs found by \citet{luhman99} is 3415--3125~K.

Such disagreement at these low effective temperatures is not unusual, given the increasing complexity of stellar spectra at $<$3000~K.  The problem is even more aggravated at young ages, when the lower surface gravities of the objects further affect their photospheric appearance.  Because of its specific pertinence to young M dwarfs, when considering the possibility below that ScoPMS~214``B'' is a member of Upper Scorpius, we will adopt the temperature scale of \citet{luhman99}.  
%At an age of $\sim$5~Myr ScoPMS~214``B'' would be expected to be much closer in surface gravity to the 1 Myr-old components of the GG Tau system than to 1--10~Gyr-old field dwarfs.  

We proceed by examining two probable scenarios: (1) a ``young'' (5~Myr) ScoPMS~214``B'' that is a member of Upper Scorpius, probably as a companion to ScoPMS~214, with $3125\leq\Teff\leq3415$~K, or (2) a ``field-aged'' (1--10~Gyr) ScoPMS~214``B'' that is simply seen in projection along the line of sight toward ScoPMS~214, with 2700~K $\leq\Teff\leq3250$~K.

\subsubsection{Is ScoPMS~214``B'' a Member of Upper Scorpius?}
\label{sec_scopms214_hr}

To decide which of the above two scenarios is valid,
%the membership of ScoPMS~214``B'' in Upper Scorpius, 
and by extension, whether ScoPMS~214A and ``B'' form a physical pair, 
we compare the locations of ScoPMS~214A and ``B'' on an HR diagram with respect to the NextGen model isochrones of \citet{baraffe_etal98}.  Mirroring the approach of \citet{luhman99}, we expect that if ScoPMS~214A and ``B'' were bound and hence co-eval, they should lie on the same isochrone.   
%This mirrors the approach of \citet{luhman99} in establishing an effective temperature scale for young M dwarfs based on the positions of the four GG~Tau components with respect to model isochrones. 
Since the temperature of ScoPMS~214``B'' is $\sim$3000~K regardless of the considered scenario, the use of the dust-free NextGen models is justified.  Indeed, the more recent DUSTY models from the Lyon group \citep{chabrier_etal00} do not extend above 3000~K, since dust is not expected to  form in stellar photospheres at such high effective temperatures.

The HR diagram analysis is illustrated in Figure~\ref{fig_scopms214_hr}.  In the ``young'' ScoPMS~214``B'' scenario we have adopted the mean distance to Upper Scorpius members, $145\pm40$~pc,
%XX change if adopting new distances
for both ScoPMS~214A and ``B''.   The bolometric luminosity of ScoPMS~214``B'' is then $\log L/\Lsun=-2.37\pm0.24$, where we have used bolometric corrections for M3--M5 dwarfs from \citet{tinney_etal93} and \citet{leggett_etal96}.
%, which range from 2.67~mag to 2.84~mag.
In this scenario, ScoPMS~214``B'' lies on the 1~Gyr isochrone (i.e., on the main sequence), in disagreement with the positioning of ScoPMS~214A above the main sequence.  Presuming that ScoPMS~214A is itself not an unresolved binary, the discrepancy indicates that the assumed distance range for ScoPMS~214``B'' is incorrect, and that probably it is not a member of the Upper Scorpius association.  While ScoPMS~214A also lies slightly beneath an extrapolation of the 5~Myr isochrone, its position is not inconsistent with the adopted age for Upper Scorpius, especially given the physical extent ($\sim$40~pc core radius) of the association.

%It is evident from Figure~\ref{fig_scopms214_hr} that in the case when both ScoPMS~214A and ``B'' are assumed to be young, they do not lie on the same NextGen isochrone.  Given that the effective temperature of ScoPMS~214''B'' was adjusted higher so that similarly young M dwarfs in Taurus lie on the same isochrone, we conclude that the assumption that ScoPMS~214``B'' is young is most probably wrong.  

In the ``field-aged'' ScoPMS~214``B'' case, the object is not a member of Upper Scorpius, and hence its heliocentric distance and bolometric luminosity are not constrained.  This case is presented by the shaded region in Figure~\ref{fig_scopms214_hr}.  Given the range of luminosities at which the shaded region intersects the main sequence, the distance to ScoPMS~214``B'' is between 70--145~pc.

%The second case, in which the effective temperature of ScoPMS~214``B'' does not adjustment because of its old age, does ostensibly fulfill the requirement that ScoPMS~214A and ``B'' lie on the same isochrone, in this case the 30~Myr line.  However, this contradicts the assumption that ScoPMS~214A and ``B'' should now have different, not identical ages.  Indeed, in this case the distance and bolometric luminosity of ScoPMS~214``B'' are indeterminate, and the HR diagram data point for ScoPMS~214``B'' probably lies at a lower luminosity.  

Therefore, we conclude that ScoPMS~214``B'' is probably not a member of Upper Scorpius and hence probably not a physical companion to ScoPMS~214.  Instead, it is likely to be a foreground M dwarf seen in projection against Upper Scorpius.  We arrive at this conclusion despite the apparent agreement between the proper motions of ScoPMS~214 and ScoPMS~214``B'' over nearly five years.  The reason for the apparent agreement is the relatively small proper motion of ScoPMS~214 (23~\masyr), which tests the precision limits of our astrometric calibration even over a five-year period.  On-going astrometric monitoring of this system and measurements of the individual radial velocities of the two components will allow us to discern the difference in their space motions.

\section{SURVEY INCOMPLETENESS AND SAMPLE BIASES} 
\label{sec_incompl_bias}

Before addressing the frequency of wide substellar companions in our sample (\S~\ref{sec_anal}), we present a brief summary of the factors that affect the completeness of our survey (\S~\ref{sec_incompleteness_short}), and discuss the various sample biases (\S~\ref{sec_biases}).  The detailed completeness analysis is relegated to the Appendix.

\subsection{Survey Incompleteness} 
\label{sec_incompleteness_short}

The principal factors that influenced the completeness of our deep survey can be divided into three categories: (1) geometrical, defined by the inner and outer working angles (IWA and OWA) of the survey (0$\farcs$55 and 12$\farcs$5, respectively) and by the distribution of heliocentric distances of the sample targets; (2) observational, defined by the flux limits of the survey relative to the predicted brightness of substellar companions; and (3) orbital, defined by the fraction of orbital phase space observed.  For the sake of simplicity in estimating the total survey incompleteness, we have assumed that the distributions of orbital semi-major axis and mass for substellar companions are $dN/d\log a \propto a^0$ and $dN/dM_2 \propto M_2^0$, respectively.  Other common distributions for these parameters are explored in \S~\ref{app_variations}, and are found not to affect the final completeness estimate by more than a factor of 1.24.

%observational, and orbital.  The geometrical incompleteness refers to the combination of limitations imposed by the inner and outer working angles (IWA and OWA) of the survey (0$\farcs$55 and 12$\farcs$5, respectively), and the distribution of heliocentric distances of the sample targets.  The observational incompleteness encompasses factors governing the sensitivity to companions of a given mass: attained imaging contrast and depth, and expected flux ratio of the binary at the age of the system.  The orbital incompleteness is a result of the arbitrariness of the orbital parameters of the secondaries, including orbital semi-major axes, inclinations, and eccentricities.  

We find that the combined completeness of the deep survey to substellar companions in 28--1590~AU %semi-major axes ranges from 64.7\% at the 0.072~\Msun\ hydrogen-burning mass limit to 43.3\% at 
semi-major axes ranges from 64.8\% at the 0.072~\Msun\ hydrogen-burning mass limit to 47.0\% at 
the 0.012~\Msun\ deuterium-burning mass limit.  The deep survey is severely incomplete ($<$30\% completeness) to companions below 0.012~\Msun\ and maximally complete (64.9\%) at and above 0.090~\Msun.  Over the combined 0.012--0.072~\Msun\ brown dwarf mass range, we estimate that the deep survey is complete to $62\%$ of substellar companions with orbital semi-major axes between 28~AU and 1590~AU (\S\S~\ref{app_orb_incompl}--\ref{app_incompl_summary}).  We combine this estimate with the observational results to obtain the underlying substellar companion frequency in \S~\ref{sec_bd_frequency}.

\subsection{Sample Biases} 
\label{sec_biases}

%Ideally, a multiplicity survey would be performed on a complete, volume-limited, unbiased sample.  Given our focus on low-mass substellar companions, such a sample is not optimal since it excludes young primaries around which our sensitivity to low mass secondaries is maximized.

Our survey sample carries an important bias against visual binaries, inherited from the FEPS target selection policy.  The FEPS sample excluded certain types of visual binaries to minimize photometric confusion in {\it Spitzer}'s 1$\farcs$5--30$\arcsec$ beam at 3.6--70~$\micron$ wavelengths \citep{meyer_etal06}.  In particular:
\begin{enumerate}
\item all FEPS sample stars were required to have no projected companions closer than 5$\arcsec$ in 2MASS, and \label{crit_2mass}
\item stars older than 100~Myr were also required to have no projected 2MASS companions closer than 15$\arcsec$, unless the companions were both bluer in $J-K_S$ and fainter at $K_S$ by $>3$~mag.\footnote{For reasons of generating a statistically significant sample size, $<$100~Myr-old stars were allowed to violate this criterion in FEPS.}
\label{crit_feps}
\end{enumerate}

The above two criteria create a non-trivial bias against stellar-mass companions in our AO sample.  Because of the seeing-limited dynamic range of 2MASS %within 5$\arcsec$ of bright stars 
\citep[$\sim4.5$~mag at 5$\arcsec$, $\sim2.5$~mag at 3$\arcsec$; see Fig.~11 in][]{cutri_etal03}, criterion \ref{crit_2mass} excludes near-equal magnitude (i.e., near-equal mass) stellar companions.  Criterion~\ref{crit_feps} then further excludes fainter, red (and hence, lower-mass) companions, although only around the $>$100~Myr-old stars.  

Therefore, any analysis of the stellar multiplicity in our survey would tend to underestimate the true stellar companion rate.  In particular, if we adopt the median distance for the complete AO sample (Table~\ref{tab_sample_stats}) and the orbital period distribution for solar mass binaries from \citet{duquennoy_mayor91}, we find that the above FEPS selection criteria have probably excluded $\sim$25\% of stellar binaries, mostly near-equal mass systems.  We do not address this bias further.  We only note in \S~\ref{sec_cmf} that it has a systematic effect on our estimate of the CMF, in the sense that we have underestimated the relative frequency of near-equal mass binaries.

An additional bias against binary stars, relevant only to the deep portion of our AO survey, is incurred by our on-the-fly selection against $\Delta K_S<4$~mag projected companions at $0\farcs8-13\farcs0$ from our deep-sample coronagraphic targets (criterion \ref{crit_dK} in \S~\ref{sec_selcrit}).  However, by keeping track of which stars were delegated to the shallow sample in this manner, we have accounted for this bias in our analysis of the CMF in \S~\ref{sec_cmf}.

Finally, our AO sample also carries a slight bias against {\it substellar} secondaries because of the second FEPS selection criterion above.
%Such bias is present to a small extent because of FEPS sample selection criterion \ref{crit_feps} above, since the fainter and redder companions that are excluded may in some cases be substellar.  However, 
%Because of the limited dynamic range ($\Delta K_S \lesssim6$~mag) of the 2MASS Point Source Catalog within 12.8$\arcsec$ (the half-width of the Palomar AO camera FOV) and 
%Because the FEPS faint-companion exclusion criterion (\ref{crit_feps} above) is applied only to stars older than 100~Myr, a bias against substellar companions exists in limited cases.  
This bias affects only 100--500~Myr-old targets in the deep sample with well-separated ($\geq$5$\arcsec$) massive brown dwarf secondaries.  Fortunately, because of the shallow depth of 2MASS ($K_S\lesssim15$~mag) and its limited dynamic range ($\Delta K_S\lesssim6$~mag) within our 12$\farcs$5 AO survey radius, the effect of this bias is negligible.  
%Since the FEPS target sample was selected to be uniformly distributed in logarithmic age between 3~Myr and 3~Gyr, less than a quarter of our sample stars fall in the 100--500~Myr age range (where 500~Myr is the upper age limit for the deep sample).  
Based on the range of assumed semi-major axis distributions for substellar companions considered in \S~\ref{app_variations}, we find that this criterion would have excluded $\leq$0.5\% of detectable $\geq$60~\Mj\ substellar companions.  Over the entire substellar companion mass sensitivity range of our survey (13--75~\Mj) the effect of this bias is negligible ($<$0.1\%).  We will therefore ignore it in the rest of the discussion.

% our discussion of the completeness of the survey to substellar companions in the Appendix.
%\S~\ref{sec_incompleteness_short} and in \S~\ref{app}.

%23\% of FEPS stars are in this age range.  0.41/1.37=30\% of 0.55--12.8$\arcsec$ companions are between 5--12.8$\arcsec$ from the primary, if uniform spacing in $\log(a)$. 66\% of 0.012--0.072\Msun brown dwarfs have masses $>0.040$\Msun, assuming $dN/d\log M \propto M^{0.7}$.  So, criterion \ref{crit_feps} misses 0.23*0.30*0.66 = 4.6\% of brown dwarf companions.

%\section{THE MULTIPLICITY OF YOUNG SOLAR ANALOGS \label{sec_multiplicity}}

\section{THE FREQUENCY OF WIDE SUBSTELLAR COMPANIONS}
\label{sec_anal}

Throughout the remainder of this paper we will use the general terms ``substellar companion'' and ``brown dwarf companion'' to refer to a 0.012--0.072~\Msun\ (13--75~\Mj) brown dwarf secondary in a 28--1590~AU orbit around a young Sun-like star, unless otherwise noted.  

%\subsection{Wide Sub-Stellar Companions \label{sec_bd_frequency}}
\subsection{Results from the Present Survey} 
\label{sec_bd_frequency}

%Based on two substellar companion detections around the 100 stars in the deep sample, and on an estimated 58\% detection rate (\S~\ref{sec_incompleteness_short}), the estimated brown dwarf companion frequency is simply $2\%/0.58=3.4\%$.  We determine the confidence limits on this estimate by deriving its probability density distribution using a Bayesian approach.

Having discovered two bona fide brown dwarf companions among the 100 stars in the deep sample, we estimate the range of true substellar companion fractions that these detections represent.  We do so by following a Bayesian approach to derive confidence ranges for the implied frequency of detectable substellar companions, and by applying the incompleteness correction derived in \S~\ref{app_incompl_summary}.

Strictly speaking, the probability of obtaining $x$ successful outcomes (e.g., brown dwarf companion detections) from a number of repetitions of an experiment with a binary outcome is governed by a binomial distribution.  In practice, the large number of experiments (100) and the small number of successful
outcomes in our case ($x=2$) mean that the probability of detecting $x$ brown
dwarfs given an expected mean rate $\mu$ is well approximated by a Poisson probability
distribution:
\begin{equation}
P(x\vert\mu) = \frac{\mu^x {\rm e}^{-\mu}}{x!}. \label{eqn_poisson}
\end{equation}
We are interested in finding what is the probability
distribution for the actual mean rate $\mu$ given $x$ detections, i.e., we seek the probability density function (p.d.f.) $P(\mu\vert x)$.

The result follows from Bayes' Theorem \citep{rainwater_wu47, papoulis84}:
\begin{equation}
P(\mu\vert x) = \frac{P(x\vert\mu) P(\mu)}{P(x)}, \label{eqn_bayes}
\end{equation}
where the $P$'s denote ``probability distributions'' rather than
identical functional forms.  $P(\mu)$ is the ``prior'' and
summarizes our expectation of the state of nature prior to the
observations.  $P(x\vert\mu)$ is the ``likelihood'' that $x$ positive outcomes are
observed given a mean of $\mu$.  $P(\mu\vert x)$, the distribution of interest, is the
``posterior'' probability that the state of nature is $\mu$, given $x$
positive outcomes.
$P(x)$ is a normalization factor equal to the sum of all
probable outcomes 
%$P(x\vert\mu^\prime)$
$P(x\vert\mu)$, given the distribution of the prior
%$P(\mu^\prime)$
$P(\mu)$:
\begin{equation}
%P(x) = \int_{0}^\infty P(x\vert\mu^\prime) P(\mu^\prime) d\mu^\prime.
P(x) = \int_{0}^\infty P(x\vert\mu) P(\mu) d\mu.
\label{eqn_norm}
\end{equation}

We assume no previous knowledge of the state of nature, and adopt a prior that minimizes the introduction of subjective information, imposing only a condition of nonnegativity: $P(\mu)=1$ for $\mu\geq0$, $P(\mu)=0$ for $\mu<0$.  That is, we assume that all positive substellar companion detection rates
% $0\leq\mu\leq100$ 
are equally probable.   
Inserting Equation~\ref{eqn_poisson} into Equations~\ref{eqn_bayes} and \ref{eqn_norm}, we obtain
\begin{equation}
P(\mu\vert x) = P(x\vert\mu) = \frac{\mu^x {\rm e}^{-\mu}}{x!}. \label{eqn_pmu}
\end{equation}
That is, the p.d.f.\ of $\mu$ is a Gamma distribution
that peaks at the observed detection rate $x$ (Fig.~\ref{fig_gamma2}).  Due to the asymmetry of the Gamma distribution, the mean value $\langle\mu\rangle$ is higher than the most likely value $\mu_{\rm ML}$: $\langle\mu\rangle = x+1 = 3 > \mu_{\rm ML}$.

We determine the confidence interval $[\mu_l, \mu_h]$ of the frequency of substellar companions $\mu$ at a desired confidence level CL by integrating $P(\mu\vert x)$ between $\mu_l$ and $\mu_h$.  We set the lower 
and upper bounds $\mu_l$ and $\mu_u$ of the confidence interval CL so that (see Fig.~\ref{fig_gamma2})
\begin{equation}
\int_{\mu_l}^{\mu_u} P(\mu^\prime \vert x) d\mu^\prime = {\rm CL}
\label{eqn_cl}
\end{equation}
and
\begin{equation}
P(\mu_l \vert x) = P(\mu_u \vert x).
\label{eqn_cl_limits}
\end{equation}
Equations (\ref{eqn_cl}) and (\ref{eqn_cl_limits}) define the minimum size confidence interval $[\mu_l, \mu_u]$ for confidence level CL \citep{kraft_etal91}.  The system of equations can not be inverted analytically, and has to be solved for $\mu_l$ and $\mu_u$ numerically.  We do so for the equivalent to the 1, 2, and 3 Gaussian sigma (68.2\%, 95.4\%, and 99.7\%) confidence
intervals.  The respective confidence ranges for $\mu$ are 0.9--3.9, 0.3--6.5, and 0.07--9.9 detectable brown dwarf companions for a survey of 100 stars.

Having thus addressed the statistical uncertainties associated with the small
number of companion detections, we now apply the estimated survey completeness 
correction (62\%) to $\mu_{\rm ML}$ and to the confidence interval limits of $\mu$. 
We find that the most likely rate of occurrence of brown dwarf companions in 28--1590~AU orbits around 3--500~Myr-old F5--K5 stars
is $\mu_{\rm ML}=2\%/0.62=3.2\%$.  The confidence intervals on this estimate are 1.5--6.3\% at the 1$\sigma$ level, 0.5--10.5\% at the 2$\sigma$ level, and 0.1--16.0\% at the 3$\sigma$ level, and are not a strong function of the prior \citep{kraft_etal91}.  The mean frequency, $3\%/0.62=4.8\%$, is higher than the most likely value, but the exact value of the mean frequency is dependent on the Bayesian prior.  The higher mean frequency of wide brown dwarf companions (6.8\%) that we reported in \citet{metchev06} was due to the inclusion of ScoPMS~214``B'' as a substellar companion.  We have now shown that ScoPMS~214``B'' is most probably an unrelated foreground star seen in projection along the line of sight towards ScoPMS~214 (\S~\ref{sec_scopms214_hr}). 

Our results for the frequency of substellar companions are built upon simple assumptions for the semi-major axis and mass distributions of substellar secondaries (\S~\ref{sec_incompleteness_short}; for greater detail, see \S~\ref{app_assumptions}).  However, our conclusions do not depend strongly on these assumptions.
% assumed functional forms of their orbital semi-major axis and mass distributions.  
As we show in \S~\ref{app_variations}, when either or both distributions are varied within empirically reasonable limits, the substellar companion frequency remains unchanged to within a factor of 1.24.  If, as we argue in \S~\ref{sec_bd_desert}, the orbital period distribution of substellar companions is the same as for stellar companions, our frequency estimate is accurate to within a factor of 1.06.

\subsection{Comparison to Previous Companion Searches} 
\label{sec_context}

%\subsubsection{Comparison to Sub-Stellar and Stellar Multiplicity \label{sec_planets_stars}}
\subsubsection{Radial Velocity Surveys} 
\label{sec_rv_comp}

Precision radial velocity surveys for extrasolar planets have revealed that brown dwarf secondaries are unusually rare ($<0.5\%$) in 0--3~AU orbits from G and K stars: a phenomenon termed ``the brown dwarf desert'' \citep{marcy_butler00}.  The dearth of brown dwarfs in radial velocity surveys is evident with respect
to the observed 0--3~AU frequencies of both extra-solar planets
\citep[5--15\%;][]{marcy_butler00, fischer_etal02} and stellar secondaries
\citep[11\%;][]{duquennoy_mayor91} around Sun-like stars.  That is, brown dwarfs are $\approx20$ times rarer than planets and stellar companions in 0--3~AU orbits.  

We found that $3.2^{+7.3}_{-2.7}$\% (2$\sigma$ confidence interval) of young Sun-like stars have 0.012--0.072~\Msun\ companions with semi-major axes between 28 and 1590~AU (\S~\ref{sec_bd_frequency}).  The much wider orbits probed in the present survey prevent a direct parallel with the radial velocity results.  Nevertheless, at face value the evidence indicates that the frequency of wide brown dwarf companions to Sun-like stars is, on average, a factor of $\sim3$ smaller than that of 0--3~AU extrasolar planets, and a factor of $\sim6$ greater than the frequency of 0--3~AU brown dwarfs.  The difference with the exoplanet frequency is not statistically significant.  The frequencies of 28--1590~AU and 0--3~AU brown dwarfs differ at the 98.6\% significance level.  That is, wide brown dwarf companions to Sun-like stars are roughly comparable in frequency to radial velocity extrasolar planets, and are probably more common than radial velocity brown dwarfs.

\subsubsection{Wide Stellar Companions}

Based on the \citet{duquennoy_mayor91} orbital period distribution and multiplicity of Sun-like stars, the frequency of 28--1590~AU stellar companions is $\approx$24\%.  Our estimated frequency
of brown dwarfs is a factor of $\sim8$ smaller, and significantly (at the $1-10^{-8}$ level) so.  Therefore, brown dwarf secondaries are indeed less common than stellar secondaries in the 28--1590~AU orbital range.  
%Combined with the conclusion from the previous comparison (\S~\ref{sec_rv_comp}), this result demonstrates that the deficiency of substellar companions at wide orbital separations from Sun-like stars is less pronounced than in the radial velocity Òbrown dwarf desert.Ó

%However, brown dwarf companions in wide orbits are significantly more common than at 0--3~AU, and may have a frequency comparable to that of 0--3~AU extrasolar planets.

\subsubsection{Other Direct Imaging Surveys for Substellar Companions} 
\label{sec_other_bd_surveys}

A large number of direct imaging surveys have been completed to date, covering a wide range in primary mass and in sensitivity to substellar companions.  Despite the disparate characteristics of these surveys, there are now enough data to analyze the ensemble of the results.

We compare our AO survey to all previously published direct imaging surveys for substellar companions to $\geq$0.2~\Msun\ primaries.  %We set the lower limit on the primary mass to avoid the increase in the fraction of $q\approx1$ binaries in very low-mass (VLM) systems \citep[][and references therein]{burgasser_etal05} that results in a high occurrence rate of brown dwarf companions to VLM stars.  Compared to low-mass ratio ($q\leq0.2$) star--brown dwarf systems, near-equal mass binaries represent a distinctly different outcome from the process of competitive accretion.  Therefore, their multiplicity and frequency do not enhance our understanding of the population properties of low-mass ratio binaries.
We include only surveys targeting $\geq$15 stars that also contain at least a cursory reference to the parent sample statistics and to the substellar companion discovery rates.   All such surveys, to our present knowledge, are listed in Table~\ref{tab_surveys}.   Additional direct imaging surveys certainly exist.  However, any published results from these have tended to report only individual detections.
%, usually individual detections, do not allow a statistical analysis.  
To this list of direct imaging surveys we have also added the radial-velocity results of \citet{marcy_butler00} for comparison. 
For each published survey, Table~\ref{tab_surveys} lists the number, median spectral type, age, primary mass, and heliocentric distance of the sample stars.  For most surveys, these values have been inferred from the description or listing of the sample in the referenced publication.  For some surveys, however, these parameters have been inferred based on the stated focus of the survey (e.g., Sun-like stars), or where appropriate, based on the properties of stars in the solar neighborhood.  Table~\ref{tab_surveys} also lists the maximum probed projected separation, the sensitivity to companion mass, the number of detected brown dwarf companions, and the rate of detection of brown dwarf companions.

Although an incompleteness analysis is crucial for the correct interpretation of survey results, the majority of published results do not contain such.  Therefore, any comparison among surveys has to be based solely on the mean or median survey sample statistics and sensitivities.  
Taking the ensemble statistics of all direct imaging surveys for substellar companions at face value, without accounting for their varying degrees of incompleteness, we find that the mean detection rate is 1.0 substellar companions per 100 stars.  Given the very low number of detections per survey (0--2), the results from all imaging companion surveys are fully consistent with each other.

We have plotted the substellar companion detection rates of all surveys on a primary mass versus outer probed separation diagram in Figure~\ref{fig_surveys}.  The outer probed separation is defined simply as the product of the survey angular radius (generally, the half-width of the imaging detector) and the median heliocentric distance for the survey sample.  The diagram reveals that the surveys with the highest detection rates of substellar companions reside in a distinct locus in the upper right quadrant of the diagram, delimited by the dotted line.  All surveys outside of this region have detection rates $\leq0.6\%$, whereas all surveys within the region have generally higher, 0.5--5.0\% detection rates.  This fact %is independent of 
transcends survey sensitivity considerations.  Some of the most sensitive companion surveys with the smallest likely degrees of incompleteness, such as the radial velocity survey of \citet{marcy_butler00} and the simultaneous differential imaging (SDI) surveys of \citet{masciadri_etal05} and \citet{biller_etal07}, lie outside of the dotted region and have detection rates well below 1\%.  That is, unless all of these highly sensitive surveys did not detect brown dwarf companions through some improbable happenstance, a significant population of brown dwarf companions apparently exists at $\gtrsim$150~AU separations from $\gtrsim$0.7~\Msun\ stars.  Brown dwarf companions appear to be less frequent both at smaller orbital separations from Sun-like stars, and at wide separations from lower-mass stars.  The dearth of brown dwarf companions to 0.2--0.6~\Msun\ stars is likely due to a combination of the lower multiplicity rate of low mass stars and the tendency of low mass binaries to exist predominantly in close-in near-equal mass systems \citep[e.g.,][and references therein]{burgasser_etal07, allen07}.   The surveys with the highest detection rates are those targeting very wide companions to $\sim1$~\Msun\ stars.  

It is important to re-iterate again that none of the detection rates for any of the surveys in Figure~\ref{fig_surveys} have been corrected for systematic or incompleteness effects.  In particular, there is a strong bias against the detection of substellar companions in narrow orbits in all direct imaging surveys because of contrast limitations.  In addition, the position of each survey along the abscissa is based on the median {\it outer} probed separation, whereas most companions are detected at smaller projected separations.  Therefore, the increase in the frequency of substellar companions to $\gtrsim$0.7~\Msun\ stars probably begins well within 150~AU.  In the \S~\ref{sec_bd_desert} we argue that the peak of the brown dwarf companion projected separation distribution may in fact occur at $\sim$25~AU, as would be expected from the \citet{duquennoy_mayor91} binary orbital period distribution.

%Published results for substellar companions to lower-mass ($\sim0.2-0.6\Msun$) stars do not exist, although in an on-going survey of solar neighborhood M dwarfs, \citet{marchal_etal03} do not report any new detections.  
%Although none of the direct imaging surveys that probe $<100$~AU median projected separations have found any substellar companions, brown dwarfs have been imaged at smaller orbital separations from Sun-like stars: e.g., HR~7672B \citep[14~AU;][]{liu_etal02}, HD~130948B/C \citep[47~AU;][]{potter_etal02}, HD~49197B \citep[43~AU;][]{metchev_hillenbrand04}.  Therefore, the $\sim$200~AU limit on the minimum orbital separation of brown dwarfs from $\sim1\Msun$ stars is by no means a hard boundary.  

\section{DISCUSSION} 
\label{sec_discussion}

\subsection{The Sub-Stellar and Stellar Companion Mass Function} 
\label{sec_cmf}

The salient characteristic of the present imaging survey is its high sensitivity
to low-mass ($M_2\leq0.1\Msun$) companions to solar analogs, i.e., to
systems with mass ratios $q\lesssim0.1$.   We found only seven 
% XX check if ScoPMS 214B, but also results from Baraffe et al. (1998) models.
such companions among 74
%was 74
 binary and one triple systems: the two brown dwarfs HD~49197B and HD~203030B, and the 0.08--0.14~\Msun\ stars HD~9472B, HE~373B, RX~J0329.1+0118B, HD~31950B, and PZ99~J161329.3--231106B (Table~\ref{tab_stellar_comps}).  A na\"{i}ve expectation from the MF of isolated objects \citep{kroupa01, chabrier01} would require approximately as many $<0.1$~\Msun\ companions as there are $>0.1$~\Msun\ companions.  Therefore, it appears that there is a dearth of widely-separated {\it both substellar and low-mass stellar} companions to Sun-like stars.

To assess the reality and magnitude of this effect we need a uniform survey of a well-characterized sample of binaries.  Unfortunately, our full survey sample is not adequate for such an analysis because the imaging depths of the deep and the shallow sub-surveys are vastly different, and because the sample is subjected to the combined effect of three different biases against binary stars (\S~\ref{sec_biases}).  We could, in principle, focus only on the deep survey of 100 young stars, for which we have a well-characterized completeness estimate.  However, doing so would not avoid any of the binarity biases.  Furthermore, the deep survey sample contains only 19 of all 75 
%was 74
binaries and triples, only six of which are in the 0$\farcs$55--12$\farcs$5 angular separation range, for which we estimated incompleteness (\S~\ref{sec_incompleteness_short}).  This number is too low for a meaningful analysis of the CMF.

Nevertheless, we can un-do some of the binarity biases and recover certain rejected stellar secondary companions by re-visiting how binaries were removed from the deep sample during survey observations.  We discuss this procedure and reconstruct a sample that is minimally biased against binaries in the following.

%that is sampled most uniformly and deeply.  

\subsubsection{Defining a Minimally Biased Sample} 
\label{sec_ad_sample}

We construct a less biased, larger sample of stars by adding to our 100-star deep sample all other $\leq$500 Myr-old stars that were initially selected to be in the deep sample 
%(based on the criteria outlined in \S~\ref{sec_selcrit}) 
but for which no coronagraphic exposures were taken because of the discovery of a close-in (0$\farcs$8--13$\farcs$0) $\Delta K_S<4$~mag companion (see \S~\ref{sec_selcrit}).  Since we did not inherit this bias from the larger FEPS program sample, but rather imposed the criteria ourselves, we knew the parent sample and were able to un-do the bias exactly.  The resulting ``augmented'' deep (AD) sample is minimally biased against binaries to the extent to which we controlled the sample generation.

There are 28 binaries excluded in this manner, that contribute to a total of 128 young stars in the AD sample.  Among these are a total of 46 
%was 46
binaries and one triple, of which 30 
% was 30
systems (including the triple) have companions between 0$\farcs$55 and 12$\farcs$5 from the primary.  Members of the AD sample are distinguished in the last column of Table~\ref{tab_stellar_comps}, where the 30 
% was 30
members with 0$\farcs$55--12$\farcs$5 companions are marked with ``AD$_{30}$''.  

We assume that the young binaries added from the shallow sample do not have additional fainter tertiary companions between $0\farcs$55--12$\farcs$5 that would have been detectable had we exposed them to the depth of the longer coronagraphic images.  Given the $\approx$10\% ratio of double to triple systems in the study of \citet{duquennoy_mayor91}, and the fact that the 28--1590~AU  orbital range ($\approx10^{4.7}-10^{7.3}$ days at 1~\Msun\ total mass) includes approximately 42\% of all companions (0--$10^{10}$-day periods) probed in \citet{duquennoy_mayor91}, we would expect that $\approx0.10\times0.42\approx4\%$ of systems have a tertiary component in a 28--1590~AU orbit.  In comparison, the $1:46\approx2\%$
%was 46:1
ratio of triples to binaries in our sub-sample indicates that such an assumption is not unreasonable: on average, we may have missed one low-mass (possibly substellar) tertiary component.  Therefore, we have potentially suffered only a small loss in completeness by including stars for which we do not have deep coronagraphic exposures.

The AD and the AD$_{30}$ samples remain biased against binaries, although mostly against near-equal mass systems (\S~\ref{sec_biases}).  Because we have placed an upper age limit of 500~Myr for membership in these samples and because of the logarithmically uniform distribution of stellar ages in the parent FEPS sample, the bias against lower mass stellar secondaries (FEPS binarity criterion~\ref{crit_feps}; \S~\ref{sec_biases}) affects less than a quarter of the stars in the AD sample: only those that are 100--500~Myr old.

The detectability of the AD$_{30}$ secondaries within the greater AD sample is  subject to the same set of target selection criteria and to the same geometrical, observational, and orbital incompleteness factors (see \S~\ref{app_incompleteness_analysis}) as for the deep survey.  Therefore we can estimate the completeness of the AD sample to the detection of secondaries in 28--1590~AU semi-major axes in the same manner as done for the deep survey.  

The completeness to 0.012--0.072~\Msun\ substellar companions in 28--1590~AU semi-major axes in the deep survey ranges from 47.0\% to 64.8\%, depending on companion mass (see %Fig.~\ref{fig_mass_incompl}a).  
\S~\ref{app_incompleteness_analysis}).
For masses $\geq$0.090~\Msun\ the deep survey is maximally complete at 64.9\%.  The integrated completeness to 0.01--1.0~\Msun\ companions is $\approx64\%$ (cf., 62\% integrated completeness to 0.012--0.072~\Msun\ substellar companions; \S~\ref{sec_incompleteness_short}).  Given the 30 
%was 30
0$\farcs$55--12$\farcs$5 binaries in the AD$_{30}$ sample, we would expect a total of $30/0.64\approx47\pm9$
% was $30/0.64\approx47\pm9$ 
binaries with 28--1590~AU semi-major axes in the 128-star AD sample, where the error on that estimate is propagated as $\sqrt{30}/0.64$.
% was $\sqrt{30}/0.64$.  
(By pure coincidence, this is 
%almost 
exactly how many multiple systems (47) are present in the AD sample.)  The incompleteness-corrected frequency of 0.01--1.0~\Msun\ companions in 28--1590~AU orbits in the AD sample is thus $47/128=37\pm7\%$.  This is somewhat higher than the 24\% integrated over the corresponding $10^{4.7}$--$10^{7.3}$-day orbital period from \citet{duquennoy_mayor91}.  Despite the bias against binaries, the higher multiplicity fraction of stars in our survey is not unexpected  because of our superior sensitivity to very low mass companions and our focus on young stars, which tend to more often be found in multiples \citep[e.g.,][]{ghez_etal93, ghez_etal97}.

\subsubsection{The Distribution of Companion Masses} 
\label{subsec_cmf}

In their G dwarf multiplicity study, \citet{duquennoy_mayor91} found that the MFs of isolated field stars and of 0.1--1.0~\Msun\ stellar companions to solar-mass primaries were indistinguishable.  We now re-visit this conclusion in light of our more sensitive imaging data and in the context of more recent determinations of the field MF.

The mass ratio distribution for our selection of 30
% was 30
 young binaries in the AD$_{30}$ sample is shown in Figure~\ref{fig_cmf}.  The distribution is fit well by a power law of the form
\begin{equation}
\log \left(\frac{d N}{d \log q}\right) = \delta \log q + b,		\label{eq_pl}
\end{equation}
equivalent to $d N/d \log q \propto q^{\delta}$, or $d N/d q \propto q^{\delta-1} \equiv q^{\beta}$.
% \equiv M_2^{-\alpha}$ is shown.  
The best-fit value for the power law index is $\delta=0.61$,
%$\delta=0.7$, 
or equivalently, $\beta=0.39\approx0.4$.
%$\beta=-0.3$.  
The reduced $\chi^2$ of the fit is adequate, 1.5, and given only three of degrees of freedom a higher-order functional fit is not warranted. The $\chi^2$ contours of $\beta$ and $b$ indicate that the parameters are correlated.  By integrating over all possible values for $b$, we find that the 68\% (one Gaussian $\sigma$) and 95\% confidence intervals for $\beta$ are $-0.75<\beta<-0.03$ and $-0.93<\beta<0.14$, 
%$-0.6<\beta<0.0$ and $-0.8<\beta<0.2$, 
respectively.

We compare this mass ratio distribution to the known MF of isolated field objects from \citet{chabrier03}.  Because the masses of the primary stars in our sample are distributed closely around 1~\Msun\ (Fig.~\ref{fig_sptype_mass_hist}b), we can directly compare the distribution of the (unitless) mass ratios to the field MF (in units of \Msun).  That is, the mass ratio distribution of our sample is essentially equivalent to the CMF in units of \Msun\ since $q=M_2/M_1 \approx M_2/\Msun$.  The power law index $\beta$ of the CMF is analogous to the linear slope $\alpha$ (Salpeter value $-$2.35) of the field MF.   As is evident from Figure~\ref{fig_cmf}, the CMF of our sample of young binaries is very different (reduced $\chi^2=7.6$) from the MF of field objects.  

A potentially more sensitive comparison between the observed mass ratio distribution and any model MFs (e.g., log-normal, power law) could be obtained using a Kolmogorov-Smirnov (K-S) test.  We do not perform Monte Carlo simulations to degrade the MF models to match the observed data, as would be necessary in the rigorous sense, but instead compare the models to the incompleteness-corrected data.  Although the K-S test is not strictly applicable with such an approach, the results from the test are nevertheless illustrative.  Thus, a one-sample Kolmogorov-Smirnov test finds only a $2\times10^{-8}$
% XXX needs to be checked
probability that the observed CMF originates from the log-normal field MF of \citet{chabrier03}.  The K-S probability that the fitted power law in Equation~\ref{eq_pl} with $\beta=-0.39$
%$\beta=-0.3$
is the correct parent CMF is 7\%.  
% XXX needs to be checked.
Ostensibly the best agreement (58\% K-S probability) with the incompleteness-corrected data is reached by a log-normal mass ratio distribution with a mean and standard deviation of 0.39.  A similar log-normal CMF was inferred independently by \citet{kraus_etal08} in their analysis of resolved binaries in Upper Scorpius.  However, we note that in our case the difference between the probabilities of the power-law and log-normal parent CMFs (7\% versus 58\%) is not statistically significant in the context of the K-S test.
%Therefore, because the K-S test is not strictly suited to analyzing incompleteness-corrected data, 
Therefore, given the already adequate reduced $\chi^2$ of the power law fit from Equation~\ref{eq_pl}, we disregard the potentially better, but statistically less well motivated, agreement with the data of the higher-order log-normal parameterization (three free parameters), in favor of the lower-order (two free parameters) power law.
%K--S prob = 7.3\% for the adopted power law ($x=-0.63, b=1.45$; or $x=-0.60, b=1.41$).  $\Chu^2_{p.d.f.}=0.55
%What about a truncated Gaussian? $\mu=0.13; \sigma=0.42$; normalization = 15.33.  K--S prob = 44.5\% ; \Chi^2_{p.d.f.} = 0.30
% Log-normal: $\mu=0.392, \log \sigma=0.394$, normalization = 10.20; K--S prob = 58.1\%. \Chi^2_{p.d.f.} = 0.10

A value near zero for our CMF power law exponent $\beta$ is consistent with the MF of $<0.1$~\Msun\ objects in the field \citep[$-1.0<\alpha\lesssim0.6$;][]{chabrier01, kroupa02, allen_etal05} and in young stellar associations \citep[$-1\lesssim \alpha\lesssim0$;][]{hillenbrand_carpenter00, slesnick_etal04, luhman04b}.  However, the monotonic rise of the CMF throughout the entire 0.01--1.0~\Msun\ mass range and the lack of a turnover near 0.1~\Msun\ disagree with MF determinations for stars in the 0.1--1.0~\Msun\ interval, where $\alpha$ ranges between $-$0.5 and $-$2 \citep{kroupa02}.  That is, in the stellar mass regime, the CMF and the MF of isolated stars are distinctly different.

We should note that the results from our companion survey may not be ideally suited for determining the CMF of both brown dwarf and stellar companions.  Indeed, we recall that our AD sample is biased against various types of visual binaries (\S~\ref{sec_biases}).  However, as we discussed in \S~\ref{sec_ad_sample}, the bias against binarity in the AD sample is mostly against near-equal mass systems, the secondaries in which would populate the highest mass ratio bin in Figure~\ref{fig_cmf}.  That is, the power-law index $\beta$ of the CMF would only further increase in value if the bias against near equal-mass binaries in our survey sample were taken into account, and the CMF would become even more disparate from the field MF.

Our conclusion counters the established view that the MF of 0.1--1.0~\Msun\ binary components is indistinguishable from the MF of isolated objects.   In arriving at the original result, \citet{duquennoy_mayor91} had compared the $0.1 < q \leq 1.0$ binary mass ratio distribution of their sample stars to an earlier form of the field MF from \citet{kroupa_etal90}.  Since the mass ratio distribution of $q>0.1$ binaries in our sample is consistent with that of \citeauthor{duquennoy_mayor91} (see Fig.~\ref{fig_cmf}b), the difference in the results stems from our superior sensitivity to $q\leq0.1$ binaries and from the recently improved knowledge of the MF of low-mass ($<0.2$~\Msun) stars in the field.

Similar conclusions were reached independently by \citet{shatsky_tokovinin02} and by \citet{kouwenhoven_etal05} in their direct imaging studies of the visual multiplicity of intermediate mass (2--20~\Msun) B and A stars.  These two surveys found that the mass ratio distribution of 45--900~AU intermediate mass binaries follows an $f(q)\propto q^{\beta}$ power law, where $\beta$ is $-$0.5 \citep{shatsky_tokovinin02} or $-$0.33 \citep{kouwenhoven_etal05}.  Our determination that $\beta=-0.39\pm0.36$
%$\beta = -0.3\pm0.3$ 
for companions to solar mass stars indicates that the shape of the CMF found by \citet{shatsky_tokovinin02} and \citet{kouwenhoven_etal05} is not specific to intermediate mass stars.
Considered together, these three sets of results provide a strong evidence for a significant difference between the MFs of wide secondaries and of isolated field objects.  That is, the mass ratio distribution of 28--1590~AU binaries is inconsistent with random pairing of stars drawn from the IMF over a vast range of primary and companion masses.  We discuss the implications of this conclusion on shaping the dearth of brown dwarf secondaries to stars below.

\subsection{The Brown Dwarf Desert as a Result of Binary Star Formation} 
\label{sec_bd_desert}

The inferred 0.01--1.0~\Msun\ CMF (\S~\ref{subsec_cmf}) naturally explains the scarcity of wide brown dwarf companions without the need to invoke formation or evolutionary scenarios specific to substellar companions.  
%A similar conclusion was drawn by \citet{kouwenhoven_etal07} for wide visual substellar companions to A stars.  
The functional form of the wide-binary CMF is also consistent with results from radial velocity studies.  Thus, \citet{mazeh_etal03} found that the CMF of K-dwarf binaries in 0--4~AU orbits is also a rising function of mass over the 0.07--0.7~\Msun\ range.  Their data are consistent with a power-law index of $\beta\approx-0.4$, in full agreement with the $-0.3\leq \beta \leq -0.5$ values for 28--1590~AU binaries found by \citet{shatsky_tokovinin02}, \citet{kouwenhoven_etal05, kouwenhoven_etal07}, and here.  

It may be argued perhaps that, given the disparate sensitivity systematics and statistical treatments in these diverse samples, such an overall agreement is merely coincidental.  Indeed,
{\it differences} in the mass ratio distributions of short- vs.\ long-period binaries within {\it single} uniform samples have been previously suggested, with dividing periods of 1000 days \citep[$\sim2$~AU;][]{duquennoy_mayor90} or 50 days \citep[$\sim0.3$~AU;][]{halbwachs_etal03}.  However, subsequent analyses by \citet{duquennoy_mayor91} and \citet{mazeh_etal03} have shown that the evidence for such discontinuities was inconclusive because of relatively small number statistics.  %Therefore, there is reason to believe that the CMF of wide visual companions and close-in spectroscopic companions may be similar.  
At the same time, the combined set of direct imaging and spectroscopic data referenced here point to an approximately uniform functional form for the CMF over 1.5 orders of magnitude in primary mass (0.6--20~\Msun), 3.3 orders of magnitude in companion mass (0.01--20~\Msun), and 4.7 orders of magnitude in physical separation (0.03--1590~AU).  That is, we see strong evidence for a universally uniform shape of the CMF.

Given such universality, it is interesting to consider whether the CMF can explain the very low frequency of brown dwarfs not only in direct imaging, but also in radial velocity surveys.  Because stellar and substellar companions to Sun-like stars appear to be derived from the same CMF (\S~\ref{subsec_cmf}), we can presume that the \citet{duquennoy_mayor91} period distribution of $\geq$0.1~\Msun\ stellar secondaries also holds for substellar companions.  Based on this period distribution, the fraction of all secondary companions in 0--3~AU orbits is $\approx$22\%.  Brown dwarfs account for $\leq0.5\%/22\%=2.3\%$ of these.  For comparison, brown dwarfs account for $\sim3.2\%/42\%=7.6\%$ of all secondaries in 28--1590~AU periods.  In the context of our inferred power-law CMF, we find that the value of the index $\beta$ would need to be as high as $0.2$ to reproduce the $\sim$3 times smaller relative frequency of radial velocity brown dwarfs.  This does not agree well with our 95\% confidence limits on $\beta$ ($-0.93<\beta<0.14$; \S~\ref{subsec_cmf}).  However, we also noted that our estimate for $\beta$ is systematically underestimated because of the bias against near-equal mass binaries in our AO sample.
  %In addition, we point out that the 0.5\% brown dwarf detection rate in radial velocity surveys is based on measurements of $M_2 \sin i$.  While it is true that because of the degeneracy with orbital inclination $i$ some of the detected brown dwarf companions may in fact be low-mass stars, it also means that some of the detected planets may in fact be brown dwarfs.  
It is therefore conceivable that the radial velocity brown dwarf desert around G stars represents just the low-mass, narrow-orbit end of a CMF that spans 3.3~dex in secondary mass and 4.7~dex in orbital semi-major axis.  {\it The problem that would need to be addressed then is not why brown dwarf companions specifically are so rare, but why the CMF differs so significantly from the MF of isolated substellar and stellar objects over all orbital ranges.}

In such a universal CMF scenario,
% the low frequency of brown dwarf companions to Sun-like stars is a consequence of the stellar binary period distribution.  
brown dwarfs would be expected to peak in frequency at semi-major axes determined by the binary period distribution: at $\approx31$~AU from solar mass stars, or at projected separations of $\approx31/1.26=25$~AU (see \S~\ref{app_assumptions} for explanation of factor of 1.26).  
At first glance, this is not consistent with the diagram of survey detection rates in Figure~\ref{fig_surveys}, where we found that (prior to correction for survey incompleteness) the highest detection rates occurred in surveys probing projected separations $\gtrsim150$~AU.  However, we pointed out that Figure~\ref{fig_surveys} compares only the median {\it outer} projected separations probed by the various surveys, whereas most companions tend to be discovered at smaller projected separations (\S~\ref{sec_other_bd_surveys}).  In addition, we need to consider that projected separations of 25~AU are usually well within the contrast-limited regime of existing direct imaging surveys of young nearby (50--100~pc) stars.  Our own survey is less than 40\% complete to objects at the hydrogen-burning mass limit in 31~AU semi-major axis orbits (see %Fig.~\ref{fig_mass_incompl}b).  
\S~\ref{app_incompleteness_analysis}).
That is, a number of $\sim30$~AU brown dwarfs around solar-mass stars may have simply been missed in direct imaging surveys because of insufficient imaging contrast.  

Unfortunately, neither of the two most sensitive direct imaging surveys that probe well within 150~AU \citep{masciadri_etal05, biller_etal07} detect any substellar companions.  However, their sample sizes are not large (54 and 28, respectively), and the null detection rates do not place significant constraints on the universal CMF hypothesis.  Conversely, the recent discovery of several probable radial velocity brown dwarfs in $>3$~AU orbits by \citet{patel_etal07} lends support to the idea that brown dwarfs are more common at wider separations, as would be inferred by extrapolation from the \citet{duquennoy_mayor91} orbital period distribution for higher-mass, stellar companions.

Finally, the $\approx$0.012~\Msun\ ($\approx$13~\Mj) deuterium-burning mass, above which we limit our analysis, does not necessarily mark the bottom of the MF of isolated objects.  Based on results from three-dimensional smoothed particle hydrodynamic simulations, \citet{bate_etal02} and \citet{bate_bonnell05} estimate that the opacity limit for gravo-turbulent fragmentation may be as low as 3--10~\Mj.  Adopting 3~\Mj\ as the limit and extrapolating the inferred CMF to $<$13~\Mj\ masses, we find that sub-deuterium-burning ``planetary-mass'' companions, if able to form through gravo-turbulent fragmentation, exist in $\geq$30~AU orbits around only $\lesssim$1\% of Sun-like stars.
 %Therefore, the lack of detection of such objects in existing surveys \citep[e.g.,][]{biller_etal07} of solar analogs is not surprising.

%\subsection{Discuss Binarity vs.\ Disks? XX}

%We don't have adequate information on the multiplicity properties of the parent FEPS sample. ``There'd be dragons there,'' says Mamajek.  But we could do an analysis similar to the one with the augmented 128-star sample: include all stars in the deep sample and stars that were thrown out of the deep sample at the telescope because of a $\Delta K_S<4$~mag companion.  Could discuss 24$\micron$, 70$\micron$, and 160$\micron$ excesses as a function of binary separation.

%Need: most up-to-date list of excesses in FEPS (retrievable from the FEPS database?).  Do any of the binaries in the 128-star sample have excesses?

\section{CONCLUSION} 
\label{sec_conclusion}

We have presented the complete results from a direct imaging survey for substellar and stellar companions to 266 Sun-like stars performed with the Palomar and Keck AO systems.  We discovered two brown dwarf companions in a sub-sample of 100 3--500~Myr-old stars imaged in deep coronagraphic observations.  Both were already published in \citet{metchev_hillenbrand04, metchev_hillenbrand06}.  In addition, we discovered 24
% was 24
 new stellar companions to the stars in the broader sample, five of which are in very low mass ratio $q\sim0.1$ systems.  %Separately, we established the physical association of a bona fide companion at the stellar/substellar boundary to the Pleiad HII~1348. 

Following a detailed consideration of the completeness of our survey, we found that the frequency of 0.012--0.072~\Msun\ brown dwarf companions in 28--1590~AU orbits around 3--500~Myr-old Sun-like stars is $3.2^{+3.1}_{-2.7}$\% (2$\sigma$ limits).  This frequency is 
%consistent with the frequency of extrasolar giant planets in 0--3~AU orbits, is 
marginally higher than the frequency of 0--3~AU radial velocity brown dwarfs, and is significantly lower than the frequency of stellar companions in 28--1590~AU orbits.  The frequency of wide substellar companions is consistent with the frequency of extrasolar giant planets in 0--3~AU orbits.
%Therefore, we conclude that wide substellar companions are less common compared to wide stellar secondaries, although the deficiency may not be as stark as in the radial velocity brown dwarf desert.  
A comparison with other direct imaging surveys shows that substellar companions are most commonly detected at $\gtrsim150$~AU projected separations from $\gtrsim0.7~\Msun$ stars.  However, because of bias against the direct imaging of faint close-in companions, brown dwarf secondaries are likely  also common at smaller projected separations.

Considering the two detected brown dwarf companions as an integral part of the broader spectrum of stellar and substellar companions found in our survey, we infer that the mass ratio distribution of 28--1590~AU binaries, and hence, the MF of 28--1590~AU secondary companions to solar-mass primaries, follows a $d N/d M_2\propto M_2^{\beta}$ power law, with $\beta=-0.39\pm0.36$ 
%$\beta=-0.3\pm0.3$
(1$\sigma$ limits).  This distribution differs significantly from the MF of isolated objects in the field and in young stellar associations, and is inconsistent with random pairing of individual stars with masses drawn from the IMF.  In this context, the observed deficiency of substellar relative to stellar companions at wide separations arises as a natural consequence of the shape of the CMF, and does not require explanation through formation or evolutionary scenarios specific to the substellar or low-mass stellar regime.

Comparing our CMF analysis to results from other direct imaging and radial velocity surveys for stellar and substellar companions, we find tentative evidence for universal behavior of the CMF across the entire 0--1590~AU orbital semi-major axis and the entire 0.01--20~\Msun\ companion mass range.  Such a universal CMF is not inconsistent with the marked dearth of brown dwarfs in the radial velocity brown dwarf desert around Sun-like stars.  That is, the properties of brown dwarf companions at any orbital separation are conceivably an extension of the properties of stellar secondaries.  Hence, we predict that the peak in semi-major axes of brown dwarf companions to solar-mass stars occurs at $\approx30$~AU.  Extrapolating the inferred CMF to masses below the deuterium burning limit, we find that if 0.003--0.012~\Msun\ ``planetary-mass'' secondaries can form through gravo-turbulent fragmentation, they should exist in $\geq30$~AU orbits only around less than 1\% of Sun-like stars.

\acknowledgments

%{\bf Acknowledgments.}  
We would like to thank Richard Dekany, Mitchell Troy, and Matthew Britton for sharing with us their expertise on the Palomar AO system, Rick Burress and Jeff Hickey for assistance with PHARO, Randy Campbell, Paola Amico, and David
Le Mignant for their guidance with using Keck AO, Keith Matthews and Dave Thompson for help with NIRC2, and our telescope operators at the Palomar Hale
and Keck~II telescopes.  We are also grateful to Keith Matthews for loaning us a pinhole mask for the astrometric calibration of PHARO, and to both Richard Dekany and Keith Matthews for key insights into the design of the calibration experiment.% and the interpretation of the results.  
Use of the FEPS Team database has proven invaluable throughout
the course of our survey.  We thank John Carpenter for building and
maintaining the database.  For the target selection, age-dating, and determination of distances to the FEPS sample stars, we acknowledge the tremendous amount of work performed by Eric Mamajek.  This publication makes use of data products from the Two Micron All Sky Survey, %which is a joint project of the University of Massachusetts and
%the IPAC/California Institute of Technology, 
funded by the NASA and the NSF.  The authors also wish to extend special
thanks to those of Hawaiian ancestry on whose sacred mountain of Mauna
Kea we are privileged to be guests.  Support for S.A.M.\ was provided by NASA through the {\it Spitzer} Legacy Program under contract 1407 and through the {\it Spitzer} Fellowship Program under award 1273192.  Research for this paper was also supported by the NASA/Origins R\&A program.

\facility{{\it Facilities:} Keck II Telescope, Palomar Observatory's 5~meter Telescope} 

\appendix

\section{INCOMPLETENESS OF THE DEEP SURVEY} 
\label{app}

Here we examine the factors affecting the sensitivity of the deep survey to substellar companions (\S~\ref{app_factors_incompl}), and, based on several assumptions about the semi-major axis and mass distributions of wide substellar companions (\S~\ref{app_assumptions}), we estimate the completeness of the survey (\S~\ref{app_incompleteness_analysis}).  We find that variations in the parameters of the semi-major axis and mass distributions have little effect (\S~\ref{app_variations}) on the final completeness estimate.  This final estimate (\S~\ref{app_incompl_summary}) is used in \S~\ref{sec_anal} in combination with the observational results from our survey to obtain the actual frequency of substellar companions.

\subsection{Factors Affecting Survey Completeness}
\label{app_factors_incompl}

Several factors need to be taken into account when estimating the
detectability of substellar companions to our stars.
These include: (i) possible sample bias against
stars harboring substellar secondaries, (ii) choice of substellar
cooling models, (iii) observational constraints (i.e., survey radius, imaging
contrast, and depth), and (iv) physical parameters of the
stellar/substellar systems (flux ratio, age, heliocentric distance,
orbit).

As discussed in
\S~\ref{sec_biases}, the deep sample is largely unbiased toward substellar companions, i.e.,
factor (i) can be ignored.  For the basis substellar cooling models (ii) we rely on the DUSTY and COND models of the Lyon group \citep{chabrier_etal00, baraffe_etal03}.  These have
been used, either alone or in parallel with the models of the Arizona group \citep{burrows_etal97}, in all other studies of substellar multiplicity.  Our choice therefore ensures that our
results will be comparable with the existing work on the
subject.  The remaining factors (iii and iv) motivate the rest of the discussion here.

\subsection{Assumptions} 
\label{app_assumptions}

We will base our incompleteness analysis on three assumptions: (1) 
that the distribution 
of semi-major axes $a$ of substellar companions to stars is flat per unit logarithmic interval of semi-major axis, $dN/d\log a \propto a^0$ (or equivalently, $dN/da \propto a^{-1}$) between 10~AU and 
2500~AU, (2) that this implies a logarithmically flat distribution in {\it projected}
separations $\rho$: $dN/d\log \rho \propto \rho^0$ (i.e., $dN/d\rho \propto \rho^{-1}$), and (3) that the mass function of substellar companions is flat per linear mass interval ($dN/dM_2 \propto M_2^{\beta}=M_2^0$) between 0.01\Msun\ and 0.072\Msun. These assumptions, albeit
simplistic, have some physical basis into what is presently known about
binary systems and brown dwarfs.  We outline the justification for each
of them in the following.

\paragraph{Assumption (1).}
Adopting a total (stellar+substellar) system mass of 1~\Msun,
the 10--2500~AU range of
projected separations corresponds approximately to orbital periods of
$10^4-10^{7.5}$~days.  This range straddles the peak (at $P=10^{4.8}$~days; $a=31$~AU), and
falls along the long-period slope of the Gaussian period distribution
of G-dwarf binaries \citep{duquennoy_mayor91}.
If we were to assume a similar formation scenario for brown dwarfs and stars,
brown dwarf secondaries
would also be expected to fall in frequency beyond $\sim$30~AU separations.  However, 
our limited amount of knowledge about brown dwarf companions suggests the opposite: brown dwarf
secondaries may appear as common as stellar secondaries at $>$1000~AU
separations \citep{gizis_etal01}, whereas a brown dwarf desert
exists at $<$3~AU semi-major axes
\citep{marcy_butler00, mazeh_etal03}.  A smattering of brown
dwarfs have been discovered in between.
A logarithmically flat distribution of
semi-major axes for substellar companions, $dN/d\log a \propto a^0$, or equivalently $dN/da \propto a^{-1}$, represents a middle ground between the known distribution of stellar binary orbits and the possible orbital distribution of known brown dwarf companions.  The
assumption is also attractive because of its conceptual and
computational simplicity.  As we discuss in \S~\ref{app_variations}, varying the linear exponent on the semi-major axis distribution between 0 and $-1$, or adopting a log-normal semi-major axis distribution as motivated by the \citet{duquennoy_mayor91} binary period distribution,
%a linearly flat ($dN/da \propto a^0$) semi-major axis distribution, or distributions such as those for sun-like binaries \citep[log-normal;][]{duquennoy_mayor91}, or for radial velocity planets \citep[$dN/da \propto a^{-0.6} for solar-mass stars;][]{cumming_etal08}, 
changes the overall completeness estimate by a factor of $\leq1.20$.

\paragraph{Assumption (2).}
For a random distribution of orbital
inclinations $i$ on the sky, the true and apparent physical separations are
related by a constant multiplicative factor: the mean value of
$\sin i$.  However, a complication is
introduced when relating the projected
separation to the true semi-major axis because of the need to consider orbital eccentricity.  Because an
object spends a larger fraction of its orbital period near the apocenter than near
the pericenter of its orbit, the ratio of the semi-major axis to the
apparent separation will tend to values $>$1.  Analytical
treatment of the problem \citep{couteau60, vanalbada68} shows that this
happens in an eccentricity-dependent manner.  Yet, when considering
the eccentricity distributions of observed binary populations
\citep{kuiper35a, kuiper35b, duquennoy_mayor91, fischer_marcy92}, both
analytical \citep{vanalbada68} and empirical Monte Carlo
\citep{fischer_marcy92} approaches yield the same
identical result: $\langle\log a\rangle \approx \langle\log \rho\rangle +
0.1$.  That is, the true semi-major
axis and the measured projected separation are, on average, related by a
multiplicative factor of 1.26, such that $\langle a\rangle=1.26 \langle\rho\rangle$.  Given assumption (1),
this then confirms the appropriateness of the current assumption that $dN/d\log \rho \propto \rho^0$.  Furthermore, it allows us to relate the projected separations of an ensemble of visual companions to their expected semi-major axes in a mean statistical sense.

\paragraph{Assumption (3).} 
The assumption for a linearly flat substellar mass distribution ($dN/dM_2\propto M_2^\beta$; $\beta=0$) parallels results from spectroscopic studies of the initial mass function (IMF) of low-mass objects in star-forming regions \citep{briceno_etal02, luhman_etal03a, luhman_etal03b, slesnick_etal04, luhman_etal04}, which 
%show a gradual decrease of the number of objects below the hydrogen-burning mass limit.  
are broadly consistent with $\alpha\sim0$ (where $\alpha$ is the exponent in $dN/dM \propto M^\alpha$).
Independently, in a recent analysis of the field substellar mass function (MF), \citet{allen_etal05} find $\alpha=0.3\pm0.6$, also consistent with zero.  Therefore, assuming that the {\it substellar} MFs in young stellar associations and in the field are representative of the MF of wide {\it substellar} companions, we adopt a linearly flat $dN/dM_2 \propto M_2^0$ CMF for our analysis.  This is consistent with our subsequent fit to the CMF in \S~\ref{subsec_cmf}, where we determine that $\beta$ is in fact $-0.39\pm0.36$ over the entire 0.01--1.0~\Msun\ substellar and stellar companion mass range.

The latter result may seem circuitous, since 
%we eventually conclude that the MF of isolated objects and the CMF are significantly different (\S~\ref{subsec_cmf}), and that 
the derivation that $\beta$ is near zero is in fact dependent on the initial assumption that $\beta$ is zero.  Nevertheless, we find that the initial guess for the CMF exponent is largely unimportant.  As we discuss in \S~\ref{app_variations}, initial values for $\beta$ ranging between $-1$ and 1 change the overall completeness estimate by $\leq 1.08$, and as a result have negligible effect on the final value for $\beta$.

\subsection{Incompleteness Analysis} 
\label{app_incompleteness_analysis}

Adopting the preceding set of assumptions, we now return to the discussion of
the remaining factors affecting survey incompleteness: factors (iii) and (iv) from
\S~\ref{app_factors_incompl}.  We address the individual
factors in three incremental steps, as pertinent to: geometrical
incompleteness, defined solely by the IWA and OWA
of the survey and by the distribution of stella heliocentric distances; observational incompleteness, defined by the flux
limits of the survey and by the predicted brightness of substellar companions; and orbital incompleteness, defined by the
fraction of orbital phase space observed.  These are the same incompleteness categories as already mentioned in \S~\ref{sec_incompleteness_short}. Throughout, we adopt the aperture-normalized r.m.s.\
detection limits determined for each star in \S~\ref{sec_survey_detlims}
and assume that the primary ages and
distances are fixed at their mean values listed in Table~\ref{tab_deep_sample}.

\subsubsection{Geometrical Incompleteness} 
\label{app_geom_incompl}

In deciding the range of projected separations that the study is
most sensitive to, we consider the full range of separations that have
been explored between the IWA and OWA of the deep survey.  For the IWA we adopt 0$\farcs$55, i.e., approximately one half width of the 0$\farcs$1 PALAO $K_S$-band PSF wider than the 0$\farcs$49 radius of the PHARO coronagraph.  For the OWA, we adopt 12$\farcs$5, which is 0$\farcs$3 less than the half-width of the PHARO FOV.  Figure~\ref{fig_app_au_limits} shows the fraction of the deep sample stars (solid line) around which successive 1~AU intervals are probed as a function of projected separation.  It is
immediately obvious that only a very narrow range of orbital separations,
between 105~AU and 125~AU, is probed around 100\% of the stars.  
All other projected separations
carry with them some degree of incompleteness that needs to be
taken into account.  From a purely geometrical standpoint, i.e., ignoring
imaging
sensitivity, the limitations imposed by the choice of IWA and OWA amount to a factor of 1.96 incompleteness (for a $dN/d\log a \propto a^0$ semi-major axis distribution) between 6~AU and 2375~AU: the projected separation
range contained between the IWA for the nearest star and the OWA for the
farthest star in the deep sample.  That is,
provided that substellar companions are detectable regardless of their
brightness anywhere between 0$\farcs$55 and 12$\farcs$5 from each
star, and provided that their distribution of semi-major axes $a$ is
logarithmically flat, only about half of the companions residing in the
6--2375~AU projected separation range would be detected.

As is evident from Figure~\ref{fig_app_au_limits}, such a wide
range of orbital separations includes regions probed around only a small
fraction of the stars.  Consideration of the full 6--2375~AU range will thus
induce a poorly substantiated extrapolation of the companion frequency.
Instead, we choose to limit the analysis to projected separations
explored around at least one-third (i.e., 33) of the stars in the deep
sample.  The corresponding narrower range, 22--1262~AU,
is delimited by the dashed lines in Figure~\ref{fig_app_au_limits}.  The
region has a geometrical incompleteness factor of 1.40 (compared to 1.96 for the
full 6--2375~AU range above).  That is, $1/1.40 =
71.4\%$ of all companions with projected separations between 22 and 1262~AU should
be recovered in our deep survey, if they are sufficiently bright.

\subsubsection{Observational Incompleteness} 
\label{app_obs_incompl}

Following an approach analogous to the one described in the preceding discussion,
we infer the projected separation range over which our survey is sensitive to a companion of a given mass.  That is, we now take into account that not all companions are sufficiently bright to be detected at all probed projected separations.  Rather their visibility is determined by their expected brightness and the attained imaging contrast.

Because mass is not an observable, we use
the absolute $K$-band magnitude of a substellar object as a proxy for
its mass, and employ the Lyon suite of theoretical models to convert between absolute magnitude and mass at the assumed stellar age.

We calculate the observational incompleteness of the deep survey for a grid of 11 discrete companion masses (0.005, 0.010, 0.012, 0.015, 0.020, 0.030, 0.040, 0.050, 0.060, 0.072, and 0.090~\Msun) and over the entire 3--500~Myr age range of our deep sample.  
We use the DUSTY models from \citet{chabrier_etal00} when the predicted companion effective temperature is above 1400~K (i.e., for spectral types L or earlier), and the COND models from \citet{baraffe_etal03} at lower effective temperatures (spectral type T).   We compare the estimated companion fluxes at the age of each of our sample stars to the corresponding flux limits for each star (see Table~\ref{tab_deep_obs}), and obtain a minimum projected separation at which a companion of a given mass would be visible around each star.  Thus, summing over all stars in the deep sample,
we estimate the observational incompleteness of the entire deep survey to companions of this mass.

The observational completeness estimates for each of the discrete set of 0.005--0.090~\Msun\ companion masses
are shown by the filled circles in Figure~\ref{fig_app_mass_incompl}a.
The geometrical completeness limit (i.e., if companion brightness were not a limiting factor)
%, which integrates to 71.4\% over the 22--1262~AU range (\S~\ref{app_geom_incompl}), 
is shown by the horizontal continuous line.  
%The vertical dotted lines mark the mass limits of sustained deuterium (D) and hydrogen (H) burning (0.012~\Msun\ and 0.072~\Msun, respectively).  
Figure~\ref{fig_app_mass_incompl}a demonstrates that the deep survey is nearly as complete as is theoretically possible to stellar-mass companions at angular separations between the IWA and OWA, since the observational completeness reaches the geometrical limit at 0.090~\Msun, just above the minimum hydrogen-burning mass. 
Figure~\ref{fig_app_mass_incompl}a also 
illustrates that the observational completeness of the deep survey is $>$50\% 
for all substellar objects above the deuterium-burning limit.  The survey completeness drops rapidly below the deuterium-burning limit because of the significantly fainter luminosities expected of non-deuterium fusing objects \citep[e.g.,][]{burrows_etal01}.

With the aim to minimize our incompleteness correction, we limit our analysis to substellar companions in the 0.012--0.072~\Msun\ range, i.e., between the deuterium- and hydrogen-burning mass limits.
The sum of the geometrical + observational completeness in this mass range is between 53.0\% and 71.3\%.  Adopting a $dN/dM_2 \propto M_2^0$ (i.e., $\beta=0$) MF for substellar companions (\S~\ref{app_assumptions}), we find that
the observational survey is 68.2\% complete to 0.012--0.072~\Msun\ substellar companions at
projected separations of 22--1262~AU from their host stars.  
%Varying the assumed power-law index $\beta$ of the secondary MF between $-1$ and 1 changes the observational completeness between 61.2\% and 67.3\%, respectively.

\subsubsection{Orbital Incompleteness} 
\label{app_orb_incompl}

The analysis so far has dealt only with the projected separation of
substellar companions.  We now consider the effect of realistic
orbital semi-major axes, inclinations, and eccentricities.

We first adopt the multiplicative factor
of 1.26 to relate the projected separation $\rho$ to the true
semi-major axis $a$: $\langle a \rangle=1.26\langle \rho \rangle$ (see \S~\ref{app_assumptions}).  That is, the orbital semi-major axes probed by the survey are on average a factor of
1.26 further from the star, at 28--1590~AU, than the range of probed projected separations.

The multiplicative transformation from $\langle \rho \rangle$ to $\langle a \rangle$ does not exhaust
the discussion of orbital incompleteness.  Because companions on orbits with
non-zero inclinations and eccentricities spend most of their
time at projected separations $\rho\neq a/1.26$, they may still be
missed in the survey.  The most likely scenarios in which this can occur
are for companions on highly inclined and/or eccentric orbits.  

With a small number of positive substellar companion detections,
orbital incompleteness issues are best addressed through Monte Carlo
simulations.  Such have been performed for a wide range of realistic
orbital inclinations and eccentricities in a study by \citet{brown04}, the
results of which we adopt here.  \citeauthor{brown04}'s work investigates the detectability of populations of
habitable extra-solar terrestrial planets with a range of orbital distributions
by the {\it Terrestrial Planet Finder--Coronagraph (TPF--C)}.
Although the angular scales and the levels of imaging contrast between the
present coronagraphic survey and the design specifications for
{\it TPF--C} are vastly different ({\it TPF--C} projections call for a
factor of $\approx2.5$ smaller IWA and $\sim10^6$ higher
contrast), the problem is conceptually the same: to determine the
completeness to orbits with a certain semi-major axis, given an opaque
coronagraph of a fixed radius.  \citet{brown04} parameterizes this problem
in terms of the ratio $\zeta$ (which he defines as $\alpha$) of the semi-major axis to the obscuration
radius, so his results are universally scalable.  His analysis
does not include treatment of imaging contrast or
limiting flux \citep[these are addressed in a follow-up
work:][]{brown05}, which makes it suitable to apply to results that have
already been corrected for these effects, as we have already done for our survey in
\S~\ref{app_obs_incompl}.

\citet{brown04} finds that the detectability of orbiting companions in a
single-visit observation, what he terms the ``single visit obscurational
completeness'' (SVOC), is a strong function of $\zeta$ between
$\zeta=1$ and 2.  The SVOC varies 
between $\approx$30\% at $\zeta=1$ and
$\approx$85\% at $\zeta=1.9$ \citep[Fig.~3 in][]{brown04}.  Higher 
SVOC, at the 95\% and 99\% levels, is achieved only for $\zeta=3.2$ and
7.1, respectively, i.e., far from the coronaghraphic edge.
The result is largely independent ($<10\%$ variation) of the assumed orbital eccentricity
$e$ for $0\leq e \leq 0.35$.

We adopt the results of \citeauthor{brown04}'s analysis and use the SVOC
values for a representative orbital eccentricity of 0.35 \citep[Table~4 in][]{brown04}---a
value near the peak of the
eccentricity distribution of G-dwarf binaries with $>10^3$~day periods
\citep{duquennoy_mayor91}.  We calculate the SVOC on the deepest image of each sample star, for each of the discrete candidate
companion masses in the 0.005--0.090~\Msun\ range considered in 
\S~\ref{app_obs_incompl}.  We define the minimum projected
separation at which a companion of a given mass is detectable
as the effective obscuration radius for that companion mass.  The results from the combined
treatment of the observational completeness (\S~\ref{app_obs_incompl})
and the SVOC are shown in Figure~\ref{fig_app_mass_incompl}a by filled triangles
and in Figure~\ref{fig_app_mass_incompl}b with the dotted lines.  The
long-dashed lines in Figure~\ref{fig_app_mass_incompl}a,b delimit the maximum
attainable SVOC, that is, when the companion brightness is not a limiting factor.  Figure~\ref{fig_app_mass_incompl}a shows that the completeness to $\geq 0.072$~\Msun\ objects is very near (64.8\%) the SVOC limit (64.9\%).  That is, the deep survey is almost maximally complete to stellar companions.  The survey is only 47.0\% complete to companions at the low end of the brown dwarf mass range at 0.012~\Msun.

The additional consideration of orbital incompleteness does not affect significantly the overall incompleteness of the survey within the posited 22--1262~AU projected separation (28--1590~AU semi-major axis) range.  Given the assumed companion
mass and orbital semi-major axis distributions (\S~\ref{app_assumptions}), the overall  (geometrical + observational + orbital) completeness becomes 62.0\%.   
%Varying the assumed power-law index $\beta$ of the CMF between $-1$ and 1 changes the completeness between 54.5\% and 60.9\%, respectively.

We note that the consideration of the SVOC, as defined by \citet{brown04}, does not address all possibilities for orbital incompleteness.  Other than being obscured by the coronagraph or lost in the glare of its host star, a companion on a highly-eccentric
orbit may fall outside the OWA, even if its semi-major axis was in the
explored range.  This additional factor, among possible
other sources of orbital incompleteness, is not taken into account here.
However, judging by the small decrease ($68.2\%-62.0\%=6.2\%$) in the overall incompleteness correction induced by the consideration of the SVOC, it is unlikely that inclusion of the remaining factors affecting orbital incompleteness will decrease the overall survey completeness below 50\%.

\subsection{Effect of Variations in the Assumed Companion Semi-major Axis and Period Distributions} 
\label{app_variations}

The above final completeness estimate is based on the assumptions for the semi-major axis and companion mass distributions adopted in \S~\ref{app_assumptions}.  These assumptions are merely guesses, and in reality the companion orbital and mass distributions may take different forms.  Indeed, in
\S~\ref{sec_bd_desert} we argue that the orbital period distribution of substellar and stellar mass companions are probably the same, while in \S~\ref{subsec_cmf} we conclude that the MFs of companions and isolated objects are different.  Both of these results are at odds with the corresponding assumptions.  It is conceivable that other initial guesses for the orbital and mass distributions of the companions may lead to different conclusions.
%, which for Sun-like stars is described by the log-normal period function of \citet{duquennoy_mayor91}.  In addition, the estimate for the power law index of the CMF in the present study is not very precise: $\beta=-0.39\pm0.36$ 
%$\beta=-0.3\pm0.3$
%(\S~\ref{subsec_cmf}).  We fixed $\beta$ at zero for our incompleteness analysis, but values in the $-1\leq\beta\leq1$ range, as found in for isolated brown dwarfs and low-mass stars, will likely lead to different results.

We therefore analyzed the completeness of the survey to substellar companions under a broader set of functional forms for the companion semi-major axis and mass distributions.  For the semi-major axis distribution we also considered: (1) the equivalent of the log-normal orbital period distribution for sun-like binary stars from \citet{duquennoy_mayor91} under the assumption that the total system mass is 1~\Msun, (2) the extrasolar planet period $dN/d\log P \propto P^{0.26}$ distribution from \citet{cumming_etal08}, which converts to $dN/da \propto a^{-0.61}$ for solar-mass primaries, and (3) a linearly flat $dN/da \propto a^0$ distribution.  For the CMF exponent $\beta$ we tested values in the $-1$ to 1 range.  

The estimates for the completeness to substellar companions for the three different semi-major axis distributions with $\beta$ fixed at zero were 62.7\%, 59.9\%, and 51.9\% respectively, all within a factor of 1.20 of the one already obtained in \S~\ref{app_orb_incompl}.  In particular, we note that the assumption of either the star-like log-normal or the planet-like period distribution altered the completeness estimate very little (by a factor of $\leq$1.04).  If we set the CMF index $\beta$ to either $-1$ or 1 but held the assumed semi-major axis distribution fixed at $dN/da\propto a^{-1}$, the completeness became 58.1\% or 64.9\%, respectively.
If we allowed both of the companion orbital and mass distributions to vary, the completeness estimates ranged from 50.2\%--66.0\%.

Overall, we found that the inferred frequency of wide substellar companions to young solar analogs in \S~\ref{sec_bd_frequency} would be affected by a factor of $\leq1.24$.  In the likely case that the orbital period distributions of substellar and stellar companions are the same, as in \citet{duquennoy_mayor91}, our inferred frequency would be accurate to a within factor of 1.06.

Such small changes to the incompleteness estimate of our survey affect the resultant CMF power law index $\beta$ only minimally.  Because the relative changes in the completeness-corrected numbers per mass bin of the CMF are much smaller than the observed trend, the variations in the fitted value for $\beta$ are well within the derived 1$\sigma$ range.

\subsection{Summary of Incompleteness Analysis and Further Considerations}
\label{app_incompl_summary}

We adopt 62\% as the final estimate for the completeness to substellar companions in our deep survey.   That is, given two detected brown dwarf companions with semi-major axes in the 28--1590~AU range, on average $0.62^{-1}=1.6$ more companions with semi-major axes in the same range have been missed.  This estimate is based on the combined consideration of the geometrical, observational, and orbital incompleteness factors described in \S~\ref{app_incompleteness_analysis}. 

\begin{figure}
\epsscale{0.7}
%\plotone{dmag_sep_new.eps}
%\plotone{dmag_sep_new_nohii1348.eps}
\plotone{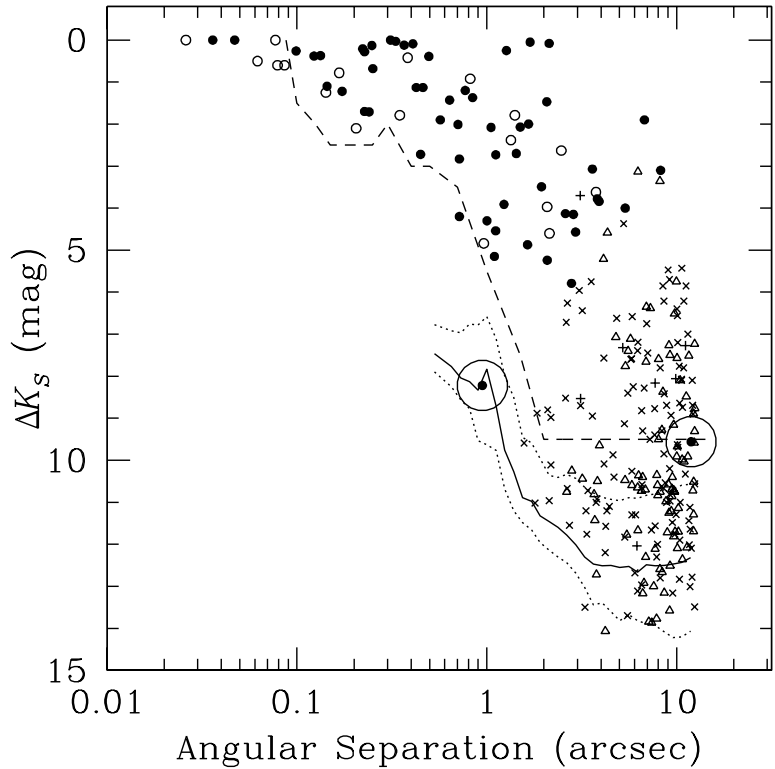}
\figcaption{Magnitude difference $\Delta K_S$ vs.\ angular separation
$\rho$ for all 
candidate companions discovered in the deep and shallow surveys.
The various symbols denote: ``$\bullet$''---astrometrically associated
companions; ``$\times$''---astrometrically unassociated background stars; and
for objects with insufficient astrometric data:
``$\circ$''---companions associated based on their $J K_S$
photometry; ``+''---objects with $J K_S$ photometry inconsistent with
association; ``$\vartriangle$''---undecided objects.
The encircled points show the two brown dwarf companions from the
survey: HD~49197B (at $\rho=0\farcs95$) and HD~203030B (at $\rho=11\farcs92$).
%(encircled with solid lines) and 1 companion at the stellar/substellar boundary (HII~1348B; encircled with a dotted line) that is not part of the unbiased companion survey.
Detection limits for the shallow (dashed
line) and deep (solid and dotted lines) components of the survey are also
shown.  The solid line shows the median contrast
$\Delta K_S$ of the deep survey, while the dotted lines delimit the 10--90 percentile region
(cf.\ Fig.~\ref{fig_ensemble_contrast}a).  Binaries with separations smaller
than the PALAO $K_S$-band diffraction limit (0$\farcs$10) were resolved
only at $J$-band.  Correspondingly, the plotted magnitude difference for
these companions is the one at $J$. 
\label{fig_dmag_sep}}
\end{figure}

\begin{figure}
%\plotone{MK_JK_new_nohii1348.eps}
\plotone{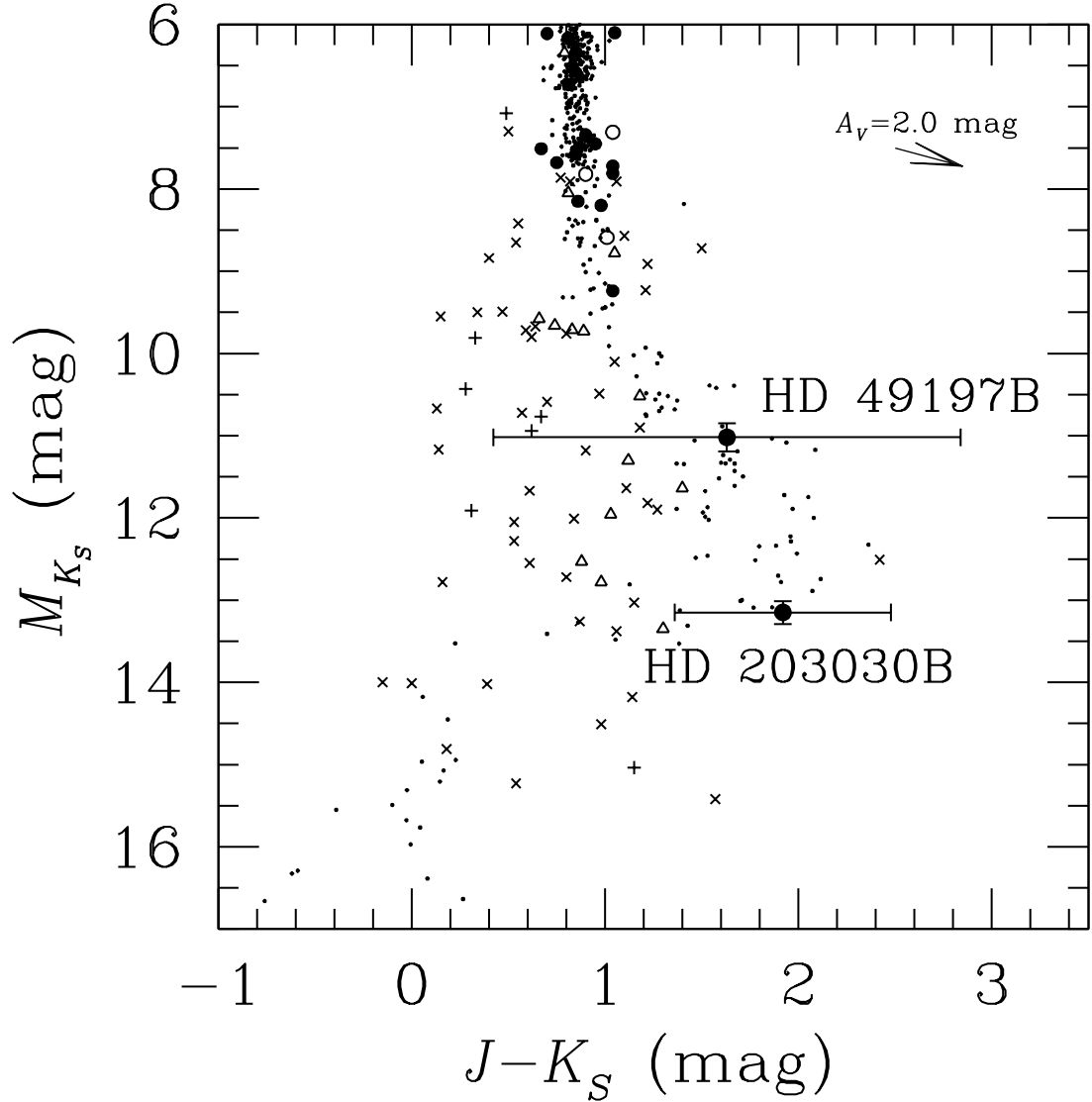}
\figcaption{$M_{K_S}$ vs.\ $J-K_S$
color-magnitude diagram of 
candidate companions for which $J$-band photometry was obtained.
The symbols are the same as in Figure~\ref{fig_dmag_sep}.  The
additional 
small dots denote M0--T8 dwarfs ($M_{K_S}\gtrsim4.5$)
with parallaxes from \citet{dahn_etal02} and \citet{vrba_etal04}.
The points with errorbars represent the two confirmed brown dwarf companions
from our survey. 
%and 1 companion at stellar/substellar boundary (HII~1348B) that does not belong to the unbiased survey for substellar companions (\S~\ref{sec_results}).  
The errorbars on HD~203030B are representative of the photometric precision for the faintest ($J\gtrsim18$~mag) objects in the survey.  Brighter objects typically had $J-K_S$ errors $<0.3$~mag, except for the large $J-K_S$ uncertainty of HD~49197B, which is unique because of its relative faintness ($\Delta J=9.6$~mag) and proximity ($\rho=0\farcs95$) to the primary.
The vector in the upper right corresponds to $A_V=2$~mag of visual extinction, equivalent to a distance of $\sim$3~kpc,
or a distance modulus of 12~mag along the galactic plane.
\label{fig_candidates_mkjk}}
\end{figure}

\begin{figure}
%\plotone{scopms214_ks_image.ps}
\plotone{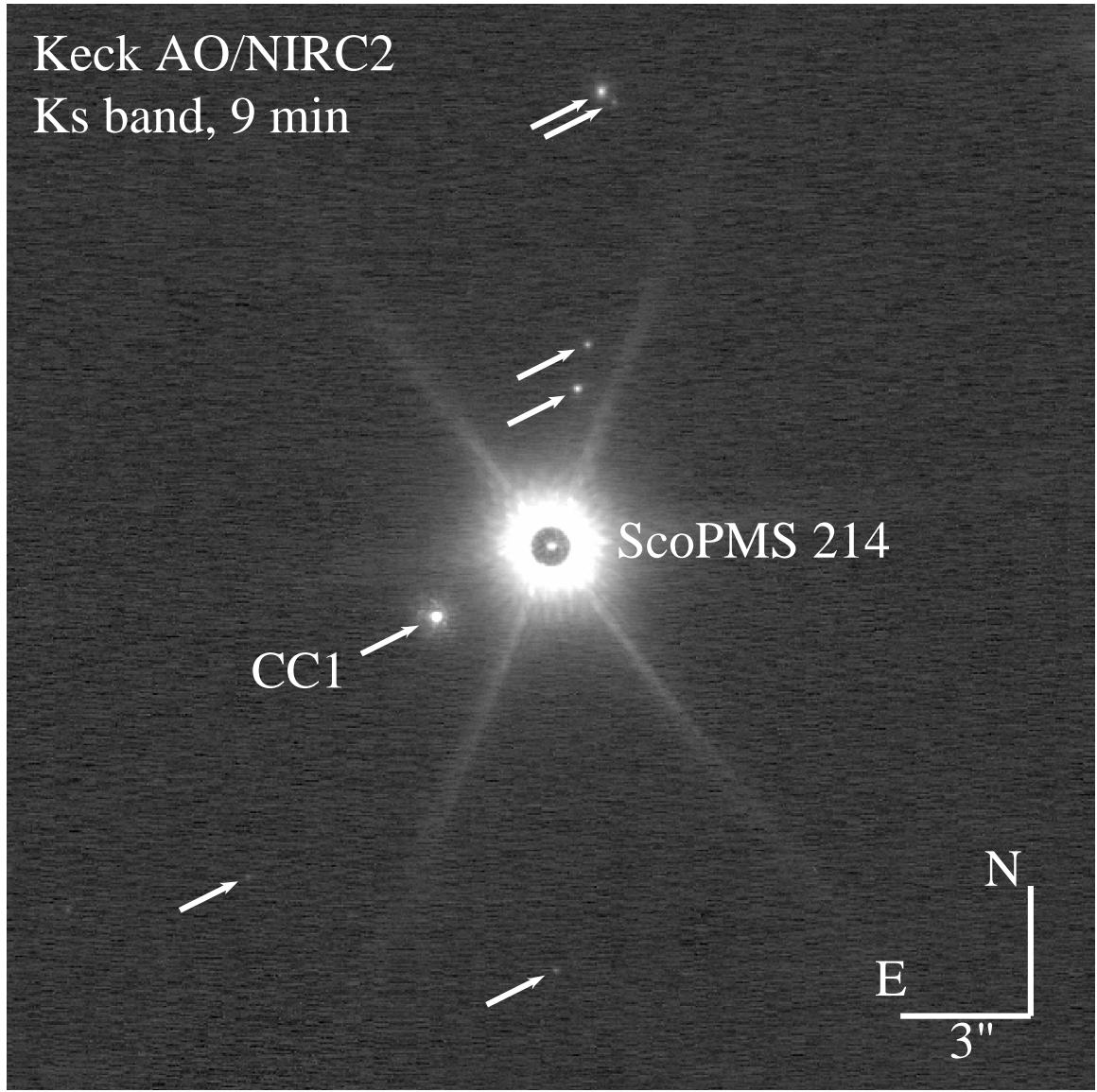}
\figcaption{$K_S$-band image of ScoPMS~214 and its candidate companions taken with NIRC2 and the Keck AO system on 5 June 2004.  The image is the median of nine 60~s exposures.  ScoPMS~214 is occulted by a partially transmissive 1$\farcs$0-diameter circular coronagraphic mask.  The seven $\rho\leq12\farcs5$ candidate companions listed in Table~\ref{tab_deep_companions} are pointed out with arrows.  The candidate proper motion companion CC1 is the brightest of the seven and closest to the star.
\label{fig_scopms214_image}}
\end{figure}

\begin{figure}
%\plotone{scopms214_pm.eps}
\epsscale{.9}
\plotone{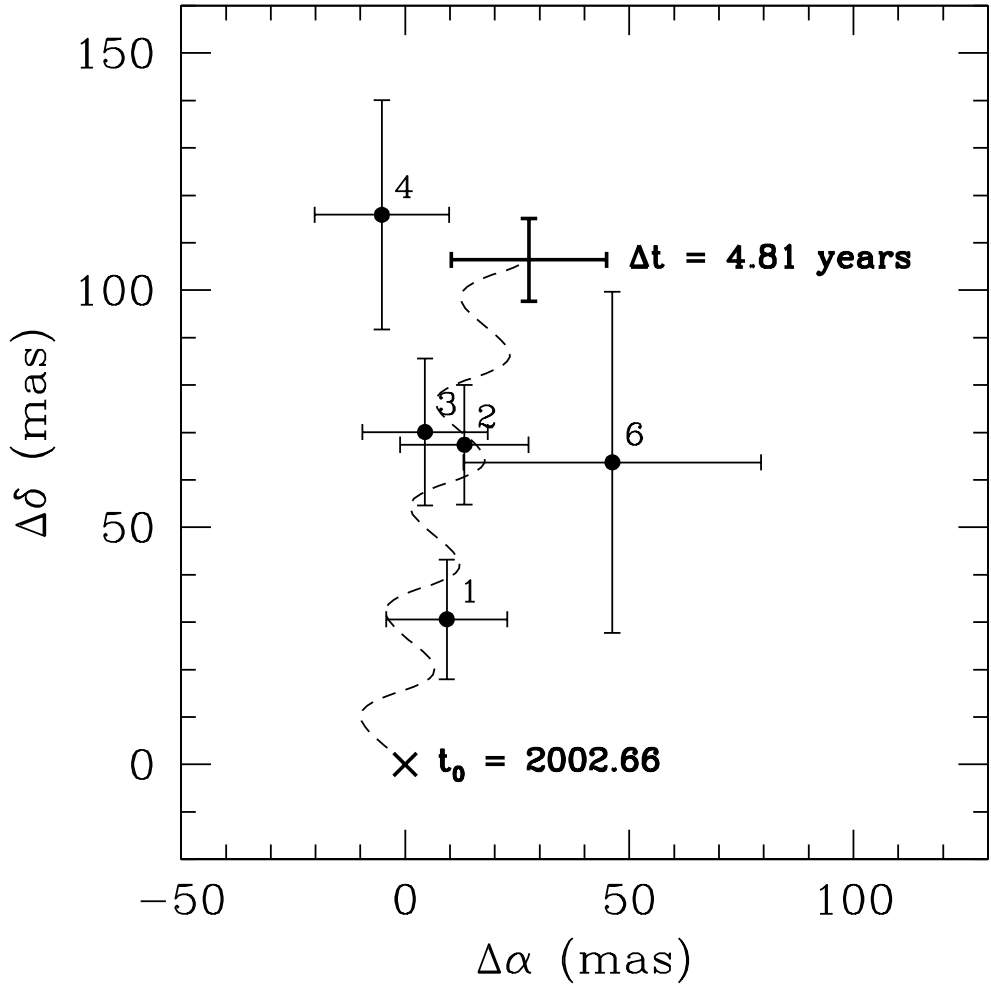}
\epsscale{1}
\figcaption{Proper motion diagram for candidate companions to ScoPMS~214, spanning the $\Delta t=4.81$~yr time period between the first and last epochs of observations, between 30 August 2002 ($t_0 = 2002.66$~yr) and 23 June 2007.  The dashed line denotes the expected relative motion of a stationary background object with respect to ScoPMS~214 between the initial epoch (marked with $\times$) and the final epoch (marked with thick errorbars without a solid point).  The solid points with thin errorbars denote the observed changes in the relative positions of candidate companions 1, 2, 3, 4, and 6. Candidate 5 was outside of the field of view of the medium ($20\arcsec\times20\arcsec$) NIRC2 camera during the last epoch of observations, and candidate 7 was below the detection limit during the initial epoch.   Candidate 1 is ScoPMS~214``B'', which shares the proper motion of ScoPMS~214 during the 4.81-year time span within 3$\sigma$ limits and is inconsistent with being a stationary background object (at the 5$\sigma$ level in declination).
Candidates 2, 3, and 4 (and 7, based on observations at intermediate epochs) are all inconsistent with being proper motion companions to ScoPMS~214 and are consistent with being background objects.  Candidates 5 and 6 are consistent with being either bona fide companions or unrelated background objects, i.e., their status is undecided. 
\label{fig_scopms214_pm}}
\end{figure}

\begin{figure}
%\plotone{scopms214b_quote_b_Kspec_wgiants.eps}
\plotone{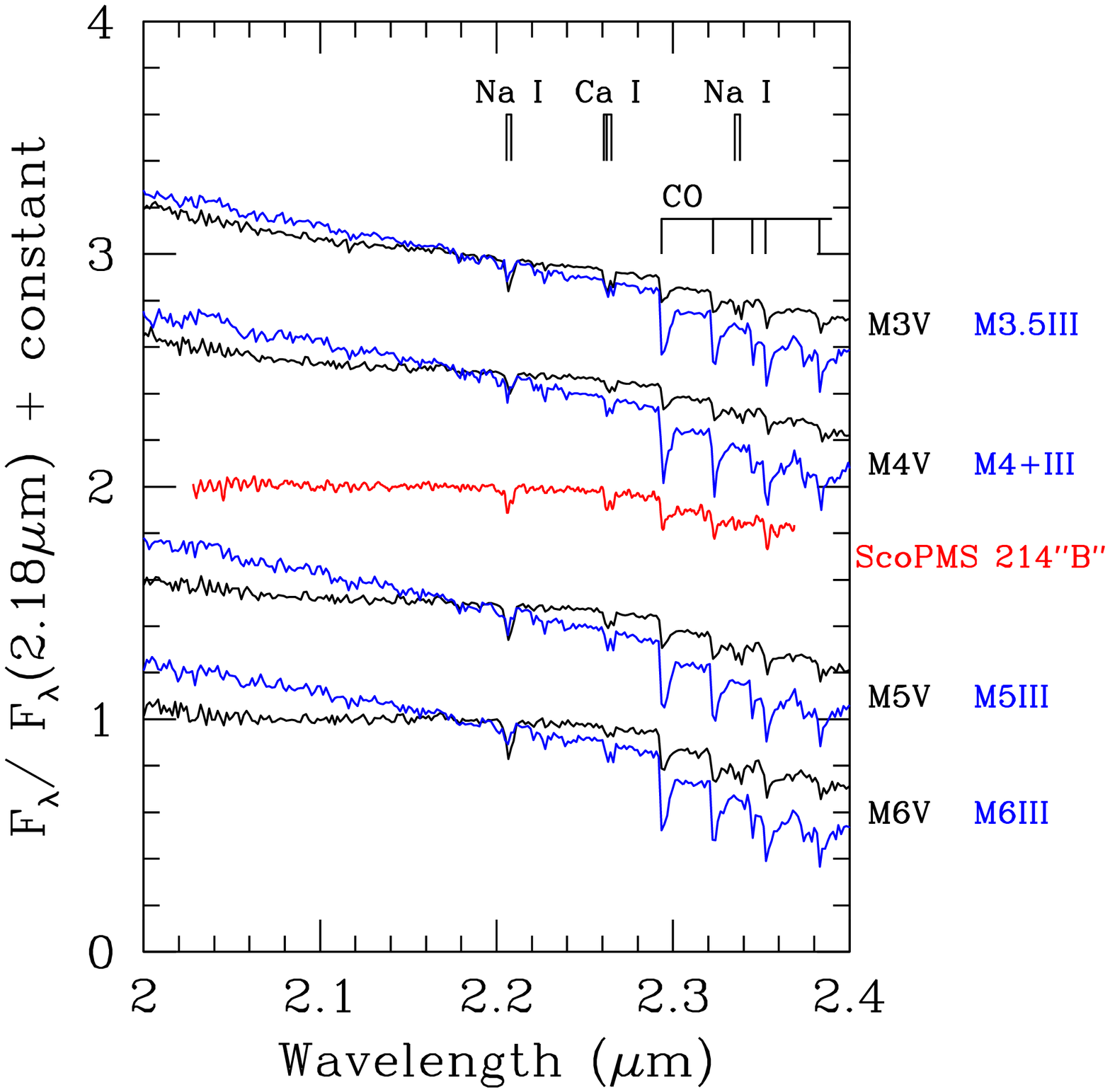}
\figcaption{$K$-band spectrum (red) of ScoPMS~214``B'' (CC1), compared to spectra of M3--M6 field dwarf (in black) and giants (in blue) from the IRTF Spectral Library \citep{cushing_etal05, rayner_etal08}, smoothed to the same $R\approx1200$ resolution.  Dominant absorption features by \ion{Na}{1} at 2.21~$\micron$ (doublet) and 2.34~$\micron$ (doublet), \ion{Ca}{1} at 2.26~$\micron$ (triplet), and CO bandheads at $\lambda\geq2.29$~$\micron$ are identified.  The comparison dwarf spectra are of Gl~388 (M3V), Gl~213 (M4V), Gl~51 (M5V), and Gl~406 (M6V), and the giants are HD~28487 (M3.5III), HD~214665 (M4+III), HD~175865 (M5III), and HD~196610 (M6III).
\label{fig_scopms214b_spec}}
\end{figure}

\begin{figure}
%\plotone{scopms214ab_LT_bcah98_luhman99temp.eps}
\plotone{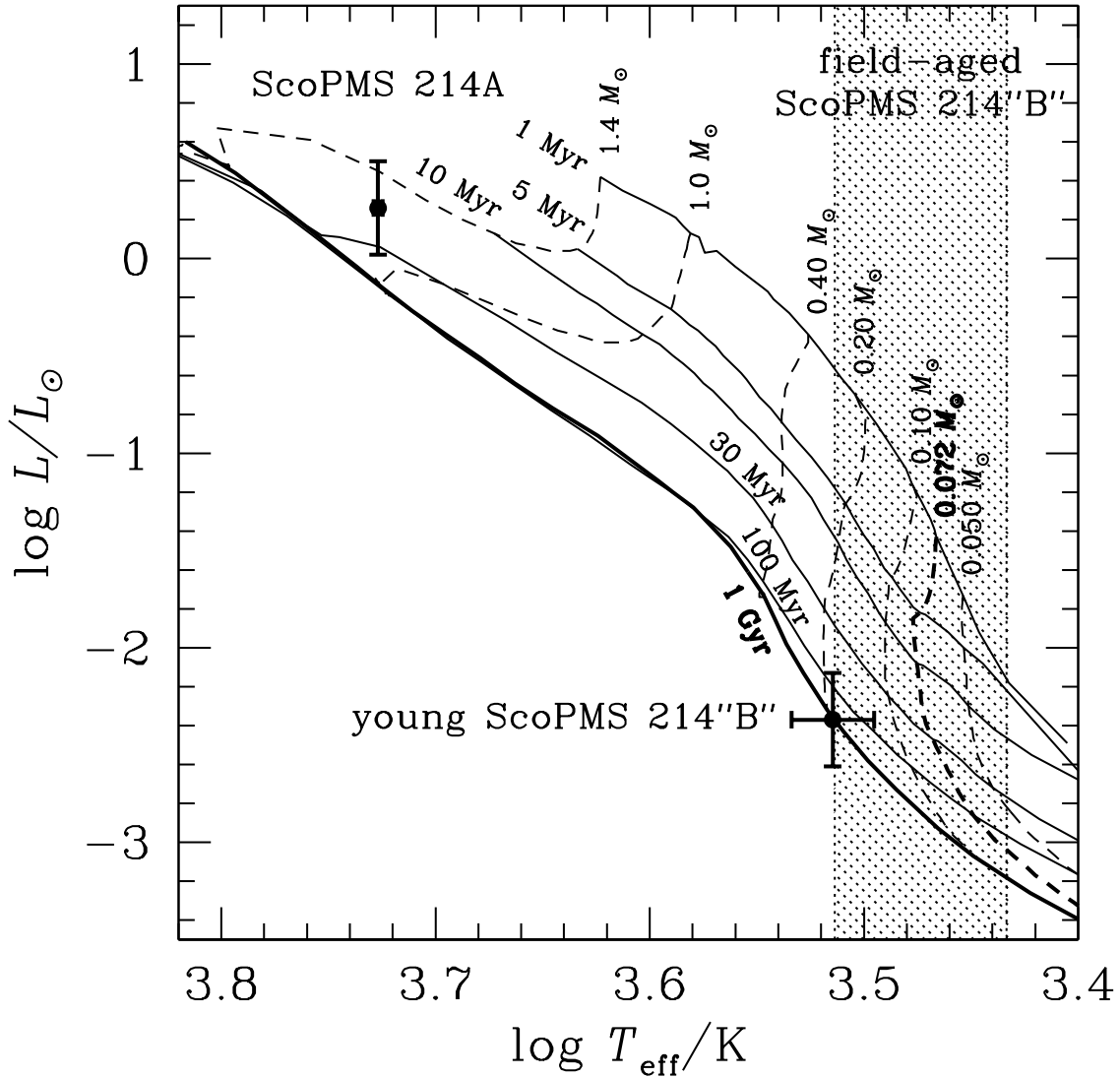}
\figcaption{HR diagram of the candidate binary ScoPMS~214A``B'' with evolutionary models for 0.05--1.4~\Msun\ objects from \citet{baraffe_etal98}.  The continuous lines are isochrones and the dashed lines are evolutionary tracks at constant mass.  The thick (1~Gyr) isochrone approximates the main sequence, and the thick evolutionary track corresponds to the minimum hydrogen-burning mass. 
The positions of ScoPMS~214A and ``B'' under the assumption of equidistance and membership in Upper Scorpius (the ``young'' ScoPMS~214``B'' scenario; \S~\ref{sec_scopms214_hr}) are shown with solid points with errorbars.  The shaded region represents the range of effective temperature allowed for ScoPMS~214``B'' if it were an unassociated field-aged (1--10~Gyr) M dwarf.  Since in the ``young'' scenario the candidate binary components do not lie on the same theoretical isochrone, ScoPMS~214``B'' is probably not a member of Upper Scorpius.  Instead, it is most likely a foreground field M dwarf. 
\label{fig_scopms214_hr}}
\end{figure}

\begin{figure}
%\plotone{gamma2_2sigma_ML.eps}
\plotone{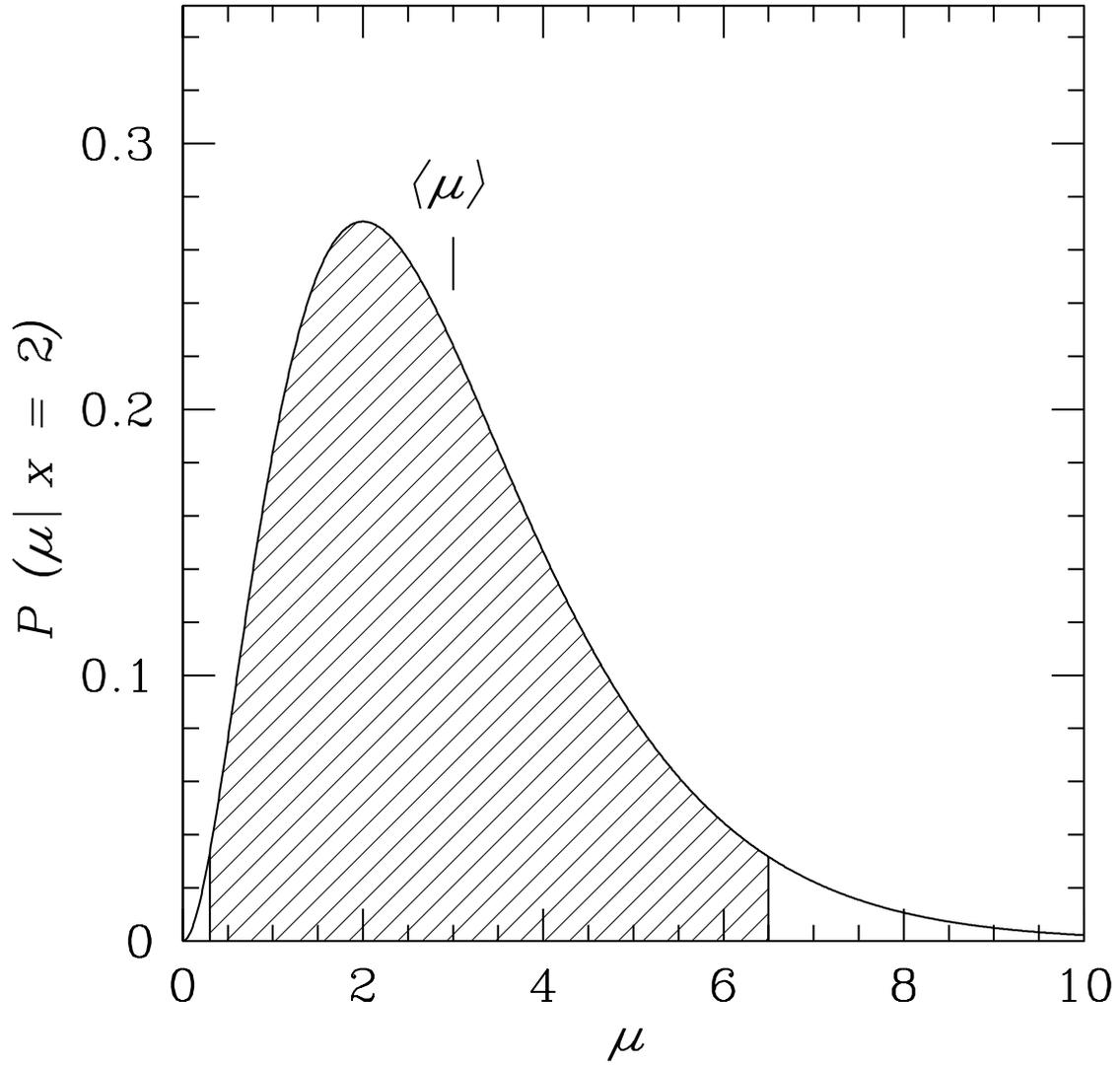}
\figcaption{Probability density distribution $P(\mu|x=2)$ for
the expected substellar companion detection rate in our survey per 100 stars, given $x=2$ detections.
The curve is a Gamma distribution (Eqn.~\ref{eqn_pmu}), with a
peak at $\mu=\mu_{\rm ML}=2$, but a mean value of $\langle\mu\rangle = x+1=3$.  The minimal 
2$\sigma$ (95.4\%) confidence interval on $\langle\mu\rangle$, $0.3<\langle\mu\rangle<6.5$ is indicated by the shaded region under the curve.
\label{fig_gamma2}}
\end{figure}

\begin{figure}
%\plotone{surveys_labeled.eps}
\plotone{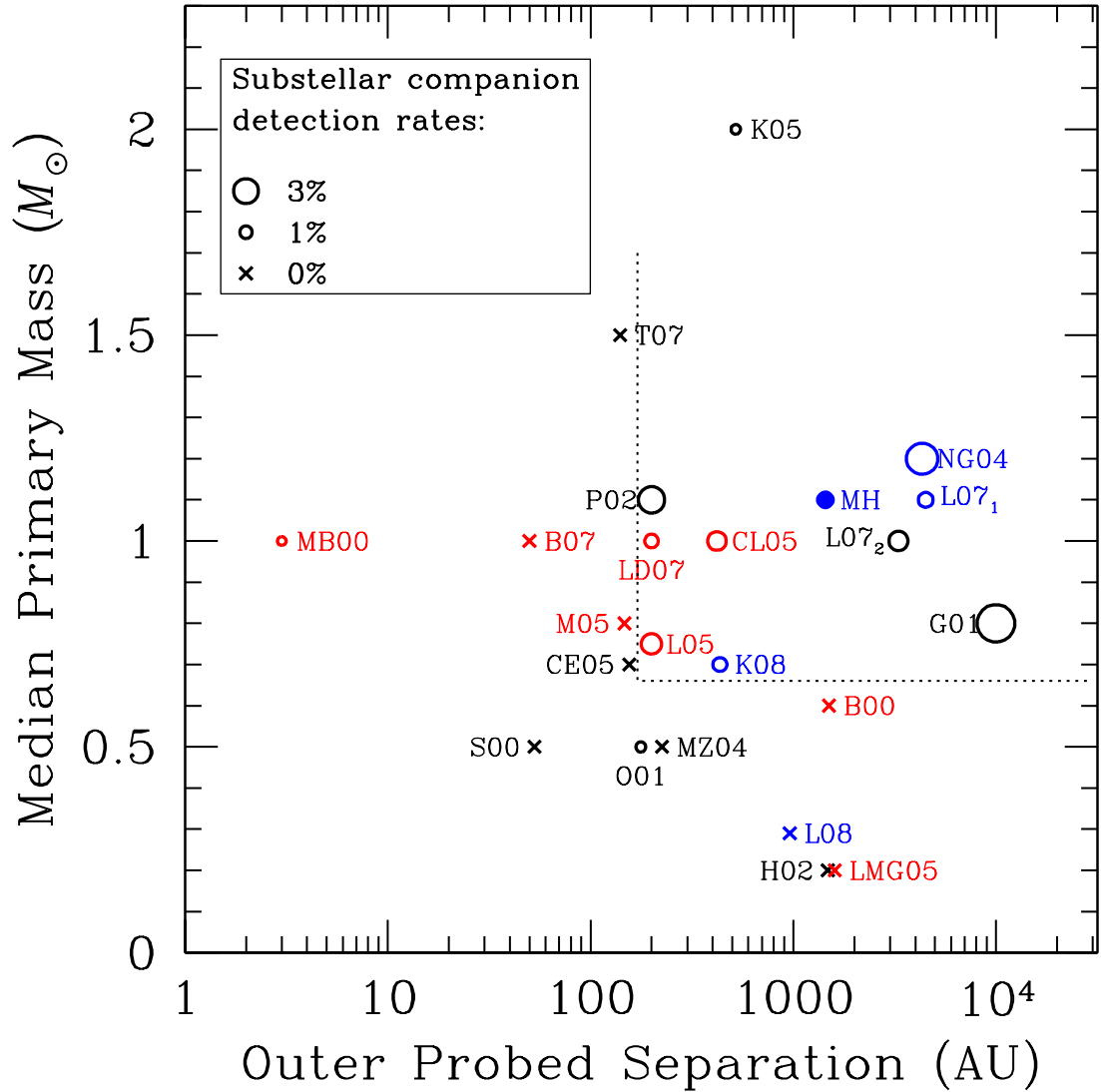}
\figcaption{Substellar companion detection rates of the published direct imaging surveys listed in Table~\ref{tab_surveys}.  Circular symbols denote surveys with at least one detection; crosses denote surveys with no detections.  The filled circle denotes the present work.  The size of the circular symbols is proportional to the survey detection rate prior to corrections for survey incompleteness.  Black symbols denote the least sensitive surveys, with $\geq30\Mj$ median companion mass sensitivity in the background-limited regime.  Blue symbols denote surveys with median companion sensitivities between 13--30~\Mj.  Red symbols mark surveys with the highest sensitivity, $<13\Mj$.  The survey labels are as  listed in the penultimate column of Table~\ref{tab_surveys}.
The locus delimited by a dotted line contains only surveys with non-zero detections, with detection rates ranging from 0.5--5\%.  All surveys outside of this region have detection rates $\leq0.6\%$.
\label{fig_surveys}}
\end{figure}

\begin{figure}
%\plottwo{cmf_logdndlogm2.eps}{q_hist_dndm2.eps}
\plottwo{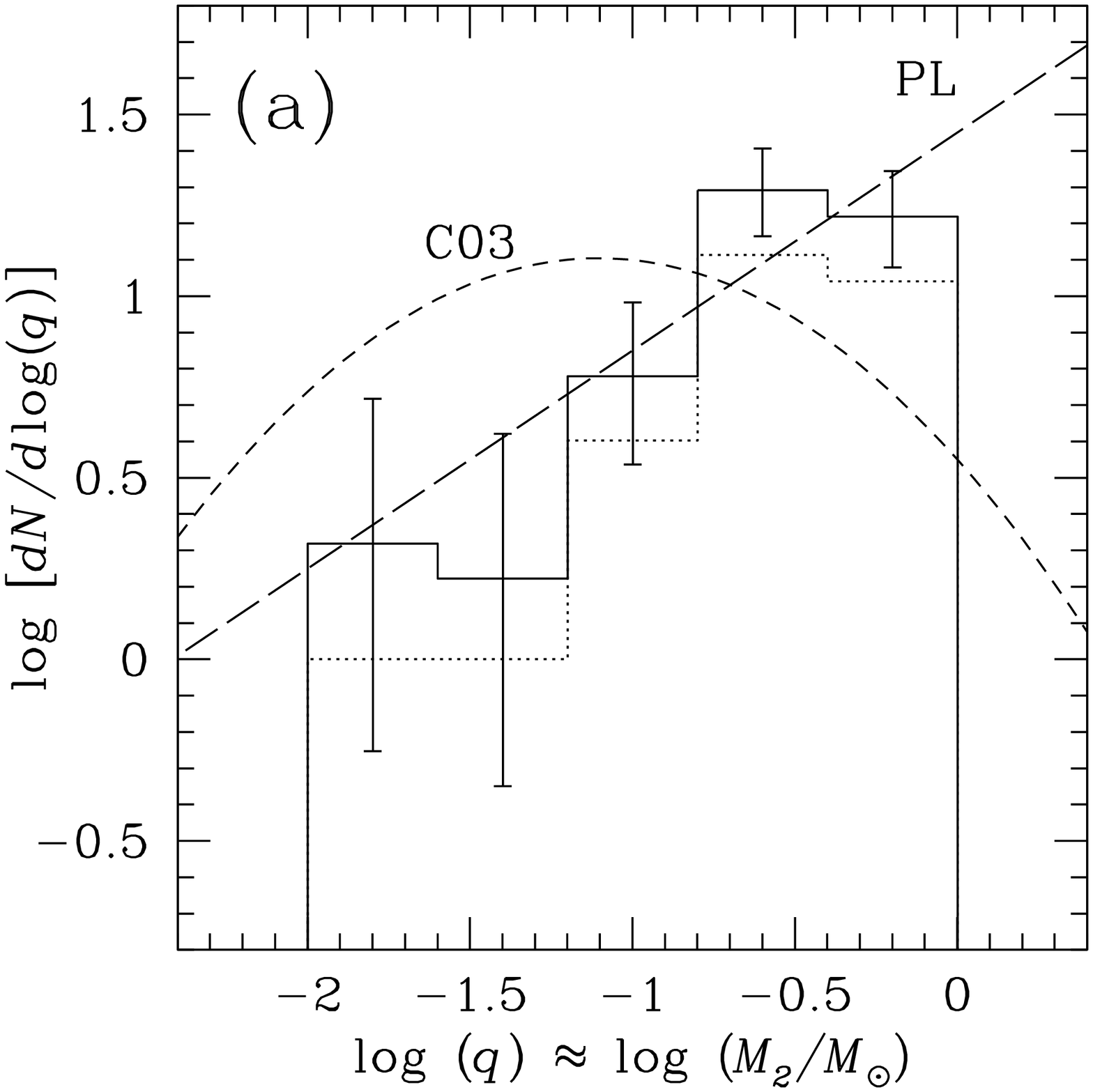}{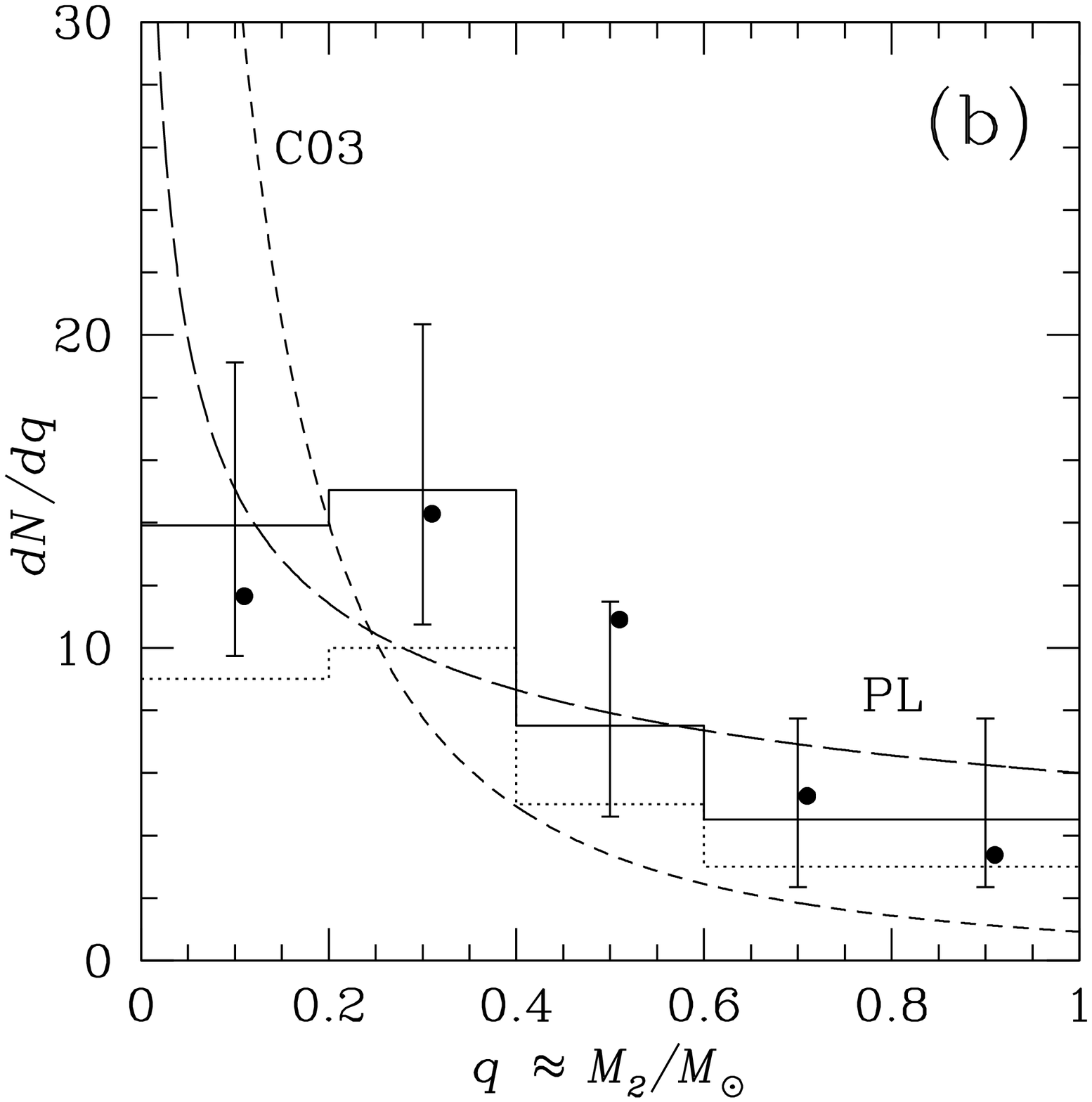}
\figcaption{Mass ratio distribution for the 30 
% was 30
$\leq$0.5~Gyr-old binaries in our AD$_{30}$ sample (see \S~\ref{sec_cmf}) in terms of $\log (d N/d \log q)$ {\bf (a)} and $d N/d q$ {\bf (b)}.  The dotted histogram traces the observed data, while the solid histogram delineates the incompleteness-corrected data.  Further incompleteness due to bias against near-equal binary systems exists in the highest mass ratio bin, but has not been taken into account in the present incompleteness correction.  The long-dashed line is a power-law (PL) fit to the data, $d N/d \log q \propto q^{\beta+1}$, with an index of $\beta=-0.39\pm0.36$ 
%$\beta=-0.3\pm0.3
(1$\sigma$ limit).  The short-dashed line represents the log-normal MF of field objects from \citet[][C03]{chabrier03} in units of \Msun, normalized to the incompleteness-corrected data.  We note that because the primary masses for stars in our sample are $\approx$1~\Msun, then  $q=M_2/M_1\approx M_2/\Msun$.   The log-normal field MF peaks at $\mu=0.08\Msun$ and has a width of $\sigma=0.69$ (in logarithmic mass units).  The Salpeter index in these units is $\alpha=-2.35$.  
The solid points in panel {\bf (b)} are the incompleteness-corrected data from \citet{duquennoy_mayor91}, normalized to our data.  The \citet{duquennoy_mayor91} data have been offset slightly to the right from ours for clarity.
\label{fig_cmf}}
\end{figure}

\begin{figure}
%\plotone{au_range.eps}
\plotone{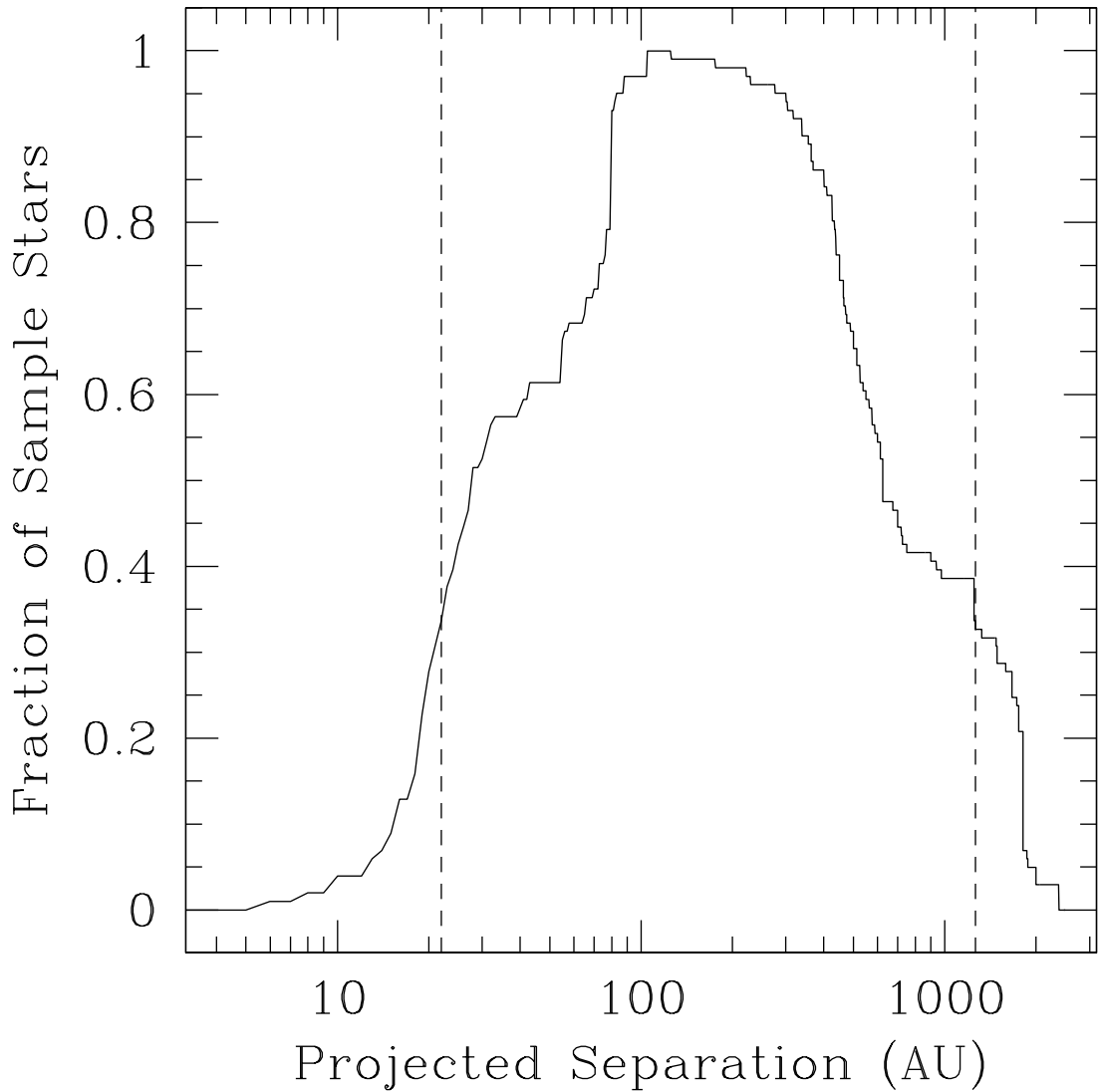}
\figcaption{Projected physical separations probed in the deep sample survey.
The vertical dashed lines delimit the region, 22--1262~AU, in which each 1~AU-wide projected separation interval was probed around at least one third of the
stars in our deep sample.  The geometrical incompleteness factor for this region is 1.40.  That is,
 $1/1.40 = 71.4\%$ of all companions in the 22--1262~AU projected separation range should have in principle been detected, had their visibility not been limited by contrast. 
\label{fig_app_au_limits}}
\end{figure}

\begin{figure}
%\plottwo{bdmass_frac_bw.eps}{au_range_svoc.eps}
\plottwo{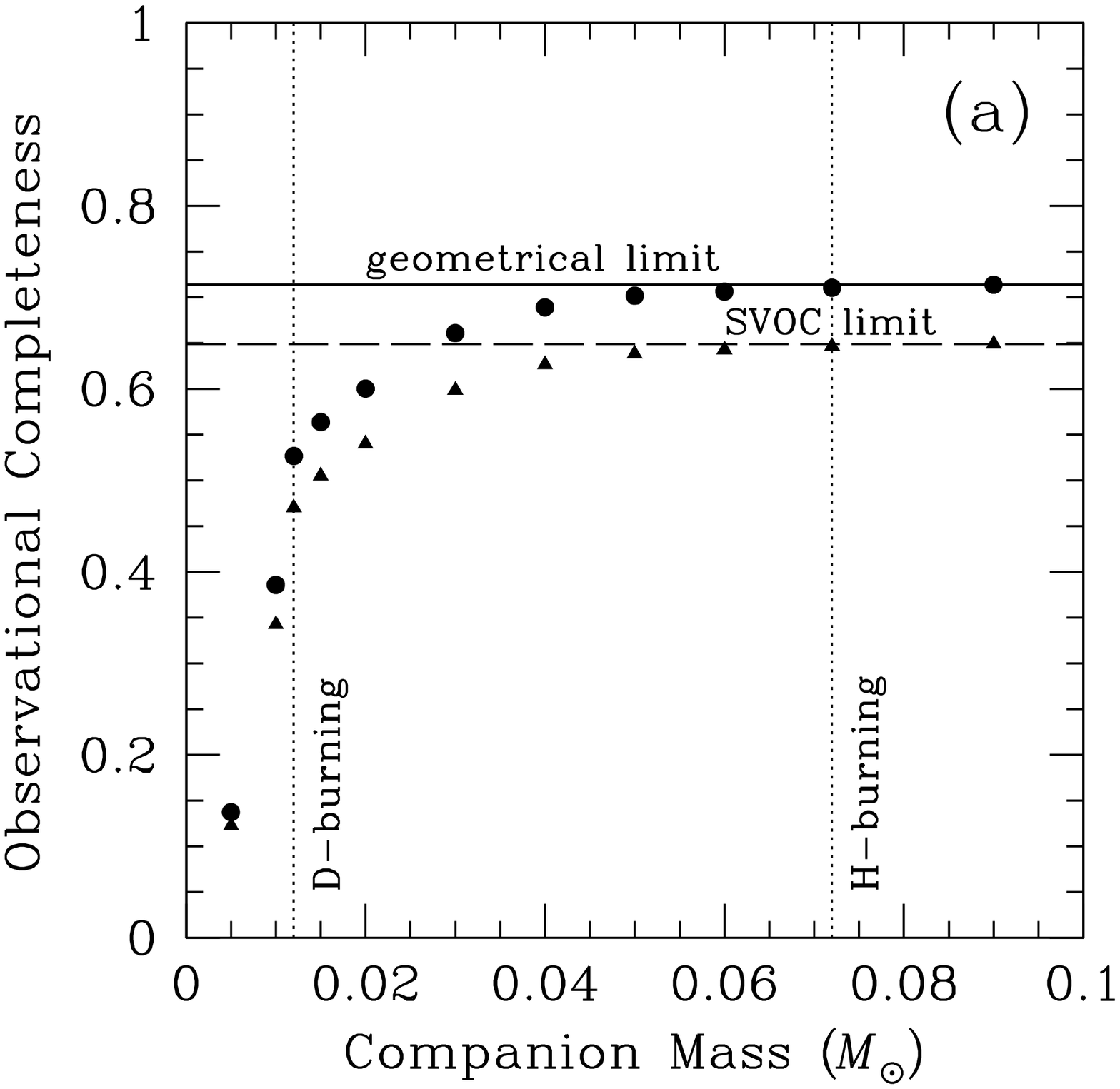}{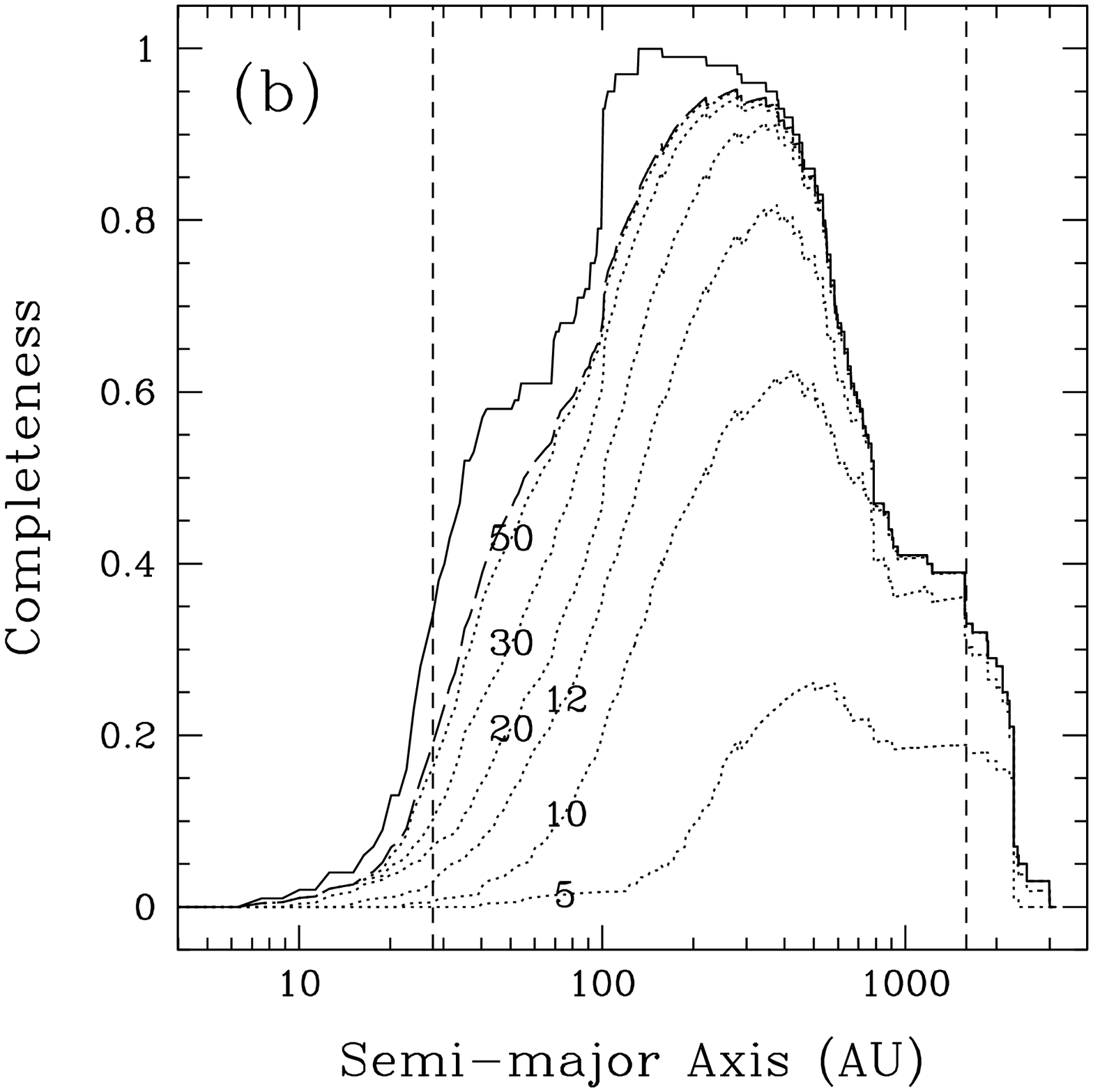}
\figcaption{{\bf (a)} Observational (circles) and total
(triangles) completeness of the deep survey as a function of companion mass.  The observational completeness at a given mass is the fraction of companions of that mass that would
be detectable within a projected separation of 22--1262~AU from all
sample stars (\S~\ref{app_obs_incompl}).  The total completeness is defined similarly, but for a 28--1590~AU range of
{\it semi-major axes}, and after consideration of orbital incompleteness 
(\S~\ref{app_orb_incompl}).
Both sets of completeness fractions are calculated assuming a logarithmically flat
distribution of companion semi-major axes $a$ (\S~\ref{app_assumptions}).  The
horizontal lines delimit the maximum possible observational (continuous
line) and orbital (long-dashed line) completeness at any given mass over 
these AU ranges.  Our definition of the orbital completeness coincides with the ``single visit obscurational completeness'' (SVOC; see \S~\ref{app_orb_incompl}) defined by \citet{brown04}.
The vertical dotted lines mark the deuterium-
(D) and hydrogen- (H) burning mass limits. {\bf (b)}  Same as
Figure~\ref{fig_app_au_limits}, but for the expected semi-major axes (rather than projected separations) of substellar companions and for a range of companion
masses.  The dotted lines are labeled with substellar masses in units of $\Msun/100$.  The solid
curve delineates the geometrical completeness limit and the long-dashed curve,
the SVOC limit (cf.~panel a).  The vertical short-dashed lines have been
adjusted from their positions in Figure~\ref{fig_app_au_limits} to correspond to the expected range of semi-major axes, 28--1590~AU, corresponding to the 22--1262~AU projected separations probed by the survey.
\label{fig_app_mass_incompl}}
\end{figure}

In closing, we recall that because the physical association status of a large fraction (31.4\%)
of candidate companions discovered in the survey remains undecided
(\S~\ref{sec_phys_assoc}), it is possible that more
bona fide substellar companions may be confirmed in this data set in the
future.  This is not very likely, given that the vast majority of the undecided
candidates are faint, reside in relatively high-density fields, and are at wide
angular separations from their candidate primaries (Fig.~\ref{fig_dmag_sep}),
i.e., they have very high probabilities of being background stars.
Because of the presently unknown and likely unimportant nature of the additional candidate companions, and for the sake of preserving statistical rigor, we have assumed that none of the remaining candidates are bona fide brown dwarfs, and that the derived
value of 62\% provides an accurate estimate of the completeness of our deep survey.  
%This is the incompleteness estimate that we have used to estimate the substellar companion frequency in \S~\ref{sec_anal}.

%\bibliography{/Users/Metchev/Work/DOCS/BIBLIOGRAPHY/adaptive_optics}

\clearpage

%

\clearpage

%\begin{figure}
%\plotone{corona.ps}
%\figcaption{Close-up images of a star behind the coronagraph, using: (a) the big Lyot stop and 0$\farcs$97 spot in PHARO, (b) the medium Lyot stop and 0$\farcs$97 spot in PHARO, and (c) the inscribed circle pupil mask and 1$\farcs$0 coronagraph in NIRC2.  Characteristic features of the PALAO PSF (e.g., a four-corner ``waffle'' pattern) and of the Keck AO PSF (diffraction spikes due to the secondary support structure) are indicated with arrows.  The medium Lyot stop in PHARO allows the formation of a Poisson spot at the location of the star, enabling consistent image registration.  The NIRC2 coronagraph is partially transmissive, and also allows image registration on the star. \label{fig_corona}}
%\end{figure}

\clearpage
\thispagestyle{empty}
\clearpage

\epsscale{1.0}

%\begin{figure}
%\plottwo{scopms214b_LT_dusty.eps}{scopms214b_LT_burrows97.eps}
%\figcaption{Luminosity---effective temperature diagrams for the low-mass companion ScoPMS~214B with evolutionary tracks from \citet[][a]{chabrier_etal00} and \citet[][b]{burrows_etal97} overlaid.  In each panel the thick solid box delimits the locus of the bolometric luminosity and the spectroscopically inferred effective temperature of ScoPMS~214B.  The thin solid and dashed lines are tracks of constant ages and mass.  The hatched regions mark the part of the luminosity-temperature locus that is consistent with the 5--10~Myr age of the Upper Scorpius association.
%XX switch axes and make like an H-R diagram.
%\label{fig_scopms214b_mass}}
%\end{figure}

%\begin{figure}
%\plottwo{q_hist_dndm_triple.eps}{cmf_logdndlogm.eps}
%\figcaption{{\bf(a)} Distribution ($d N/d q$) of companion mass ratios $q$ among the 30 $\leq0.5$~Gyr-old binaries in our augmented deep sample (\S~\ref{sec_cmf}).  The dotted histogram traces the observed data, while the solid histogram delineates the incompleteness-corrected data.  The solid points are data from \citet{duquennoy_mayor91}.  The long-dashed Gaussian curve ($\mu=0.23\Msun, \sigma=0.42\Msun$) is one of the preferred fits to the 0.1--1.3~\Msun\ field MF from \citet{kroupa_etal90}.  The short-dashed line is an updated fit to the 0.01--1.0~\Msun\ field MF from \citet{chabrier03}: a log-normal function with $\mu=0.08\Msun$ and $\sigma=0.69$ (in logarithmic mass units), normalized to our data.  {\bf (b)} Distribution ($\log (d N/d \log M_2)$) of the secondary masses.  The data are nearly identical to the mass ratio distribution in panel (a) because of the predominance of $\approx1\Msun$ stars in our sample.  In addition to the MFs from panel (a), we have shown a power-law fit (dot-dashed line) to the data.  The best-fit value of the index is $x=-0.7\pm0.5$ (2$\sigma$ limit).  The Salpeter index in these units is 1.35.
%\label{fig_cmf}}
%\end{figure}

\end{document}